\documentclass[longauth]{aa} 

\usepackage{booktabs}
\usepackage{graphicx}
\usepackage{lscape}
\usepackage{multirow}
\usepackage{xcolor}
\usepackage{graphicx}
\usepackage{ftnxtra}
\usepackage{fnpos}
\usepackage{txfonts}
\usepackage{soul}
\setstcolor{red}
\usepackage{supertabular}

\usepackage{natbib,twoopt}
\usepackage[colorlinks, citecolor=blue, linkcolor=blue]{hyperref}
\makeatletter
\renewcommand*\aa@pageof{, page \thepage{} of \pageref*{LastPage}}
    \def\instrefs#1{{\def\scsep{\def\scsep{,}}\@for\w:=#1\do{\scsep\ref{inst:\w}}}}
    \renewcommand{\inst}[1]{\unskip$^{\instrefs{#1}}$}
\makeatother

\begin{document}

   \title{The CARMENES search for exoplanets around M dwarfs}
   \subtitle{Guaranteed time observations Data Release 1 (2016-2020)\thanks{Full Tables 1 and 2 are only available at the CDS via anonymous ftp to \url{cdsarc.u-strasbg.fr (130.79.128.5) or via http://cdsarc.u-strasbg.fr/viz-bin/cat/J/A+A/}} }

   \author{I.~Ribas\inst{ice,ieec} \and             
   A.~Reiners\inst{iag} \and                        
   M.~Zechmeister\inst{iag}                         
   J.\,A.~Caballero\inst{cabesac,lsw} \and          
   J.\,C.~Morales\inst{ice,ieec} \and               
   S.~Sabotta\inst{lsw,tls} \and                    
   D.~Baroch\inst{ice,ieec} \and                    
   P.\,J.~Amado\inst{iaa} \and                      
   A.~Quirrenbach\inst{lsw} \and                    
   M.~Abril\inst{iaa} \and                          
   J.~Aceituno\inst{caha,iaa} \and                  
   G.~Anglada-Escud\'e\inst{ice,ieec} \and          
   M.~Azzaro\inst{caha} \and                        
   D.~Barrado\inst{cabesac} \and                    
   V.\,J.\,S.~B\'ejar\inst{iac,ull} \and            
   D.~Ben\'itez~de~Haro\inst{caha} \and             
   G.~Bergond\inst{caha} \and                       
   P.~Bluhm\inst{lsw} \and                          
   R.~Calvo~Ortega\inst{iaa} \and                   
   C.~Cardona~Guill\'en\inst{iac,ull} \and          
   P.~Chaturvedi\inst{tls} \and                     
   C.~Cifuentes\inst{cabesac} \and                  
   J.~Colom\'e\inst{ice,ieec} \and                  
   D.~Cont\inst{iag} \and                           
   M.~Cort\'es-Contreras\inst{cabesac} \and         
   S.~Czesla\inst{tls,hs} \and                      
   E.~D\'iez-Alonso\inst{oviedo,ucm} \and           
   S.~Dreizler\inst{iag} \and                       
   C.~Duque-Arribas\inst{ucm} \and                  
   N.~Espinoza\inst{stsci,johns,mpia} \and          
   M.~Fern\'andez\inst{iaa} \and                    
   B.~Fuhrmeister\inst{hs} \and                     
   D.~Galad\'i-Enr\'iquez\inst{caha} \and           
   A.~Garc\'ia-L\'opez\inst{isdefe,cabesac} \and           
   E.~Gonz\'alez-\'Alvarez\inst{cabinta} \and       
   J.\,I.~Gonz\'alez~Hern\'andez\inst{iac,ull} \and 
   E.\,W.~Guenther\inst{tls} \and                   
   E.~de~Guindos\inst{caha} \and                    
   A.\,P.~Hatzes\inst{tls} \and                     
   Th.~Henning\inst{mpia} \and                      
   E.~Herrero\inst{ice,ieec} \and                   
   D.~Hintz\inst{tucson-LPL,hs} \and                
   \'A.\,L.~Huelmo\inst{iaa} \and                   
   S.\,V.~Jeffers\inst{mps,iag} \and                
   E.\,N.~Johnson\inst{iag,mps} \and                
   E.~de~Juan\inst{caha} \and                       
   A.~Kaminski\inst{lsw} \and                       
   J.~Kemmer\inst{lsw} \and                         
   J.~Khaimova\inst{iag} \and                       
   S.~Khalafinejad\inst{lsw,mpia} \and              
   D.~Kossakowski\inst{mpia} \and                   
   M.~K\"urster\inst{mpia} \and                     
   F.~Labarga\inst{ucm} \and                        
   M.~Lafarga\inst{warwick,ice,ieec} \and           
   S.~Lalitha\inst{birmingham,hs} \and              
   M.~Lamp\'on\inst{iaa} \and                       
   J.~Lillo-Box\inst{cabesac} \and                  
   N.~Lodieu\inst{iac,ull} \and                     
   M.\,J.~L\'opez~Gonz\'alez\inst{iaa} \and         
   M.~L\'opez-Puertas\inst{iaa} \and                
   R.~Luque\inst{chicago,iac,ull,iaa} \and          
   H.~Mag\'an\inst{iaa,caha} \and                   
   L.~Mancini\inst{roma,torino,mpia} \and           
   E.~Marfil\inst{iac,ull,cabesac,ucm} \and         
   E.\,L.~Mart\'in\inst{iac,ull,csic} \and          
   S.~Mart\'in-Ruiz\inst{iaa} \and                  
   K.~Molaverdikhani\inst{lmu,origins,lsw,mpia} \and
   D.~Montes\inst{ucm} \and                         
   E.~Nagel\inst{hs,tls} \and                       
   L.~Nortmann\inst{iag,iac,ull} \and               
   G.~Nowak\inst{iac,ull} \and                      
   E.~Pall\'e\inst{iac,ull} \and                    
   V.\,M.~Passegger\inst{iac,ull,hs,oklahoma} \and  
   A.~Pavlov\inst{mpia} \and                        
   S.~Pedraz\inst{caha} \and                        
   V.~Perdelwitz\inst{kimmel,hs} \and               
   M.~Perger\inst{ice,ieec} \and                    
   A.~Ram\'on-Ballesta\inst{iaa} \and               
   S.~Reffert\inst{lsw} \and                        
   D.~Revilla\inst{iaa,ucm} \and                    
   E.~Rodr\'iguez\inst{iaa} \and                    
   C.~Rodr\'iguez-L\'opez\inst{iaa} \and            
   S.~Sadegi\inst{lsw} \and                         
   M.\,\'A.~S\'anchez~Carrasco\inst{iaa} \and       
   A.~S\'anchez-L\'opez\inst{leiden,iaa} \and       
   J.~Sanz-Forcada\inst{cabesac} \and               
   S.~Sch\"afer\inst{iag} \and                      
   M.~Schlecker\inst{tucson-AS,mpia} \and           
   J.\,H.\,M.\,M.~Schmitt\inst{hs} \and             
   P.~Sch\"ofer\inst{iaa,iag} \and                  
   A.~Schweitzer\inst{hs} \and                      
   W.~Seifert\inst{lsw} \and                        
   Y.~Shan\inst{oslo,iag} \and                      
   S.\,L.~Skrzypinski\inst{ucm} \and                
   E.~Solano\inst{cabesac} \and                     
   O.~Stahl\inst{lsw} \and                          
   M.~Stangret\inst{padova,iac,ull} \and            
   S.~Stock\inst{lsw} \and                          
   J.~St\"urmer\inst{lsw,chicago} \and              
   H.\,M.~Tabernero\inst{cabinta} \and              
   L.~Tal-Or\inst{ariel,iag} \and                   
   T.~Trifonov\inst{mpia,sofia} \and                
   S.~Vanaverbeke\inst{astrolab,vvs,leuven} \and    
   F.~Yan\inst{hefei,iag} \and                      
   M.\,R.~Zapatero~Osorio\inst{cabinta}             
          }
   \authorrunning{I. Ribas et al.}
   \titlerunning{CARMENES Data Release 1}
   \institute{\label{inst:ice}Institut de Ci\`encies de l'Espai (ICE, CSIC),
            Campus UAB, c/~Can Magrans s/n, E-08193 Bellaterra, Barcelona, Spain\\
            \email{iribas@ice.cat}
         \and
            \label{inst:ieec}Institut d'Estudis Espacials de Catalunya (IEEC), c/ Gran Capit\`a 2-4, E-08034 Barcelona, Spain
         \and
            \label{inst:iag}Institut f\"ur Astrophysik und Geophysik, Georg-August-Universit\"at, Friedrich-Hund-Platz 1, D-37077 G\"ottingen, Germany
        \and
            \label{inst:cabesac}Centro de Astrobiolog\'ia (CAB), CSIC-INTA, Campus ESAC, Camino Bajo del Castillo s/n, E-28692 Villanueva de la Ca\~nada, Madrid, Spain
        \and
            \label{inst:lsw}Landessternwarte, Zentrum f\"ur Astronomie der Universit\"at Heidelberg, K\"onigstuhl 12, D-69117 Heidelberg, Germany
        \and
            \label{inst:tls}Th\"uringer Landessternwarte Tautenburg, Sternwarte 5, D-07778 Tautenburg, Germany
        \and
            \label{inst:iaa}Instituto de Astrof\'{\i}sica de Andaluc\'{\i}a (IAA-CSIC), Glorieta de la Astronom\'{\i}a s/n, E-18008 Granada, Spain
        \and
            \label{inst:caha}Centro Astron\'omico Hispano en Andaluc\'ia (CAHA), Observatorio de Calar Alto, Sierra de los Filabres, E-04550 G\'ergal, Almer\'ia, Spain
         \and
            \label{inst:iac}Instituto de Astrof\'isica de Canarias (IAC), E-38200 La Laguna, Te\-ne\-ri\-fe, Spain
        \and
            \label{inst:ull}Departamento de Astrof\'isica, Universidad de La Laguna, E-38206 La Laguna, Te\-ne\-ri\-fe, Spain
        \and
            \label{inst:hs}Hamburger Sternwarte, Universit\"at Hamburg, Gojenbergsweg 112, D-21029 Hamburg, Germany
        \and
            \label{inst:oviedo}Instituto Universitario de Ciencias y Tecnolog\'ias Espaciales de Asturias, c/~Independencia 13, E-33004 Oviedo, Spain
        \and
            \label{inst:ucm}Departamento de F\'isica de la Tierra y Astrof\'{i}sica and IPARCOS-UCM (Instituto de F\'{i}sica de Part\'{i}culas y del Cosmos de la UCM), Facultad de Ciencias F\'isicas, Universidad Complutense de Madrid, E-28040 Madrid, Spain
        \and
            \label{inst:stsci}Space Telescope Science Institute, 3700 San Martin Drive, Baltimore, MD 21218, United States of America
        \and
            \label{inst:johns}Department of Physics and Astronomy, Johns Hopkins University, Baltimore, MD 21218, United States of America
        \and
            \label{inst:mpia}Max-Planck-Institut f\"ur Astronomie, K\"onigstuhl 17, D-69117 Heidelberg, Germany
        \and
            \label{inst:isdefe}{Isdefe, Beatriz de Bobadilla 3, E-28040 Madrid, Spain}
         \and
            \label{inst:cabinta}Centro de Astrobiolog\'ia (CAB), CSIC-INTA, Carretera de Ajalvir km~4, E-28850 Torrej\'on de Ardoz, Madrid, Spain
       \and
            \label{inst:tucson-LPL}Lunar and Planetary Laboratory, University of Arizona, 1629 East University Boulevard, Tucson, AZ 85721, United States of America
        \and
            \label{inst:mps}Max-Planck-Institut f\"ur Sonnensystemforschung, Justus-von-Liebig-Weg 3, D-37077 G\"ottingen, Germany
        \and
            \label{inst:warwick}Centre for Exoplanets and Habitability and Department of Physics, University of Warwick, Coventry, CV4 7AL, United Kingdom
        \and
            \label{inst:birmingham}School of Physics and Astronomy, University of Birmingham, Edgbaston, Birmingham B15 2TT, United Kingdom
        \and
            \label{inst:chicago}Department of Astronomy and Astrophysics, University of Chicago, 5640 South Ellis Avenue, Chicago, IL 60637, United States of America
        \and
            \label{inst:roma}Dipartimento di Fisica, Universit\`a degli Studi di Roma ``Tor Vergata'', Via della Ricerca Scientifica 1, I-00133, Rome, Italy
        \and
            \label{inst:torino}INAF -- Osservatorio Astrofisico di Torino, via Osservatorio 20, I-10025, Pino Torinese, Italy
        \and
            \label{inst:csic}Consejo Superior de Investigaciones Cient\'ificas, E-28006 Madrid, Spain
        \and
            \label{inst:lmu}Universit\"ats-Sternwarte, Ludwig-Maximilians-Universit\"at M\"unchen, Scheinerstrasse 1, D-81679 M\"unchen, Germany
        \and
            \label{inst:origins}Exzellenzcluster Origins, Boltzmannstrasse 2, D-85748 Garching, Germany
        \and
            \label{inst:oklahoma}Homer\,L. Dodge Department of Physics and Astronomy, University of Oklahoma, 440 West Brooks Street, Norman, OK 73019, United States of America
        \and
            \label{inst:kimmel}Kimmel fellow, Helen Kimmel Center for Planetary Science, Weizmann Institute of Science, Rehovot, Israel
        \and
            \label{inst:leiden}Leiden Observatory, Leiden University, Postbus 9513, 2300 RA, Leiden, The Netherlands
        \and
            \label{inst:tucson-AS}Department of Astronomy/Steward Observatory, The University of Arizona, 933 North Cherry Avenue, Tucson, AZ 85721, United States of America
        \and
            \label{inst:oslo}Centre for Earth Evolution and Dynamics, Department of Geosciences, Universitetet i Oslo, Sem S{\ae}lands vei 2b, 0315 Oslo, Norway
        \and
            \label{inst:padova}INAF -- Osservatorio Astronomico di Padova, Vicolo dell’Osservatorio 5, 35122 Padova, Italy
        \and
            \label{inst:ariel}Department of Physics, Ariel University, Ariel 40700, Israel
        \and
            \label{inst:sofia}Departament of Astronomy, Sofijski universitet ,,Sv. Kliment Ohridski'', 5 James Bourchier Boulevard, 1164 Sofia, Bulgaria
        \and
            \label{inst:astrolab}AstroLAB IRIS, Provinciaal Domein ``De Palingbeek'', Verbrandemolenstraat 5, 8902 Zillebeke, Ieper, Belgium
        \and
            \label{inst:vvs}Vereniging Voor Sterrenkunde, Oude Bleken 12, 2400 Mol, Belgium
        \and
            \label{inst:leuven}Centre for Mathematical Plasma Astrophysics, Katholieke Universiteit Leuven, Celestijnenlaan 200B, bus 2400, B-3001 Leuven, Belgium
        \and
            \label{inst:hefei}Department of Astronomy, University of Science and Technology of China, Hefei 230026, China
            }

   \date{Received 4 September 2022 / Accepted dd Month 2022}

  \abstract
   {The CARMENES instrument, installed at the 3.5 m telescope of the Calar Alto Observatory in Almer\'ia, Spain,
   was conceived to deliver high-accuracy radial velocity (RV) measurements with long-term stability to search
   for temperate rocky planets around a sample of nearby cool stars. Moreover, the broad wavelength coverage
   was designed to provide a range of stellar activity indicators to assess the nature of potential RV signals
   and to provide valuable spectral information to help characterise the stellar targets.}
   {We describe the CARMENES guaranteed time observations (GTO), spanning from 2016 to 2020, during which
   19\,633 spectra for a sample of 362 targets were collected. We present the CARMENES Data Release 1 (DR1), which
   makes public all observations obtained during the GTO of the CARMENES survey. }
   {The CARMENES survey target selection was aimed at minimising biases, and about 70\,\% of all known M
   dwarfs within 10\,pc and accessible from Calar Alto were included. The data were pipeline-processed, and
   high-level data products, including 18\,642 precise RVs for 345 targets, were derived.
   Time series data of spectroscopic activity indicators were also obtained.}
   {We discuss the characteristics of the CARMENES data, the statistical properties of the stellar sample, and the
   spectroscopic measurements. We show examples of the use of CARMENES data and provide a contextual view of
   the exoplanet population revealed by the survey, including 33 new planets, 17 re-analysed planets, and 26
   confirmed planets from transiting candidate follow-up. A subsample of 238 targets was used to derive updated
   planet occurrence rates, yielding an overall average of $1.44\pm0.20$ planets with 1\,M$_\oplus
   < M_\mathrm{pl} \sin i < 1000$\,M$_\oplus$ and 1\,d $< P_\mathrm{orb} < 1000$\,d per star, and indicating that
   nearly every M dwarf hosts at least one planet. All the DR1 raw data, pipeline-processed data, and high-level
   data products are publicly available online.}
   {CARMENES data have proven very useful for identifying and measuring planetary companions. They are
   also suitable for a variety of additional applications, such as the determination of stellar fundamental
   and atmospheric properties, the characterisation of stellar activity, and the study of exoplanet
   atmospheres.}


   \maketitle

\section{Introduction} \label{sec:introduction}

M-type dwarfs provide some advantages with respect to Sun-like stars in the search for exoplanets,
particularly those with low masses. Their relatively small sizes and masses result in stronger planetary
signals. Furthermore, their low intrinsic luminosities imply that temperate planets orbiting within their
liquid-water habitable zone have shorter orbital periods, of the order of tens of days \citep{Kopparapu2013}.
In addition, they constitute an abundant stellar population, comprising the majority of stars (78.5\,\%) in the
solar neighbourhood \citep{Reyle2021A&A...650A.201R}. The main drawbacks of M dwarfs as targets for
exoplanet searches are their intrinsic faintness and the fact that a relatively large fraction of them show
magnetic activity phenomena, especially the later spectral types \citep{Reiners2012}. A number of efforts have
successfully exploited the so-called M-dwarf opportunity for planet detection over the past few decades
\citep[e.g.][]{Delfosse1999,Endl2003,Wright2004,Bonfils2005,Nutzmann2008,Zechmeister2009,Johnson2010,
Ricker2015,Affer2016,Seifahrt2018,Bayliss2018}.

The CARMENES\footnote{Calar Alto high-Resolution search for M dwarfs with Exoearths with Near-infrared and
optical \'Echelle Spectrographs; \url{https://carmenes.caha.es}.} instrument and survey were specifically
conceived to search for temperate rocky planets around a sample of nearby cool stars \citep{Quirrenbach2014}.
The spectrograph was designed to provide high-accuracy radial velocity (RV) measurements with long-term
stability in a broad wavelength interval where M-dwarf stars have the peak of their spectral energy
distribution. Moreover, such wide coverage provides a range of stellar activity indicators to assess
the nature of potential RV signals as well as valuable spectral information that can be used to characterise
the stellar targets.

The CARMENES instrument is installed at the 3.5 m telescope of the Calar Alto Observatory in Almer\'{\i}a,
Spain (37$^\circ$13$\arcmin$25$\arcsec$N, 2$^\circ$32$\arcmin$46$\arcsec$W). It provides nearly continuous
wavelength coverage from 520\,nm to 1710\,nm from its two channels: the visual channel (VIS), with a
spectral resolution of $R = 94\,600$, covers the range $\lambda$ = 520--960\,nm, while the near-infrared
channel (NIR) yields a resolution of $R = 80\,400$ within a wavelength interval $\lambda$ = 960--1710\,nm
\citep{Quirrenbach2016}. Both channels are coupled to the telescope by optical fibres, with a projection
of $1\farcs5$ on the sky.

A sample of about 350 M dwarfs across all M spectral subtypes comprises the targets of the main survey.
A total of 750 useful nights were reserved as guaranteed time observations (GTO) for the CARMENES consortium,
and these ran for five years, from 1 January 2016 to 31 December 2020.

The present publication accompanies the release of the observations acquired with the CARMENES VIS channel
over the course of the RV survey within the GTO programme, which we have dubbed the CARMENES
Data Release 1 (DR1). This includes raw data, calibrated spectra, and high-level data products, such as RVs
and  spectroscopic indicators. The paper is structured as follows. Section~\ref{sec:survey} describes the
design and execution of the CARMENES survey. In Sect.~\ref{sec:sample} we present the CARMENES GTO target
sample and provide a description of its statistical distribution. Section~\ref{sec:obs} describes the
observations collected within the GTO and the processing data flow from raw frames to calibrated RVs and
ancillary data products. In Sect.~\ref{sec:results} we discuss the properties of the CARMENES DR1 regarding
internal and external precision, we provide information regarding the presence of periodic signals in the
data, and we present and discuss the sample of exoplanets in the surveyed targets. Furthermore, we present
revised planet occurrence rates considering all publicly released data. Finally, Sect.~\ref{sec:conclusions}
provides the summary and conclusions of the work.

\section{The CARMENES survey} \label{sec:survey}

The initial goal of the GTO survey was to collect approximately 70 spectra for each of the foreseen 300 targets
\citep{Garcia-Piquer2017A&A...604A..87G}, which would have yielded a grand total of $\sim$21\,000 spectra.
During the survey, we identified a number of targets with high-amplitude RV variations (RV scatter
$>$10\,m\,s$^{-1}$ and $v \sin i >$ 2\,km\,s$^{-1}$), which we classified as RV-loud \citep{TalOr2018}.
For each of them, we obtained about 11 observations and monitoring was subsequently discontinued. A
similar approach was followed for spectroscopic binaries, for which we acquired a number of measurements
just enough to derive reliable orbital solutions \citep{Baroch2018,Baroch2021A&A...653A..49B}. For some
of the binaries with the longest periods, however, monitoring at very low cadence has been extended over
time to constrain better the orbital and physical parameters of the components.

Despite the discontinued targets, some time into the survey it was realised that reaching 70
observations per star would not be possible, mostly because of the large number of measurements
needed to characterise newly discovered exoplanets as a consequence of the measured astrophysical
jitter and also because of the telescope and instrument overhead times being somewhat longer than
initially considered. Furthermore, with the launch of the Transiting Exoplanet Survey Satellite
(TESS) mission in 2018 \citep{Ricker2015}, the CARMENES Consortium agreed to invest approximately
50 useful GTO nights in following up TESS transiting planet candidates with M-dwarf hosts
(CARMENES-TESS follow-up programme). As a consequence of the new circumstances, it was decided
that the survey should aim at acquiring a minimum of 50 observations per target, which would yield
plenty of planet detections and provide meaningful constraints on planet occurrence rates. At the
same time, we redefined the relative priorities of the sample to favour stars of spectral type
M4\,V and later to exploit optimally the CARMENES capabilities in a relatively unexplored range
of stellar host masses. Such a decision implied that the faint end of the M2\,V and M3\,V targets
in the sample would have lower chances of being scheduled because of the employed criteria
\citep{Garcia-Piquer2017A&A...604A..87G}.

\begin{table*}[!t]
    \centering
    \caption{\label{tab:props} Basic properties and number of measurements for the CARMENES DR1 target sample.}
    \begin{tabular}{@{}llcccccclll@{}}
        \hline
        \hline
        \noalign{\smallskip}
Karmn & Star name & $N_{\rm AVC}$  & $N_{\rm RVC}$ & $M$ (M$_{\odot}$) & $R$ (R$_{\odot}$) & $T_{\rm eff}$ (K) &
$P_{\rm rot}$ (d) & Ref. $P_{\rm rot}$ & Survey & Comments\\
        \noalign{\smallskip}
        \hline
        \noalign{\smallskip}
J00051+457  &                     GJ 2  &   52  &   53  & 0.49 & 0.49 & 3773 &  15.4 &     DA19 &    GTO \\
J00067-075  &                  GJ 1002  &   89  &   91  & 0.11 & 0.12 & 3169 &       &          &    GTO \\
J00162+198E  &              LP 404-062  &   18  &   18  & 0.27 & 0.28 & 3329 & 105.0 &     DA19 &    GTO \\
J00183+440  &                   GX And  &  216  &  223  & 0.39 & 0.40 & 3603 &  45.0 &     SM18 &    GTO \\
J00184+440  &                   GQ And  &  193  &  196  & 0.16 & 0.18 & 3318 &       &          &    GTO \\
J00286-066  &                  GJ 1012  &   50  &   53  & 0.34 & 0.35 & 3419 &       &          &    GTO \\
J00389+306  &                Wolf 1056  &   58  &   60  & 0.41 & 0.41 & 3551 &  50.2 &     DA19 &    GTO \\
J00403+612  &  2MASS J00402129+6112490  &   40  &   41  & 0.47 & 0.47 & 3709 &       &          &   TESS \\
J00570+450  &                G 172-030  &   16  &   16  & 0.33 & 0.34 & 3488 &       &          &    GTO \\
J01013+613  &                    GJ 47  &   10  &   10  & 0.37 & 0.37 & 3564 &  34.7 &     SM18 &    GTO \\        \noalign{\smallskip}
        \hline
    \end{tabular}
     \tablefoot{This is a sample list. The full
table can be downloaded from CDS via \url{http://cdsarc.u-strasbg.fr/viz-bin/cat/J/A+A/}.}
    \tablebib{
    DA19: \citet{DiezAlonso2019},
    SM18: \citet{Suarez2018}.
}
\end{table*}

\begin{figure*}
    \centering
    \includegraphics[width=\linewidth]{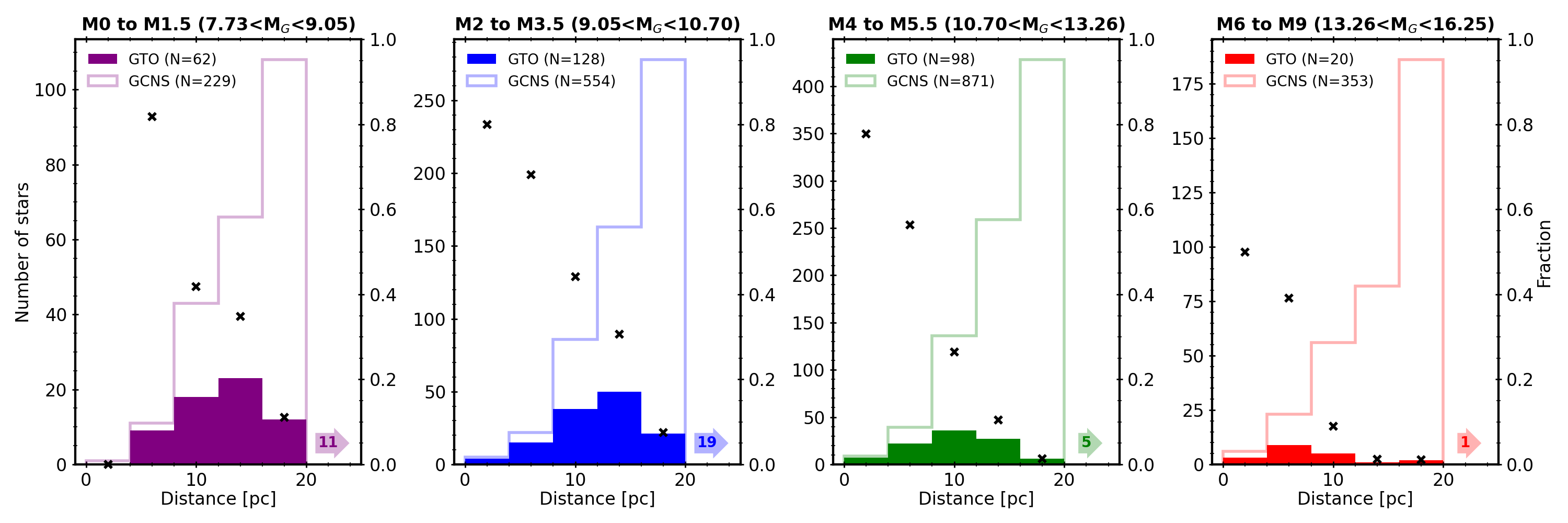}
    \caption{Distribution of the CARMENES GTO target sample (excluding the SB2 and ST3 systems) as a function
    of distance ($d <$ 20\,pc) for different spectral types or absolute {\it Gaia} $G$-band intervals. Some
    stars in the sample are at greater distances, and this number is provided inside the right-pointing arrow.
    One of the targets at the K--M spectral type boundary has an $M_G$ value below 7.73\,mag and is not
    included; hence, the total number of stars plotted is 344. The distance distribution of the GCNS for the
    same intervals is also shown, and the ratios between the two are depicted as black crosses with the scale
    in the right $y$ axis.}
    \label{fig:completeness}
\end{figure*}

At the end of the GTO survey in 2020, the minimum number of 50 measurements had not been reached for
all surveyed targets. The CARMENES DR1 therefore contains unequal number of observations, with a median
of 30 observations per star. However, some of the targets, such as RV standards and stars with suggestive
planetary signals, were observed up to a few hundred times. About two thirds of the targets have time
series of at least three years, and for almost half of the targets the observations cover at least four years.
Only 10\% of the targets are observed for less than a year. The cadence is random and non-uniform, not
only because of observability but also for scientific reasons (e.g. priority increased when a planet
candidate signal required more detailed sampling). In 2020, a proposal was submitted to the competitive
Calar Alto Legacy projects call, and an additional 300 nights were awarded to the CARMENES Consortium
to complete the survey during 2021--2023 and, hence, fulfil the goal of attaining at least 50
observations per target.

\section{CARMENES GTO sample}\label{sec:sample}

The CARMENES GTO sample of M dwarfs is generally composed of the brightest stars of every spectral
subtype that are visible from Calar Alto ($\delta > -23^\circ$), as described in \citet{Alonso2015}.
Effectively, this means that about 70\,\% of the full sky is observable by the CARMENES survey.
We only excluded stars that are known members of visual binaries at separations closer than 5\,$\arcsec$.
We explicitly did not bias our sample with regard to age, metallicity, or magnetic activity, nor did
we exclude stars with planets that were already known. More information on the selection criteria was
provided by \citet[hereafter Rei18b]{Reiners2018} and references therein. The sample described by Rei18b
was composed of 324 stars. Throughout the survey we added nine additional targets as a result of
supervening circumstances such as new exoplanet announcements, interesting targets (e.g. in the TESS
continuous viewing zone), and revised spectroscopic classification. Furthermore, we added 18 targets
from the CARMENES-TESS follow-up programme. As opposed to Rei18b, we also included in our current
analysis double- and triple-line spectroscopic binaries and triples (SB2 and ST3, respectively) and
some visual binaries. We have found 17 of such binaries in the sample, 11 of which are new additions
to Rei18b, but six were present there because they had not yet been identified as SB2, ST3, or visual
binaries.

Table~\ref{tab:props} presents a selection of relevant properties of the 362 targets in the CARMENES
GTO sample. The different columns list basic stellar parameters ($M$, $R$, $T_{\rm eff}$), rotation
periods ($P_{\rm rot}$), and the number of measurements in the release, both in the form of
pipeline-produced RVs ($N_{\rm RVC}$) and zero-point-corrected RVs ($N_{\rm AVC}$). Descriptions of
these two data products are provided in Sect.~\ref{sec:obs}. The basic stellar parameters were taken
from the latest version of Carmencita, which is the CARMENES input catalogue \citep{Caballero2016},
and from the series of papers on the characterisation of the CARMENES GTO sample
\citep{Alonso2015,Cortes2017,Jeffers2018,DiezAlonso2019,Cifuentes2020A&A...642A.115C,Perdelwitz2021A&A...652A.116P}.
In the case of targets where more than one set of lines are visible in the spectra (SB2, ST3, and visual
binaries), the basic parameters are not listed (as they are ill-defined) and the column $N_{\rm RVC}$
provides the total number of CARMENES observations released. The penultimate column indicates if the target
is part of the blind GTO survey or if it is a TESS exoplanet candidate. An asterisk marks targets already
tabulated by Rei18b. We are not discussing here the statistical distribution of the target sample regarding
brightness and spectral type. The general properties are equivalent to those in Figs.~2 and~3 of Rei18b,
which already comprised most of the sample presented here ($>$90\,\%).

The volume completeness of the CARMENES GTO sample can be investigated by comparing distance distributions
with the {\em Gaia} Catalogue of Nearby Stars \citep[GCNS;][]{Gaia2021A&A...649A...6G}, which is assumed to
be complete at the brightness cuts and spectral types of interest. In Fig.~\ref{fig:completeness}, we show a
collection of histograms as a function of distance out to 20\,pc for several spectral type and {\em Gaia}
$G$-band absolute magnitude ($M_G$) intervals. To allow for a comparison, spectral types of GCNS stars were
estimated from $M_G$ following the corresponding relationship by \citet{Cifuentes2020A&A...642A.115C}. The
ratio between the number of stars in the CARMENES sample and the number of known stars in the GCNS is also
shown. The ratio, that is, the sample completeness, decreases with M subtype (from 27\,\% for early Ms to 6\,\%
for late Ms), as expected due to brightness limitations. The global completeness of the CARMENES sample at
20\,pc, including all spectral types, is 15\,\%. If we consider distances to 10\,pc, then the ratio of sample
stars to known stars exceeds 50\,\% in all intervals except for the latest Ms, where the ratio is 28\,\%.
Altogether, the CARMENES GTO sample contains nearly half (48\,\%) of all known M dwarfs within 10\,pc of the
Sun \citep{Reyle2021A&A...650A.201R}, and about 70\,\% of those accessible from the Calar Alto Observatory.
Most nearby M dwarfs that are not in the sample have close companions at less than 5\,$\arcsec$.

\section{Observations} \label{sec:obs}

The observations of the CARMENES GTO survey were collected in a signal-to-noise (S/N) limited fashion.
That is, using the number of counts from the exposure meter of the NIR channel ($c_{\rm EM}$) and a
calibrated relationship  -- S/N $\propto \sqrt{c_{\rm EM}}$, -- the integration was continued until reaching
a S/N ratio of 150 at order 50 ($\sim$1200\,nm) of the CARMENES NIR channel, or was interrupted after an
integration of 1800\,s to avoid excessive contamination from cosmic rays and line broadening due to Earth's
rotation. According to the calculations by \citet{Reiners2020}, a spectrum with S/N = 150 at 1200\,nm for
an early- to mid-type  M dwarf produces a typical uncertainty of 1\,m\,s$^{-1}$ in RV from photon shot
noise, which was the required value for the survey.

A total of 19\,633 spectra were acquired as part of the GTO programme. However, a small fraction of them do
not have sufficient quality for precise RV work and were not considered in our subsequent analysis. They were
flagged by the processing pipeline because of low S/N, high S/N implying saturation risk, contamination by
twilight, Moon, or stray light. The total number of spectra yielding useful RV measurements is 19\,161. The
discarded 472 spectra are still accessible from the Calar Alto
archive\footnote{\url{http://caha.sdc.cab.inta-csic.es/calto}.} in raw format but are not part of the CARMENES
DR1.

The processing of the data was done automatically with a pipeline, including the reduction of raw frames,
the extraction and calibration of spectra, the determination of RVs using a template-matching algorithm, and
the calculation of cross-correlation function (CCF) products. Full details on the applied procedure are
provided below. The data for SB2 and ST3 targets were only processed up to the extraction and calibration of
spectra, and were not analysed to determine precise RVs because our methodology is not suitable when more than
one set of stellar lines is present in the spectra. Finally, we provide the full set of data products for
18\,642 out of the 19\,161 good spectra.

\subsection{Processing pipeline}

The observations were reduced with the {\tt caracal}\footnote{CARMENES Reduction And CALibration.} 
pipeline, with the data flow being described by \citet{Caballero2016}. The extraction pipeline is based on the
{\tt reduce} package of \citet{Piskunov2002A&A...385.1095P} but many routines have been revised.
In particular, we developed the flat-relative optimal extraction \citep[FOX,][]{Zechmeister2014A&A...561A..59Z}
and wavelength calibration scripts, which combine spectra from hollow-cathode lamps (HCLs) and Fabry-P\'erot
(F-P) \'etalons \citep{Bauer2015A&A...581A.117B}. The data release in this work is based on {\tt caracal~v2.20}.

\subsection{Radial velocities}

The RVs for the CARMENES DR1 were computed with {\tt serval}\footnote{SpEctrum Radial Velocity
AnaLyser. \url{https://github.com/mzechmeister/serval}. Based on the version committed on
2022\nobreakdash-01\nobreakdash-26 \url{https://github.com/mzechmeister/serval/tree/a348b4c}.}
\citep{Zechmeister2018} and {\tt raccoon}\footnote{Radial velocities and Activity indicators from
Cross-COrrelatiON with masks. \url{https://github.com/mlafarga/raccoon}.} \citep{Lafarga2020A&A...636A..36L}.
Both software packages were specifically developed for data coming from the CARMENES instrument, although they
can process spectroscopic data from other precise RV instruments as well
\citep[e.g.][]{Stefansson2020,Hoyer2021,Wang2022,Turtelboom2022}.

The {\tt serval} code implements a data-driven approach, where both RVs and templates are derived from the
observations themselves via a least-squares fitting procedure similar to the Template-Enhanced Radial velocity
Re-analysis Application \citep[{\tt TERRA};][]{Anglada2012}. The co-adding is performed by cubic B-spline
regression. For the barycentric correction, the default option is the Python implementation
{\tt barycorrpy}\footnote{\url{https://github.com/shbhuk/barycorrpy}.}
\citep{Wright2014PASP..126..838W,Kanodia2018RNAAS...2....4K}.
The RVs for each spectral order are produced, and the global RV of the spectrum is subsequently computed as a
simple weighted mean over the spectral orders. By default, the ten bluest and the ten reddest spectral orders
are not used. In those regions, the instrument efficiency decreases. Furthermore, the red end is strongly
affected by telluric contamination and dichroic cutoff. For faint late M dwarfs additional blue orders may be
omitted because of low S/N. Since the present data release contains the order-wise RVs, a more sophisticated
recalculation of RV values (robust means, re-weighting using a posteriori information) employing a detailed
chromatic analysis is also possible \citep[e.g.][]{Zechmeister2019}. Finally, corrections for instrumental drift
and secular acceleration \citep{Kuerster2003} are applied to the global RV, yielding the so-called RVC (Radial
Velocity Corrected) velocities.

The RV error bar is calculated as the weighted mean of the order-wise RVs \citep[see Eq.~15 in][]{Zechmeister2018}
and takes into account photon noise, readout noise, and model mismatch. The last contribution quantifies
the difference between the spectrum and the template (caused, for example, by cosmic rays, telluric contamination
or detector artefacts), and the excess scatter of the averaged individual orders (caused, for example, by telluric
contamination affecting specific orders or a chromatic trend). Thus, formal RV uncertainties are based
on the quality of the template fit and not on estimates of any physical effects during observation or calibration
(e.g. modal noise). The CARMENES instrument was designed to minimise all such effects
\citep{Seifert2012,Sturmer2014} but, if anyway present to some extent, they will result in excess noise
(instrumental jitter).

A further RV data product is provided, namely AVC (Average Velocity Corrected) velocities. These are obtained
from RVCs by correcting for nightly zero points (NZPs; see Sect.~\ref{sec:NZP}). AVC RVs are not calculated if
no instrumental drift value is available. The total error bar of each AVC RV considers the uncertainties of
the RVC and the corrections added in quadrature. In addition to RVs, {\tt serval} provides a further set of
useful parameters. These include the chromatic index (CRX; a measure of the wavelength dependence of the RVs),
the differential line width (dLW), and spectral line indices (e.g. H$\alpha$, Ca\,{\sc i}, Ca\,{\sc ii}~IRT
(infrared triplet), and Na\,{\sc i}~D), which are valuable activity indicators \citep{Fuhrmeister2019,Schoefer2019}.
A full description of these {\tt serval} products and their calculation methodologies is provided in
\citet{Zechmeister2018}.

The {\tt raccoon} code is based on the CCF concept \citep{Baranne1996}, whereby a weighted binary mask is used to
calculate the convolution with each observed spectrum. In our implementation, we derived the mask from
the {\tt serval} template of the target itself. One of the outputs is the RVs, which are known to be less
precise than values coming from template matching for M dwarfs \citep{Perger2017}, but still allow for a
cross-check with {\tt serval}. Other relevant CCF parameters produced by {\tt raccoon} are the contrast (CON),
the full width at half maximum (FWHM), and the bisector inverse slope (BIS). These parameters can be regarded
as moments of the CCF that carry information on the characteristics of the stellar lines and, therefore, can
be used to assess variability coming from astrophysical sources. Further details can be found in
\citet{Lafarga2020A&A...636A..36L}.

The RVs in the CARMENES DR1 may differ from RVs that have appeared in previous CARMENES publications. This
is because {\tt serval} and {\tt raccoon} are steadily maintained and new upgrades are continuously made.
In addition, all parameters are recalculated when new spectra of a target are considered (i.e. producing
a new template) and, thus, slightly different values may result. Finally, the NZP corrections can vary when
new data are considered, also impacting on the final velocities. In any case, any differences with published
data are generally minor.

\subsection{Telluric contamination correction} \label{sec:tellurics}

The Earth atmosphere imprints spectral features from its molecular and atomic components (mostly H$_2$O
and O$_2$ in the VIS domain), called telluric lines (or tellurics, for short), onto the stellar spectrum.
{\tt serval} handles this contamination by simply masking telluric lines when computing the
RVs (during co-adding, telluric lines are strongly down-weighted, and severely contaminated template
regions are masked as well during RV computation). Masking lines is a straightforward and robust first-order
approach. The default mask of {\tt serval} flags regions where the telluric line depth is typically about 5\,\%
or greater. Various tests using different thresholds and resulting mask widths showed this value
to provide optimal results by trading off wavelength coverage (i.e. RV precision) and systematic effects
from telluric contamination. While the telluric mask is static in the detector frame, it moves in
the stellar rest frame because of Earth's yearly barycentric motion. To ensure that identical spectral
regions are used for RV determination throughout the observing season, an alternative approach would be to
mask out the full barycentric velocity range around each telluric feature. However, we preferred not to use
such a procedure because it significantly diminishes the available wavelength range and, thus,
the amount of RV information.

There may be cases where the residual telluric RV content (due to high airmass or micro-telluric
contamination) may still be significant. Such residuals can most likely affect cases where the RV internal
precision is very high (e.g. high S/N observations) and where the stellar RV signal is weak (e.g. fast
rotators). Residual telluric contamination can result in spurious RV periodicities, mostly yearly signals
or their aliases \citep{Damasso2022}. Hence, caution is advised in the interpretation of those typically
long-period, low-amplitude signals. Improvements may be made by re-weighting spectral orders, reprocessing
with more conservative masks, or employing a more sophisticated telluric modelling scheme
\citep[e.g.][]{Nagel2019PhD}.

\subsection{Nightly zero points} \label{sec:NZP}

\begin{figure*}
    \centering
    \includegraphics[width=\linewidth]{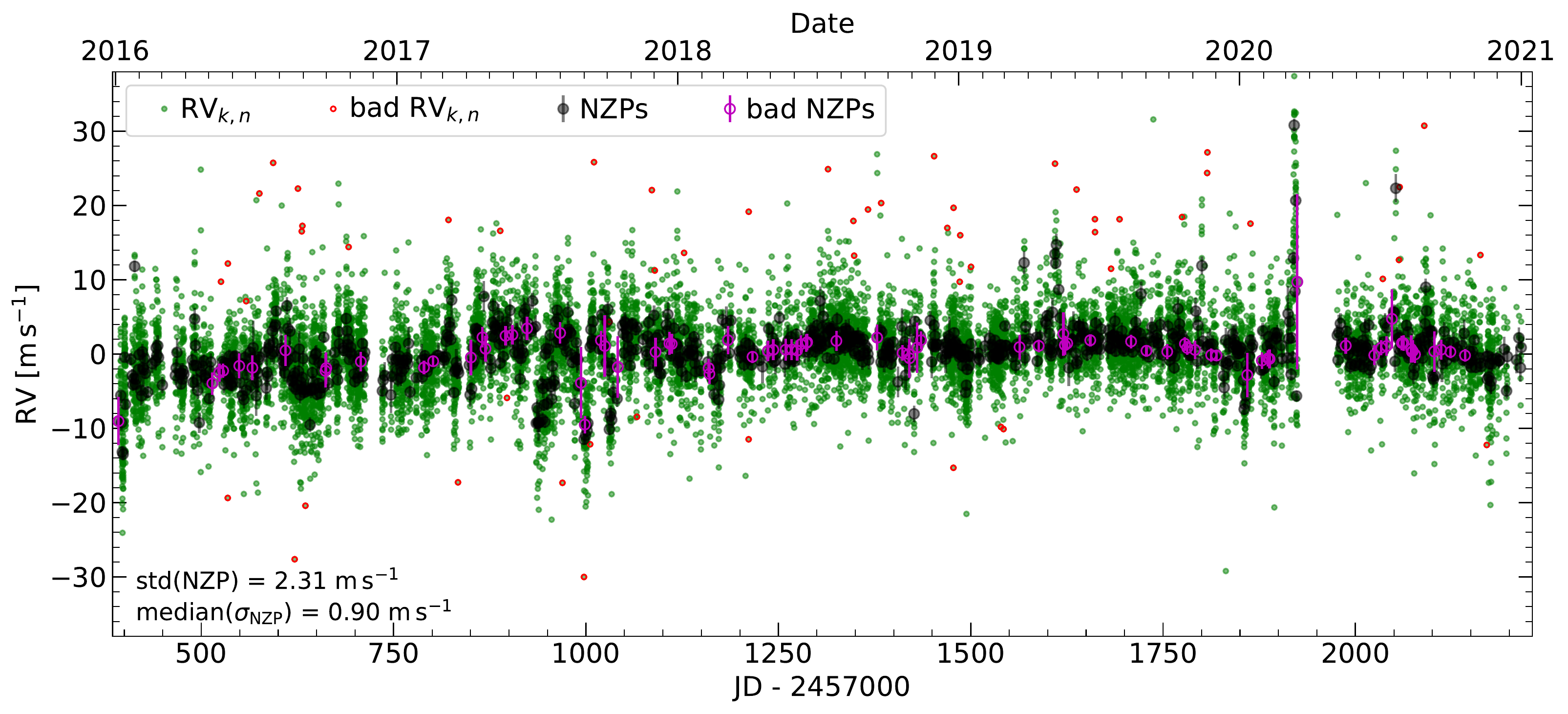}
    \caption{Nightly zero points (NZPs) of CARMENES VIS. The RVs of all RV constant stars
    (small green points) are zero-centred. A mean (the NZP, solid black circles) is computed for each night,
    in which identified outliers (small red points) are omitted. NZPs with fewer than three RV standard
    observations (open magenta circles) are replaced by a local mean.}
    \label{fig:NZP}
\end{figure*}

Although the CARMENES spectrograph is usually wavelength calibrated each afternoon and nightly instrumental
drifts are measured with the F-P \'etalon, stellar RVs from the same night often share common systematic
effects, which produce NZP offsets generally of a few m\,s$^{-1}$ with a median error bar of
0.9~m\,s$^{-1}$ (see Fig.~\ref{fig:NZP}). We employ RV-constant stars (${\rm rms} < 10$\,m\,s$^{-1}$) to
calculate NZPs, with the exact procedure being described in more detail by \citet{Trifonov2018A&A...609A.117T}.
The resulting values are subsequently subtracted from each of the {\tt serval} RV measurements. To avoid
self-biasing the measurements, the zero point of RV-constant stars is calculated by removing the target
itself from the calibration pool \citep{TalOr2019MNRAS.484L...8T}. Tests revealed that NZP-corrected
RVs improve the statistical significance of the best-fit models of CARMENES exoplanet discoveries, thus
illustrating the benefits of the correction procedure. The same algorithm was applied by
\citet{TalOr2019MNRAS.484L...8T} to archival HIRES (High Resolution Echelle Spectrometer) Keck RVs and
by \citet{Trifonov2020A&A...636A..74T} to reprocessed RVs from HARPS (High Accuracy Radial velocity Planet
Searcher) spectra. In both cases the studies revealed and corrected systematic effects in those instruments.

Table~\ref{tab:NZP} provides NZP values for all the CARMENES GTO nights. Reasons explaining the nightly
offsets can be various, including a drift of the F-P, degraded quality of aged HCLs, strong instrument
drifts during the $\sim$15 min calibration sequence (F-P and HCL calibration frames cannot be taken
simultaneously), and different injection of calibration light coupled with insufficient scrambling. We were
able to reduce some fraction of the night-to-night variability found during the initial CARMENES operations
through hardware configuration changes and by employing a different strategy when acquiring the daily
calibration sequences. As a result, the NZP scatter diminishes slightly after two years of operation
(Fig.~\ref{fig:NZP}). In addition to the night-to-night offsets, we also performed a correction for
intra-night drift. The correction was found to be significant early in the survey and related to a
temperature effect of the F-P subsystem. A hardware upgrade on 6 September 2017 (BJD 2458003) greatly
decreased the temperature coupling and eliminated the need for such a correction. In any case, the effects
of self bias and intra-night drift correction are small. Further details on the instrument performance
are provided in \citet{Bauer2020}.

\begin{table}[!t]
    \centering
    \caption{\label{tab:NZP} Nightly zero points for CARMENES. }
    \begin{tabular}{@{}ccccc@{}}
        \hline
        \hline
        \noalign{\smallskip}
        JD & NZP   & $\sigma_{\rm NZP}$ & $N_{\rm RV}$ & Flag\\
           & [m\,s$^{-1}$] & [m\,s$^{-1}$] &  & \\
        \noalign{\smallskip}
        \hline
        \noalign{\smallskip}
2457390 & $-$6.35 & 1.23 & 0 & 1\\
2457391 & $-$8.06 & 2.85 & 0 & 1\\
2457392 & $-$9.04 & 3.25 & 1 & 1\\
2457393 & $-$9.04 & 3.24 & 0 & 1\\
2457394 & $-$6.73 & 4.53 & 0 & 1\\
2457395 & $-$5.57 & 0.56 & 34 & 0\\
2457396 & $-$9.23 & 1.08 & 3 & 0\\
2457397 & $-$13.18 & 0.80 & 31 & 0\\
2457398 & $-$13.48 & 0.84 & 23 & 0\\
2457399 & $-$6.73 & 4.09 & 0 & 1\\
\noalign{\smallskip}
        \hline
    \end{tabular}
    \tablefoot{Listed are the Julian date (valid from UT12:00 to UT12:00 next day), the velocity of the
    nightly zero point (NZP), its uncertainty estimate $\sigma_{\rm NZP}$, the number of RV-quiet star
    RVs used to calculate the NZP and a quality flag (where 0 indicates no issue with the calculation and
    1 means that the NZP could not be calculated, in which case the NZP is replaced by a moving NZP average
    from adjacent nights). This is a sample list. The full table can be downloaded from CDS via
    \url{http://cdsarc.u-strasbg.fr/viz-bin/cat/J/A+A/}.
    }
    \end{table}

\section{Results} \label{sec:results}

The CARMENES DR1 provides raw spectroscopic data for the total sample of 362 targets but only full data
products (including RVs, spectroscopic indices, and CCF parameters) for 345 targets, that is,
excluding 17 SB2 and ST3 systems. Precise RVs of the components of 12 of these spectroscopic multiple
systems and full orbital and physical analyses were presented by \citet{Baroch2018,Baroch2021A&A...653A..49B}.
The remaining five binary systems, namely J05084$-$210, J06396$-$210, J09133+668, J16343+571 (\object{CM~Dra}), and
J23113+085, do not have a publication using CARMENES data yet. The procedure described in Sect.~\ref{sec:obs}
was applied to the 18\,642 suitable spectra and these produced the same number of RV determinations and
associated data products. However, a fraction of those measurements lack a velocity drift calculation because
of the poor quality of the simultaneous F-P spectrum. As a consequence, the number of drift-corrected RV
measurements is 17\,749, and these correspond to 344 targets. Only the faint target J16102$-$193
(\object{K2--33}) is not in the final sample because all spectra were taken without simultaneous F-P.
All data products associated with the CARMENES DR1 and ancillary files are available
online\footnote{\label{fn:url_cab}\url{https://carmenes.cab.inta-csic.es}.}.

\begin{figure}[!t]
    \centering
    \includegraphics[width=\linewidth]{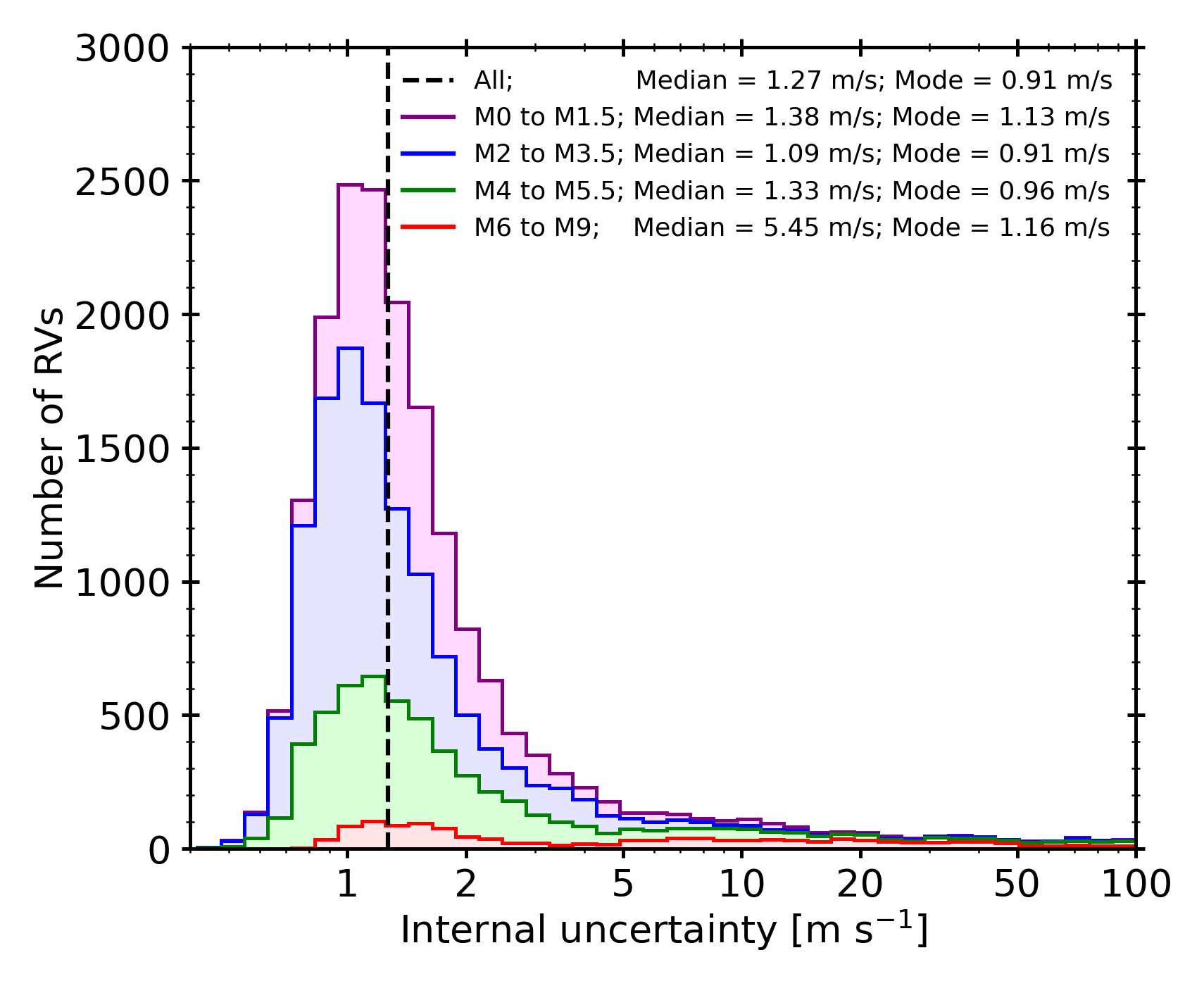}
    \caption{Stacked histograms and statistical parameters (median and mode) of the internal precision
    (formal uncertainties) of the 18\,642 precise RV measurements in the CARMENES DR1.}
    \label{fig:precision}
\end{figure}

\begin{figure*}
    \centering
    \includegraphics[width=\linewidth]{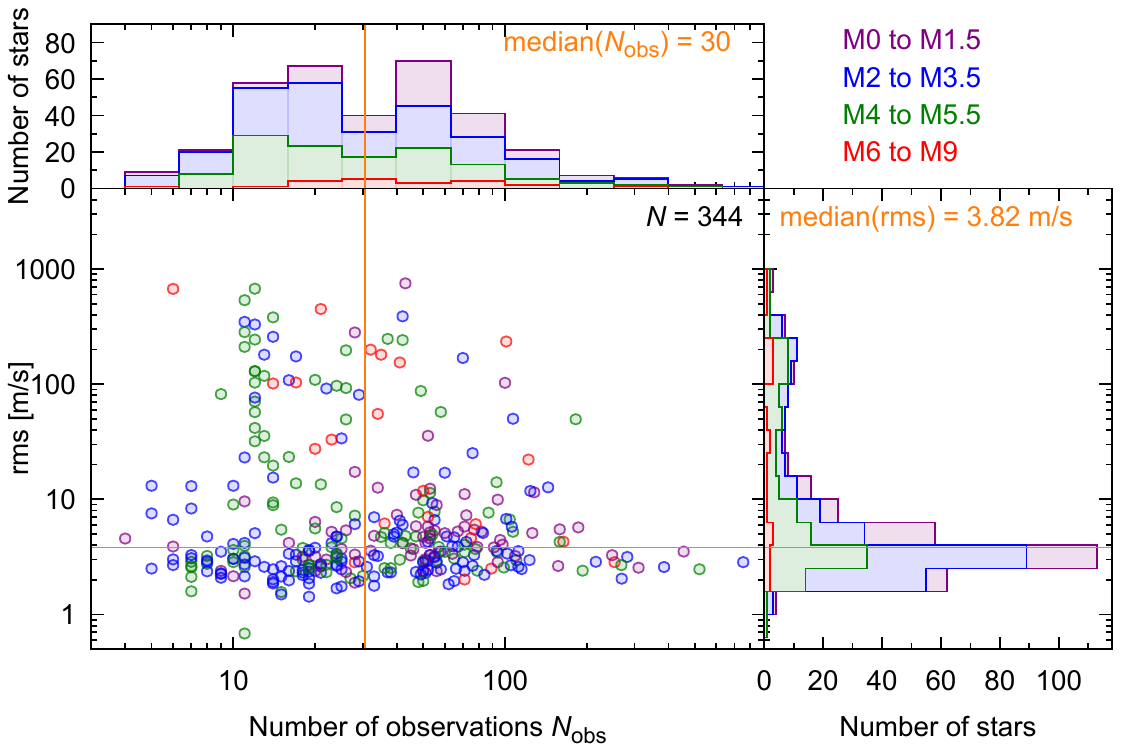}
    \caption{Distribution of observations. The rms is computed for each of the 344 targets (circles colour-coded
    by spectral type) from their {\tt serval} RV time series (NZP-corrected). No periodic signals
    (activity or planetary) were removed. The faint target J16102$-$193 (K2--33) is excluded, since all its
    spectra were taken without F-P.}
    \label{fig:sample_rms}
\end{figure*}

In Fig.~\ref{fig:precision} we illustrate the distribution of the formal uncertainties of the RV measurements
(internal precision). Targets are grouped into four spectral-type bins using the same criteria as in
Fig.~\ref{fig:completeness}. Brighter targets have typical uncertainties of $\sim$1\,m\,s$^{-1}$, as their
S/N at 1200\,nm reached 150, but fainter targets have larger uncertainties due to the larger photon noise.
The median value of the internal precision is 1.27\,m\,s$^{-1}$, with the maximum of the distribution (mode)
at 0.91\,m\,s$^{-1}$.

The distribution of observations and their dispersion are illustrated in Fig.~\ref{fig:sample_rms}, also
grouped in spectral-type bins. The scatter plot depicts the rms of the RV time series for each of the 344
NZP-corrected targets as a function of the number of observations, $N_{\rm obs}$. Histograms of rms and
number of observations are shown along both axes of the plot. The weighted rms of $n=1...N_k$ observations
of each target $k$ is calculated as
\begin{equation}
{\rm rms} = \sqrt{\frac{1}{\sum_n w_n}\sum_n w_n({\rm RV}_n-{\rm RV})^2},
\end{equation}
where the epoch weights $w_n$ include a jitter term $\sigma_{j}$, which is added in quadrature to the formal RV
uncertainties $\sigma_{{\rm RV},n}$
\begin{equation} w_n = \frac{1}{\sigma_{{\rm RV},n}^2+\sigma_{j}^2}.
\end{equation}
The $\sigma_{j}$ and the re-weighted mean RVs were obtained self-consistently for each star via a maximum likelihood
optimisation\footnote{See function {\tt mlrms} in
\url{https://github.com/mzechmeister/python/blob/master/wstat.py}.}.
For the calculation we used the RVs as measured, with NZP correction, and no known signals of
any nature (activity, planets) were subtracted. The median and mode of the distributions are 3.9\,m\,s$^{-1}$
and 3.3\,m\,s$^{-1}$, respectively. These values can be compared to those characterising the internal
precision in Fig.~\ref{fig:precision} to conclude that the RVs are most likely dominated by jitter
(and signal) from astrophysical sources, which is statistically estimated to have a median contribution of
$\sim$3.5\,m\,s$^{-1}$. No obvious rms trends as a function of spectral type are observed except for a much
higher rms (27.5\,m\,s$^{-1}$) for the latest bin due to a large fraction of low S/N measurements.

\subsection{High-resolution spectroscopic time series data}

We compiled data tables of the time series of the RVs as described in Sect. \ref{sec:obs}, as well as additional
ancillary parameters, such as stellar activity indices, for each of the 345 M dwarfs (excluding SB2 and ST3) in
the CARMENES sample. The dataset includes RV, CRX, dLW, and chromospheric line indices (H$\alpha$, Ca\,{\sc i},
Ca\,{\sc ii}~IRT, and Na\,{\sc i}~D) from {\tt serval}, and the CCF RV, BIS, FWHM, and CON obtained with
{\tt raccoon}. Data from each spectral order are provided separately. For the RVs, both the values produced by
{\tt serval} (RVC) and those obtained after applying NZP corrections (AVC) are included. Furthermore, the
exposure time and airmass of the observations, and the instrumental drift, the barycentric Earth RV,
and the secular acceleration corrections applied to calculate RVCs are also provided.

\begin{figure*}
    \centering
    \includegraphics[width=\linewidth]{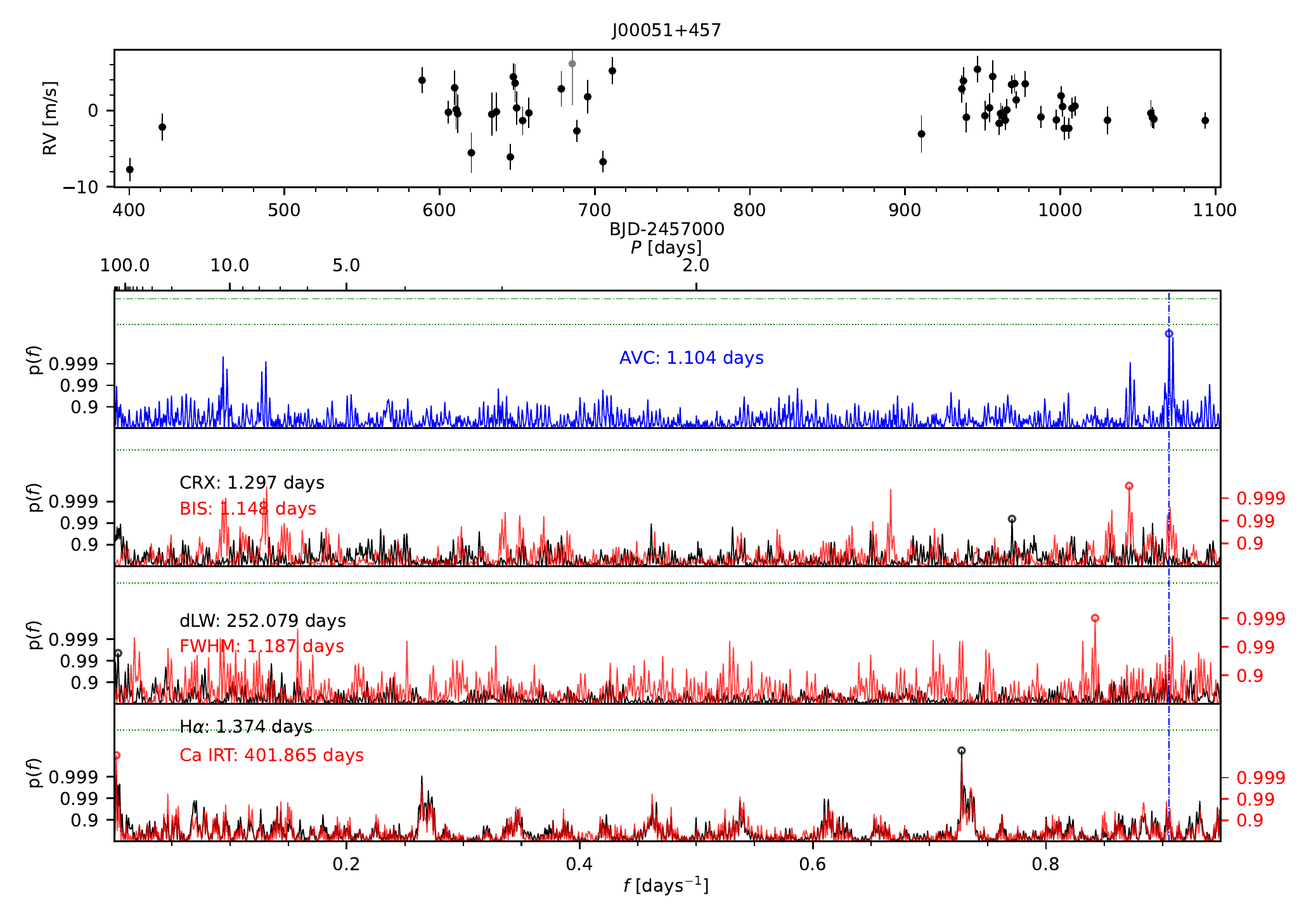}
    \caption{Time series data and periodograms for target J00051+457 (GJ~2). {\it Top panel}: NZP-corrected
    RV time series. Outlier measurements (grey symbols; see text) are excluded in the periodogram
    calculation. {\it Bottom panels} (from top to bottom): GLS periodograms of the NZP-corrected RVs (AVC, blue),
    the chromatic index (CRX, black) and CCF bisector span (BIS, red), the differential line width (dLW,
    black) and the CCF FWHM (red), and the chromospheric activity indices corresponding to the H$\alpha$
    (black) and averaged Ca\,{\sc ii}~IRT lines (red). The dash-dotted blue vertical line in all panels
    marks the position of the most significant peak in the RV periodogram, while the circles in the
    sub-panels highlight the position of the strongest signal in each periodogram, with the period
    given in the legend. The 0.1\,\%, 1\,\%, and 10\,\% FAP levels are shown as horizontal dashed, dash-dotted,
    and dotted  green lines, respectively, and are visible if they fall within the displayed vertical
    range. Vertical axes in panels that show two different datasets are scaled such that the
    FAP levels are identical. The colour code allows the graph to be related with the axis labels and legends.}
    \label{fig:sample_periodograms}
\end{figure*}

Graphical representations of the time series of {\tt serval} RVs corrected for NZPs as described in
Sect.~\ref{sec:NZP} have been produced for all targets with at least five valid NZP-corrected RV values and are
available online\textsuperscript{\ref{fn:url_cab}}. An example is provided in
Fig.\,\ref{fig:sample_periodograms} for the target J00051+457 (\object{GJ~2}). Periodogram analyses
of the RVs and several relevant activity indices are also presented. Before computing the periodogram, we
applied a clipping criterion to the measured values of the RVs and indices to avoid obvious outliers and
poor-quality measurements. All data points deviating by more than $3 \sigma$ from the mean were eliminated and so
were measurements with error bars greater than the average value plus $3 \sigma$. Nightly averages were
computed for the targets J00183+440 (\object{GX\,And}), J00184+440 (\object{GQ\,And}), and J07274+052
(\object{Luyten's Star}), as they had observations at higher cadence.

We subsequently computed the generalised Lomb-Scargle (GLS) periodogram \citep{GLS2009} of the RVs, the
CRX, dLW, H$\alpha,$ and Ca\,{\sc ii} IRT indices, and the CCF parameters BIS and
FWHM. We grouped the indices and CCF parameters in pairs according to their expected sensitivity to the
same activity phenomena, according to the analysis of \citet{Lafarga2021}. Therefore, three panels with
equivalent activity indicators are provided in Fig. \ref{fig:sample_periodograms}, namely CRX \& BIS, dLW
\& FWHM, and H$\alpha$ \& Ca\,{\sc ii} IRT. We considered periods ranging from twice the time span of the
observations (to identify long-term variations) to the Nyquist frequency as computed from the closest RV
measurement pairs of each dataset. However, an upper frequency limit of 0.95\,day$^{-1}$ was set to avoid
daily aliases, except for targets with known short-period periodicities (close-in transiting planets and
very fast-rotating stars). We calculated the false alarm probability (FAP) by running 10$^{5}$ bootstrap
realisations of the datasets. From the bootstrapped data, we also computed the probability of each
periodogram peak by assessing the number of times that the real periodogram at a given frequency is above
all the realisations. This probability is related to the GLS power \citep{Zechmeister2009} and is used in
the graphical representation.

\subsection{Exoplanets in the CARMENES GTO sample}

The CARMENES sample was designed to preserve completeness as much as possible. Therefore, the initial
target selection did not explicitly exclude known planet hosts. The inaugural CARMENES survey paper by
\citet{Trifonov2018A&A...609A.117T} analysed the CARMENES data for a sample of seven targets known to
host 12 planets. In this study, a hitherto unknown second, long-period planet orbiting J11417+427
(\object{GJ~1148}) was reported, qualifying as the first exoplanet discovered by CARMENES. Shortly after,
\citet{Reiners2018a} published the first exoplanet detected from data collected solely from the CARMENES
survey. Since then, a succession of announcements has been made using data from the CARMENES blind survey,
totalling 33 newly discovered planets in 28 planetary systems at the time of writing this paper. The new
CARMENES planets are marked with a `d' in column `Type' of Table~\ref{tab:planets}. In some cases,
CARMENES data were combined with precise RVs from other instruments (such as HARPS, HARPS-N, ESPRESSO,
HIRES, IRD, MAROON-X, etc.) to enhance the statistical significance of the measurements. Moreover, over
the course of the survey, five already announced exoplanets were re-analysed using CARMENES data. Together
with the 12 known planets in \citet{Trifonov2018A&A...609A.117T}, this makes up a total of 17 planets
that are marked with an `r' in Table \ref{tab:planets}.

As explained in Sect.~\ref{sec:introduction}, the CARMENES Consortium decided to invest a fraction of the
GTO time in following up on transiting planet candidates. Some of the targets came from the K2 mission
\citep{Howell2014}, but most of them were provided by the ongoing TESS mission. The campaign has been
fruitful, and CARMENES has led or contributed to the confirmation of 26 such planet
candidates and helped measure their masses. The CARMENES planets resulting from follow-up activities
are marked with an `f' in Table \ref{tab:planets}. The TESS planet candidate around J11044+304 (TOI-1806)
has been followed up and validated with CARMENES; however, its parameters are not listed in the table
because they are not yet sufficiently significant.

The columns in Table~\ref{tab:planets} provide basic information on the targets and their planets. The
parameters are taken from each of the quoted references. $N_{\rm CAR}$ and $N_{\rm other}$ are the number
of RVs from CARMENES and other instruments, respectively, that were used in the corresponding publication.
$N_{\rm CAR}$ may differ from the number of measurements in the DR1 release. Cases where $N_{\rm CAR}$ is
greater than the number in DR1 correspond to recent publications that include observations taken after 31
December 2020, as part of the new CARMENES Legacy+ survey, while cases where $N_{\rm CAR}$ is below the
number of measurements in DR1 are those where additional measurements within the CARMENES GTO were taken
after the quoted publications. For four such planets we present new parameters in Table~\ref{tab:planets}
considering all the measurements in DR1. The four revised planets are J06548+332\,b (\object{GJ~251}\,b;
\citealt{Stock2020b}) J08413+594\,b,c (\object{GJ~3512}\,b,c; \citealt{Morales2019}), and J16167+672S\,b
(\object{HD~147379}\,b; \citealt{Reiners2018a}).

\setcounter{table}{2}
\begin{table}[!t]
    \centering
    \caption{CARMENES survey targets with confirmed or claimed exoplanets that have no dedicated publications
with CARMENES data.}
    \label{tab:planets_nocarm}
    \footnotesize
    \begin{tabular}{@{}llcl@{}}
        \hline
        \hline
        \noalign{\smallskip}
Karmn & Star name & $N_{\rm CAR}$ & Ref. \\
        \noalign{\smallskip}
        \hline
        \noalign{\smallskip}
J04219+213    & LP 415-17 (K2-155)   &   4 & Hir18\\
J04520+064    & GJ 179                 &  10 & How10\\
J04538$-$177  & GJ 180                 &  25 & Tuo14\\
J05019$-$069  & LP 656-038           &   8 & AD7\\
J06105$-$218  & HD 42581 A             &  54 & Tuo14\\
J07274+052    & Luyten's Star          & 756 & AD17\\
J08409$-$234  & LP 844-008           &  27 & AE12\\
J10023+480    & BD+48 1829             &  23 & Hob19\\
J11477+008    & FI Vir                 &  58 & Bon18\\
J12388+116    & GJ 480                 &   7 & Fen20\\ 
J13119+658    & PM J13119+6550         &  12 & Dem20\\
J16102$-$193  & K2-33                 &  27 & Dav16\\
J16254+543    & GJ 625                 &  33 & SM17\\
J16303$-$126  & V2306 Oph              &  94 & Wri16\\
J16581+257    & BD+25 3173             &  55 & Joh10\\
J17355+616    & BD+61 1678C            &  26 & Pin19\\
J17364+683    & BD+68 946 AB           &  41 & Bur14\\
J18353+457    & BD+45 2743             &  16 & GA21\\
J19206+731S   & 2MASS J19204172+7311434&  22 & Cad22\\
J22096$-$046  & BD--05 5715           &  61 & But06\\
J23064$-$050  & TRAPPIST-1            &  17 & Gil16\\
        \noalign{\smallskip}
        \hline
    \end{tabular}
    \tablefoot{$N_{\rm CAR}$ is the number of measurements in CARMENES DR1.}
    \tablebib{
    AD17: \citet{2017A&A...602A..88A},
    AE12: \citet{2012ApJ...746...37A},
    Bon18: \citet{2018A&A...613A..25B},
    Bur14: \citet{2014ApJ...789..114B},
    But06: \citet{2006PASP..118.1685B},
    Cad22: \citet{2022AJ....164...96C},
    Dav16: \citet{2016Natur.534..658D},
    Dem20: \citet{2020A&A...642A..49D},
    Fen20: \citet{Feng2020}; 
    GA21: \citet{2021A&A...649A.157G},
    Gil16: \citet{2016Natur.533..221G},
    Hir18: \citet{2018AJ....155..124H},
    Hob19: \citet{2019A&A...625A..18H},
    How10: \citet{2010ApJ...721.1467H},
    Joh10: \citet{Johnson2010},
    Pin19: \citet{2019A&A...625A.126P},
    SM17: \citet{2017A&A...605A..92S},
    Tuo14: \citet{2014MNRAS.441.1545T},
    Wri16: \citet{Wright2016ApJ...817L..20W}.
}
\end{table}

In addition to the publications using CARMENES RVs, the DR1 includes measurements of targets for which there
have been exoplanet detections or claims in the literature but do not have a specific publication using
CARMENES data at the time of writing. These are listed in Table \ref{tab:planets_nocarm}, along with the
number of released CARMENES epochs. Besides planet detections and confirmations, there are some targets in
our sample for which planets have been announced and are listed in exoplanet catalogues but could not
be confirmed or are controversial given the data obtained with CARMENES or other instruments. The list of
such planets is provided in Table~\ref{tab:challenge}. We are not including a planet around J00183+440
(GX~And\,b) because the CARMENES observations now seem to support a planetary scenario for the 11.44-day
signal (Trifonov et al., in prep.), in contrast to the initial CARMENES data
\citep{Trifonov2018A&A...609A.117T}, which were casting doubt on its nature.

\begin{table*}
    \centering
    \caption{Exoplanets challenged by CARMENES.}
    \label{tab:challenge}
    \begin{tabular}{@{}lllllll@{}}
        \hline
        \hline
        \noalign{\smallskip}
        Karmn & Star name &  Planet  & $P_{\rm p}$ [d] & Ref. & Alternative & Ref.\\
        \noalign{\smallskip}
        \hline
        \noalign{\smallskip}
        J01125$-$169& YZ Cet & b & 1.98 & Ast17 & 2.02\,d (alias) & Sto20a\\
        J02002+130  & TZ Ari & c & 242 & Fen20 & Spurious & Qui22\\
        J02222+478  & GJ 96 & b & 73.9 & Hob18 & Spurious & This work\\
        J04429+189  & GJ 176  & b & 10.2 & End08 & Spurious, new 8.78 d & For09, But09, Tri18\\
        J09561+627  & GJ 373  & b & 17.8 & Tuo19, Fen20 & Rotation & This work\\
        J10196+198  & AD Leo  & b & 2.23 & Tuo18 & Rotation & Car20, Rob20, Kos22 \\
        J10564+070  & CN Leo  & c & 2.69 & Tuo19 & Rotation & Laf21\\
        J11033+359  & Lalande 21185  & b & 9.9 & But17 & 12.9\,d & Dia19, Sto20b\\
        J11302+076  & K2-18   & c & 9.0 & Clo17 & Rotation & Sar18\\
        J11509+483  & GJ 1151 & b & 2.02 & Mah21 & 390\,d & Per21, Bla22\\
        J16303$-$126& V2306 Oph& b & 4.89 & Wri16 & 1.27\,d (alias) & Sab21\\
        \noalign{\smallskip}
        \hline
    \end{tabular}
    \tablebib{
        Ast17: \citet{Astudillo2017A&A...605L..11A};
        Bla22: \citet{BlancoPozo2022};
        But09: \citet{Butler2009};
        But17: \citet{Butler2017AJ....153..208B};
        Car20: \citet{Carleo2020};
        Clo17: \citet{Cloutier2017A&A...608A..35C};
        Dia19: \citet{Diaz2019A&A...625A..17D};
        End08: \citet{Endl2008};
        Fen20: \citet{Feng2020};
        For09: \citet{Forveille2009};
        Hob18: \citet{Hobson2018A&A...618A.103H};
        Kos22: \citet{Kossakowski2022};
        Laf21: \citet{Lafarga2021};
        Mah21: \citet{Mahadevan2021ApJ...919L...9M};
        Per21: \citet{Perger2021A&A...649L..12P};
        Qui22: \citet{Quirrenbach2022};
        Rob20: \citet{Robertson2020};
        Sab21: \citet{Sabotta2021A&A...653A.114S};
        Sar18: \citet{Sarkis2018AJ....155..257S};
        Sto20a: \citet{Stock2020a};
        Sto20b: \citet{Stock2020b};
        Tri18: \citet{Trifonov2018A&A...609A.117T};
        Tuo18: \citet{Tuomi2018};
        Tuo19: \citet{Tuomi2019arXiv190604644T};
        Wri16: \citet{Wright2016ApJ...817L..20W}.
    }
\end{table*}

\begin{figure*}
    \centering
    \includegraphics[width=\linewidth]{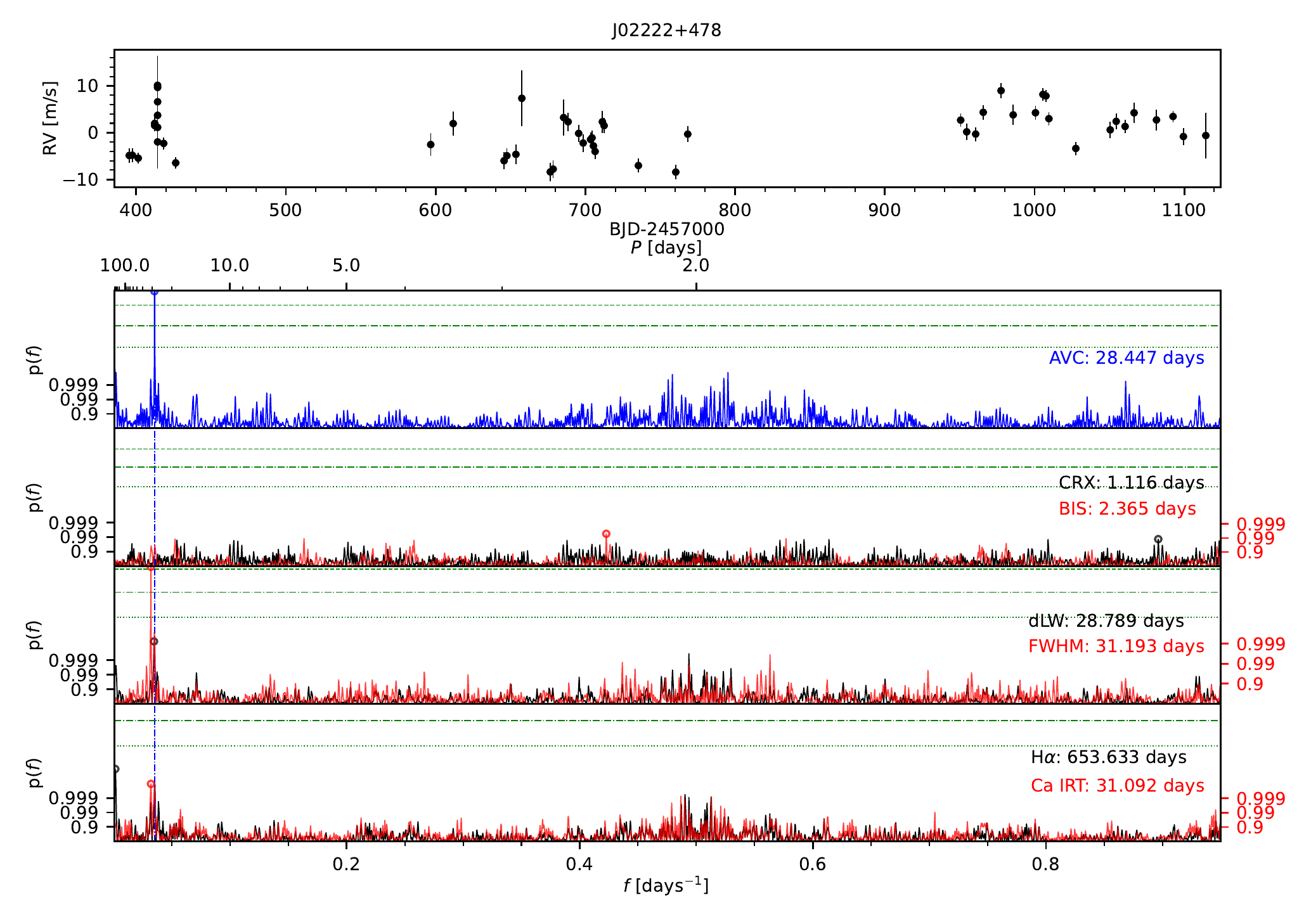}
    \caption{Same as Fig.~\ref{fig:sample_periodograms}, but for J02222+478 (GJ 96). No planetary signal at a
    period of 73.9\,d, as claimed by \cite{Hobson2018A&A...618A.103H}, is visible in the CARMENES data. The only
    significant periodicity is at $\sim$28.5\,d and seems to be related to activity given the counterparts in
    some of the indicators. Phase-folded plots are shown in Fig. \ref{fig:J02222+478_2}.}
    \label{fig:J02222+478}
\end{figure*}

\begin{figure}
    \centering
    \includegraphics[width=4.4cm]{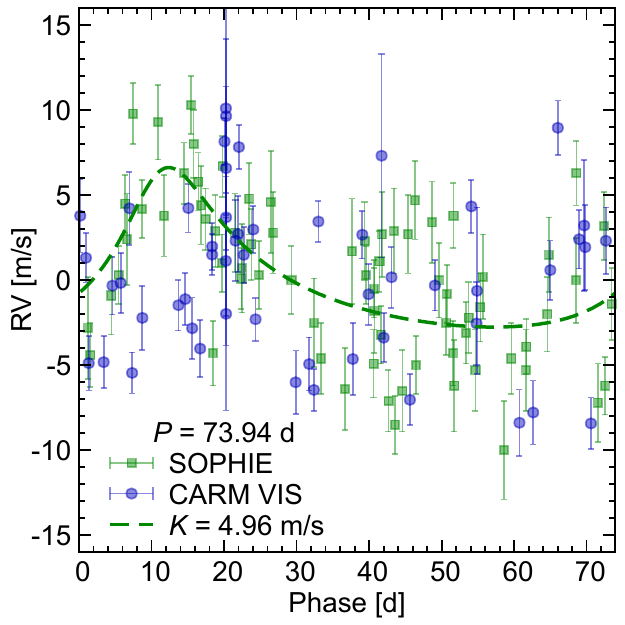}
    \includegraphics[width=4.4cm]{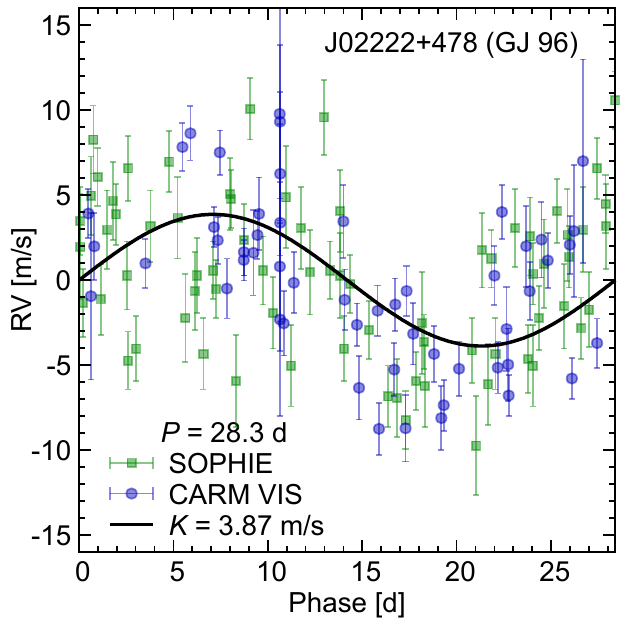}
    \caption{Phase-folded RV data of J02222+478 (GJ 96). {\em Left}: SOPHIE (green squares)
    and CARMENES VIS (blue circles) phase-folded to a period of 73.94\,d. The planetary Keplerian signal (dashed
    green line) proposed by \citet{Hobson2018A&A...618A.103H} is based on the SOPHIE data alone and disfavoured
    by the CARMENES measurements. {\em Right}: Same as the left panel, but phase-folded to the periodicity of 28.3\,d,
    which is most likely associated with stellar activity.}
    \label{fig:J02222+478_2}
\end{figure}

\begin{figure*}
    \centering
    \includegraphics[width=\linewidth]{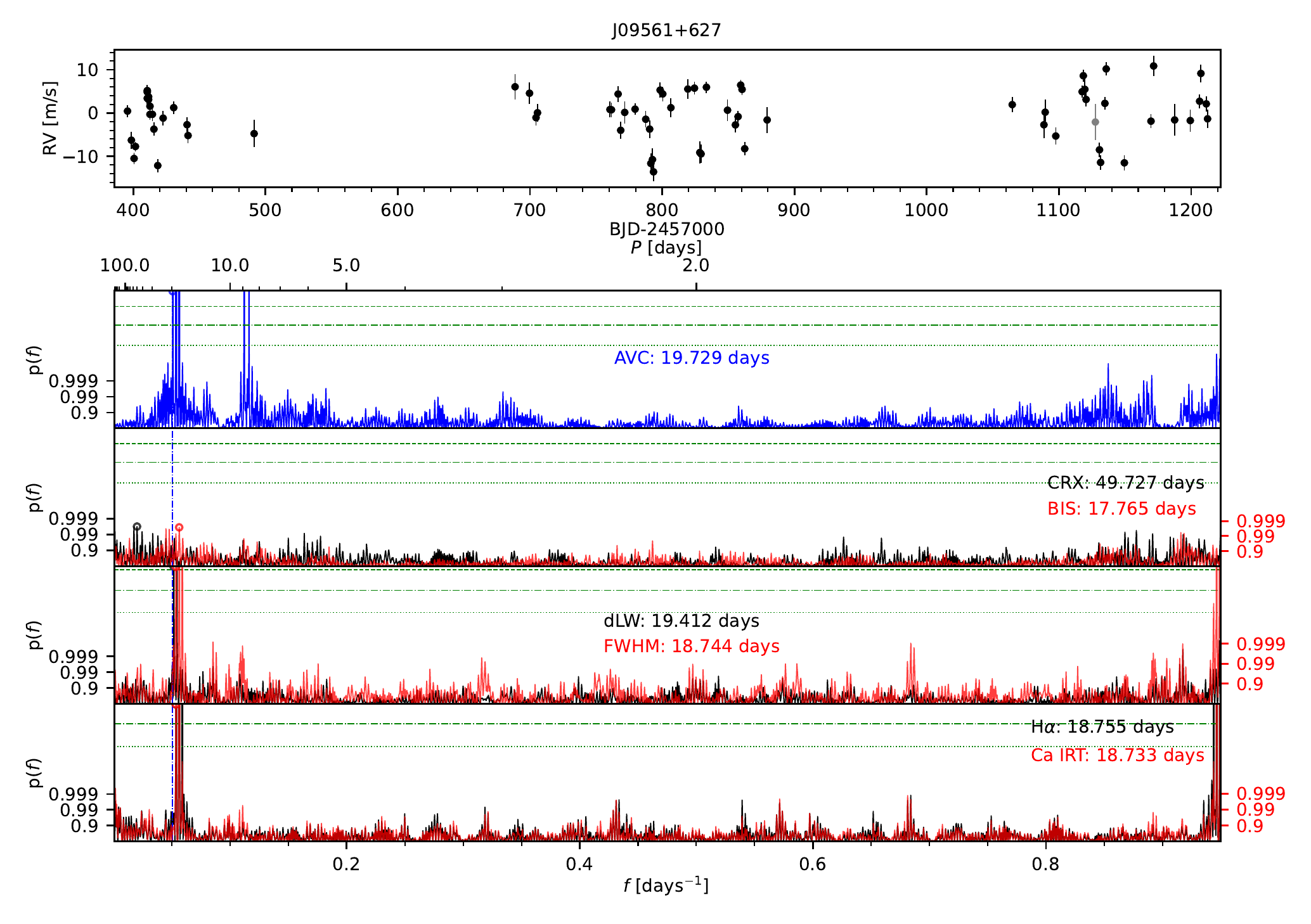}
    \caption{Same as Fig.~\ref{fig:sample_periodograms}, but for J09561+627 (GJ 373). The periodic signal at
    17.8\,d attributed to a planet by \citet{Tuomi2019arXiv190604644T} and \citet{Feng2020} most
    likely arises from stellar activity in view of the counterparts in most of the activity indicators.}
    \label{fig:J09561+627}
\end{figure*}

Some of the challenged planets were already discussed in dedicated publications, as listed in
Table~\ref{tab:challenge}. In two other cases, we carried out the analysis as part of the present work.
Particularly, the candidates announced around J02222+478 (GJ 96) and J09561+627 (GJ 373) can be quite
confidently ruled out as planets. For GJ 96, \cite{Hobson2018A&A...618A.103H} announced a planet
candidate based on 72 SOPHIE RVs\footnote{The paper quotes 79 RVs, but only 72 RVs are published
in \url{https://cdsarc.cds.unistra.fr/ftp/J/A+A/618/A103/tablea1.dat}.}. The periodograms in
Fig.\,\ref{fig:J02222+478} show that the 75-day signal present in SOPHIE data is absent in the 53 RVs
from CARMENES. Instead, the dominant signal is at 28.5\,d, which \cite{Hobson2018A&A...618A.103H} already
attributed to stellar activity. Indeed, this period is also present in the dLW, Ca\,{\sc ii} IRT, and
H$\alpha$ time series of the CARMENES data, thus ruling out its planetary nature. Phase-folded plots
are shown in Fig. \ref{fig:J02222+478_2}. The signal in GJ 373 at about 17.8\,d announced as a planet by
\citet{Tuomi2019arXiv190604644T} and \citet{Feng2020} can most definitely be attributed to stellar
rotation modulation since it appears strongly in CARMENES activity indicators such as dLW and H$\alpha$
and some CCF parameters, as can be seen in Fig.\,\ref{fig:J09561+627}.

\begin{figure*}[!t]
    \centering
    \includegraphics[width=\linewidth]{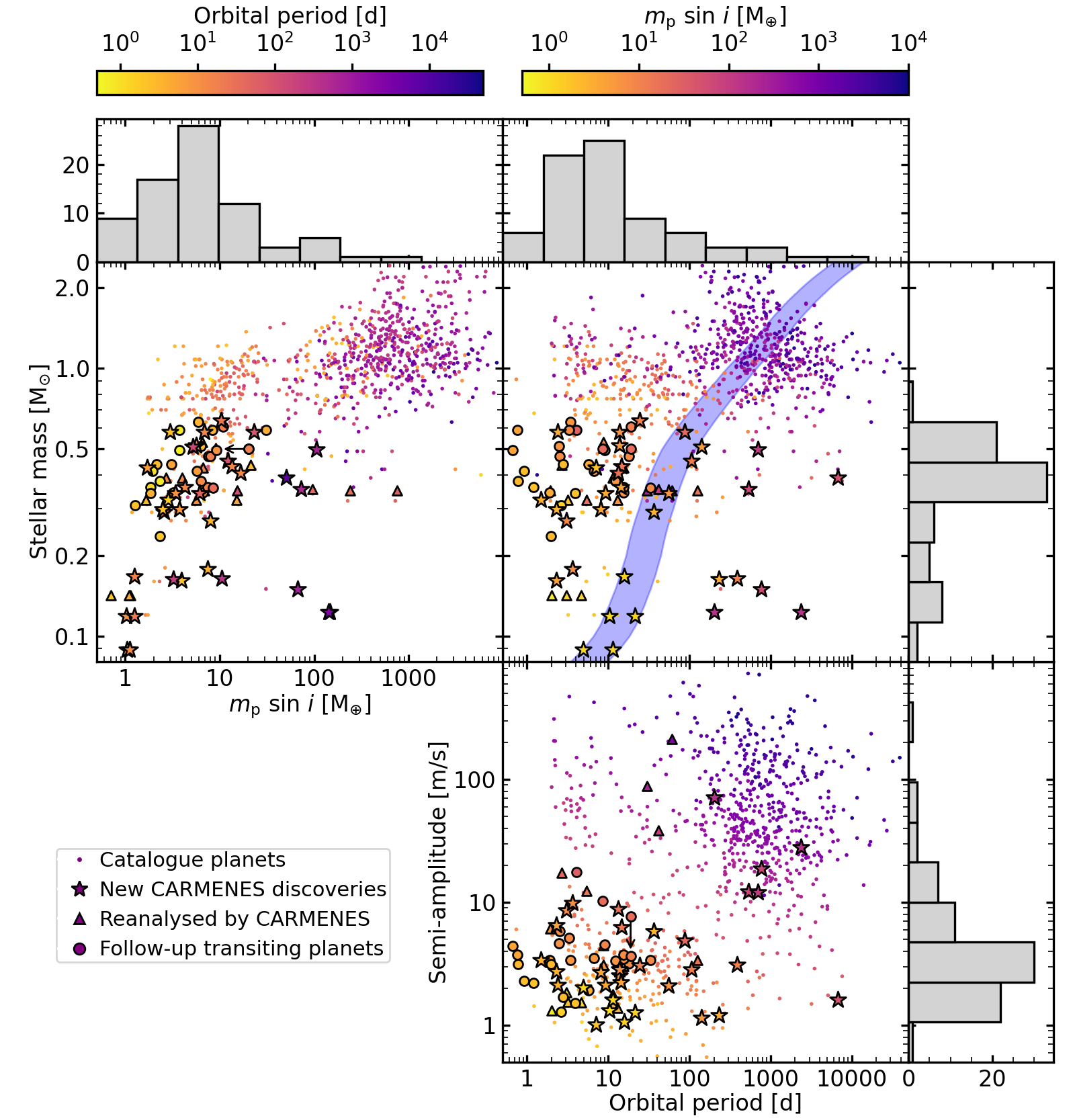}
    \caption{Scatter plots of the CARMENES DR1 exoplanet sample compared to the complete sample of catalogued
    planets in the NASA Exoplanet Archive detected via RVs (903; small dots). Different symbols indicate planets
    newly detected from the CARMENES blind survey (33; stars), planets confirmed from transit follow-up (26;
    circles), and known planets re-analysed with CARMENES data (17; triangles). The three panels correspond to
    pairs of different relevant parameters, with the complementary colour scale introducing a third dimension.
    The histograms along the axes show distributions of the corresponding parameters for the CARMENES planet
    sample. The blue shaded band in the top-right panel represents the liquid-water habitable zone
    with limits defined by the `runaway greenhouse' and `maximum greenhouse' criteria \citep{Kopparapu2013}.}
    \label{fig:planet_dist}
\end{figure*}

The sample of planets in Table \ref{tab:planets} is represented graphically in Fig. \ref{fig:planet_dist}.
Scatter plots combine the stellar mass, minimum planetary mass, orbital period, and RV semi-amplitude. The
planet samples in the diagrams comprise those with CARMENES analyses (showing `d', `r', and
`f' separately in Table \ref{tab:planets}) and those coming from the NASA Exoplanet
Archive\footnote{\label{fn:url_exoplanetarchive}\url{https://exoplanetarchive.ipac.caltech.edu}, accessed
on 1 July 2022.}. The latter correspond only to RV-detected planets (i.e. planets discovered through
photometric transits are excluded). Also, histogram distributions of each of these quantities for the planets
analysed with CARMENES data are depicted as side plots.

A few features in Fig. \ref{fig:planet_dist} are worth discussing. Regarding stellar mass, a majority of
CARMENES planets have host stars with masses between 0.25\,M$_{\odot}$ and 0.65\,M$_{\odot}$, which
constitute the bulk of the sample. Remarkably, half of the 24 RV planets with stellar hosts below
0.25\,M$_{\odot}$ known to date have been discovered by CARMENES, a testament of the advantage offered by
a red-optimised RV spectrometer in the late-type host regime. In terms of planetary mass, the majority
of the CARMENES planets are in the super-Earth to the Neptune-mass domain, although several Earth-mass
planets have been detected orbiting some of the lower-mass targets in our sample. Remarkable cases are
two systems, J02530+168 (\object{Teegarden's Star}; \citealt{Zechmeister2019}) and J00067--075
(\object{GJ~1002}; \citealt{Suarez2022}), each with two Earth-mass planets within the
liquid-water habitable zones of their stars. Also, CARMENES has discovered six Saturn- and Jupiter-mass
planets, some of them around very low-mass primaries, thus defying canonical planet formation models
\citep{Morales2019}, which predict very low occurrence rates of giant planets around M-type dwarfs
\citep[e.g.][]{Schlecker2022}. As expected for detectability reasons, most of the CARMENES planets have
orbital periods from a few days to a few tens of days. Although not discussed here, the CARMENES GTO survey
also announced \citep{Baroch2021A&A...653A..49B} two brown dwarf candidates on very long-period orbits
($P\gtrsim3000$\,days), around 10504+331 (\object{GJ 3626}) and J23556$-$061 (\object{GJ 912}).

\subsection{Planet occurrence rates}

Using the CARMENES DR1 sample we calculated planet occurrence rates in a similar way as was done by
\citet[hereafter Sab21]{Sabotta2021A&A...653A.114S}. In that work, preliminary occurrence statistics
were calculated using a subsample of 71 targets having at least 50 CARMENES RV measurements. The
re-analysis in the present work applies similar target selection criteria. From the initial 362 targets in
the CARMENES GTO sample we excluded 124 targets because of several possible reasons: ($i$) 
they were added later for transit follow-up (mostly TESS candidates; 20 targets); ($ii$) 
they are spectroscopic binaries and triples (23 targets); ($iii$) 
they are part of the RV-loud sample as defined by \citet{TalOr2018} (52 targets); or ($iv$) 
we obtained fewer than ten RV measurements (29 targets). The sample therefore comprises a total of 238
targets, including 69 of the 71 targets in Sab21. For the two targets not included in the CARMENES
DR1, one was excluded after being classified as a late-K dwarf (J18198$-$019, \object{HD~168442}), and
the other one was subsequently classified as a resolved binary (J23113+085, \object{NLTT~56083}).

For the planetary sample, we re-ran the signal retrieval and vetting algorithm from Sab21 (see the results in
Table \ref{tab:signals}). The only change that we made was the period limit used for the long period
planets. Sab21 included every signal if the time baseline was longer than two orbital periods, while here
we include every signal with time coverage of at least 1.5 times the orbital period. If we considered the
more conservative period limit of Sab21, we would exclude several giant planets from the planet sample and
that would therefore reduce the statistical soundness of our analysis. As a result, we regard the new
criterion as a better balance between being too conservative but still making sure that the signal is indeed
periodic. Using this criterion, we identified 37 planets that can be confirmed with
CARMENES data alone and three additional planet candidates (around J05033$-$173, J17033+514, and
J18409$-$133). We also include 13 planets with fewer than 50 RVs that were detected using data from
other surveys (mainly HADES, HArps-n red Dwarf Exoplanet Survey, \citealt{Pinamonti2022}, and HARPS,
\citealt{Bonfils2013}) if they induce an RV semi-amplitude of $K > 2$\,m\,s$^{-1}$. We assume that we
would have detected such planets if we had not terminated the observations because of our independent
knowledge. We obtained those targets from a comparison with the two exoplanet databases on The Extrasolar
Planets Encyclopaedia\footnote{\url{http://exoplanet.eu}.} \citep{Schneider2011}
and the NASA Exoplanet Archive\textsuperscript{\ref{fn:url_exoplanetarchive}}. In Table~\ref{tab:signals},
we mark planets that are listed in one of the databases and are well below our detection limits, planets
with fewer than 50 RVs that are included in our planet sample, and archive planets that are not supported
by CARMENES data. In this way we increase the planetary sample in Sab21 by 26 planets (from 27 to 53) for
the recalculation of the occurrence rates. The total number of 53 planets reside in 43 planetary systems.

\begin{figure}[!t]
    \centering
    \includegraphics[width=\linewidth]{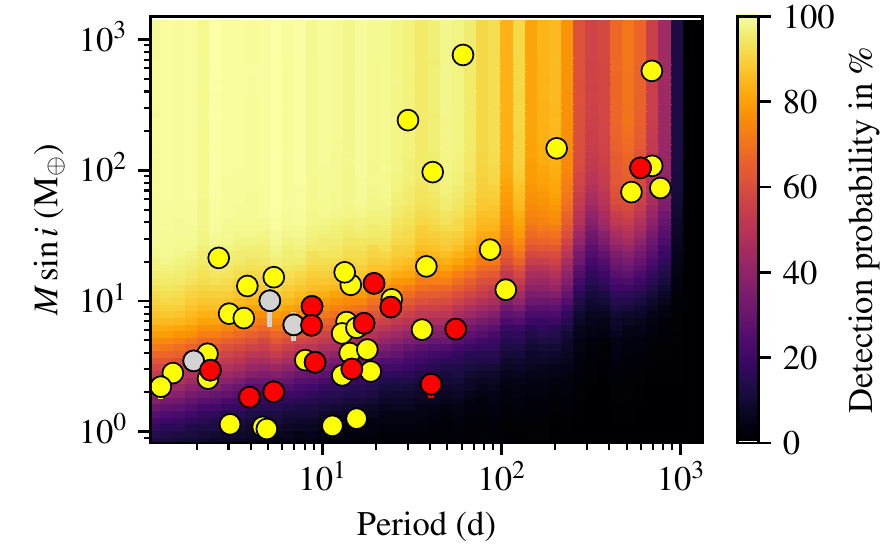}
    \caption{CARMENES detection probability map in the period-mass plane derived from the observations of
    238 targets. The detection probability is calculated using each target mass, and the average
    of all detection probabilities for the individual grid points is shown in the period-planet mass plane
    as a colour map. For example, a 10\% detection probability for 238 targets indicates that the survey data
    can detect the respective planet in approximately 24 targets. Yellow circles indicate CARMENES planet
    detections and planet detections from other instruments that are confirmed by CARMENES data; grey circles
    planet candidates; and red circles planet detections from other surveys for which no sufficient CARMENES
    measurements are available to confirm them. The five planets in the low-probability detection region
    ($<$10\%) are Teegarden's Star b and c, YZ Cet c and d, and Wolf 1069 b, all of which are Earth-mass
    planets with host stars of less than 0.2~M$_{\odot}$, thus suggesting that such planets are very abundant.}
    \label{fig:planet_occurrence}
\end{figure}

We calculated individual planet detection maps for all targets following the procedure described
by Sab21. The numerical and graphical results are available online\textsuperscript{\ref{fn:url_cab}}.
The global detection probabilities across the period-mass plane and the planets mentioned above are
shown in Fig.~\ref{fig:planet_occurrence}, with the colour scale indicating the average of
all detection probabilities for the individual grid points. There are five planets in a low-probability
region, which means that we can only detect such planets for a small fraction of our sample.
Unsurprisingly, these are Teegarden's Star~b and~c, YZ Cet~c and~d, and Wolf~1069 b, all of which are Earth-mass
planets with very low-mass stellar hosts.

\begin{figure}[!t]
    \centering
    \includegraphics[width=\linewidth]{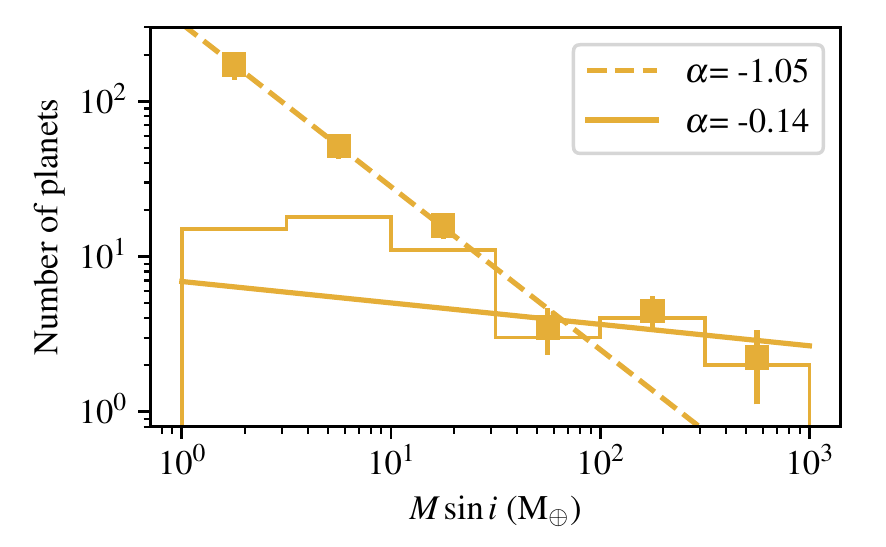}
    \caption{Histogram of the number of planets detected in six $M_\text{pl} \sin i$ bins. The squares
    indicate the number of planets corrected for the survey sensitivity averaged over periods of up to
    240\,days. The dashed line and the solid line indicate the power law fit for $M_\text{pl} <$
    32\,M$_\oplus$ and $M_\text{pl} >$ 32\,M$_\oplus$, respectively.}
    \label{fig:power_law}
\end{figure}

Using the same method as in Sab21, we obtained the power-law distribution in $M_\text{pl} \sin i$
for the occurrence rate estimate (Fig.~6 in Sab21). The updated power-law with
$N_\text{pl, corrected}= a~(M_\text{pl} \sin i)^\alpha$ is only slightly shallower, with
the slope changing from $\alpha=-1.14 \pm 0.16$ to $\alpha=-1.05 \pm 0.01$ for planets with masses
below 30\,M$_\oplus$, and from $\alpha = -0.26 \pm 0.17$ to $\alpha = -0.14 \pm 0.25$ for higher-mass
planets (see Fig.~\ref{fig:power_law}). For our occurrence rate determination, we used this power-law
as an initial assumption on the $M_\text{pl} \sin i$ distribution, instead of assuming a log-uniform
distribution\footnote{The code and combined maps used to calculate occurrence rates are found in
\url{https://github.com/ssabotta/calculate_occurrence_rate}.}. The results are summarised in
Table~\ref{tab:occurrence_all}. We report both the number of planets per star ($\overline{n}_\text{pl}$)
and the frequency of stars with planets ($F_\text{h}$). To obtain the latter, we repeated the analysis
but instead of including all planets, we reduced the planet sample and took  only the single planet
with the highest $K$ amplitude in the system. We then inspected the complete period-mass plane with
periods of 1\,d to 1\,000\,d and $M_\text{pl} \sin i$ of 1\,M$_\oplus$ to 1\,000\,M$_\oplus$. In this
parameter range, we determined an overall occurrence rate of $\overline{n}_\text{pl} = 1.44\pm0.20$
planets per star and $F_\text{h} = 94^{+4}_{-9}\,\%$ stars with planets. This means that the planet
multiplicity in our sample is around 1.5 planets per system.

\begin{table}
\caption{Planet occurrence rates for a sample of 238 M dwarfs in the CARMENES DR1 sample, including candidates
and planet detections from other surveys. }
\label{tab:occurrence_all}
\centering
\begin{tabular}{@{}l c c c c@{}}
\hline\hline
& \multicolumn{4}{c}{$P$ (d)} \\
& 1--10 & 10--100 & 100--1000 & 1--1000 \\
\hline
\noalign{\smallskip}
\multicolumn{5}{c}{{\em (a) Planets with} 100\,M$_\oplus \sin i < M_\mathrm{pl} < 1000$\,M$_\oplus$} \\
\noalign{\smallskip}
$N_\text{pl,det}$ & 0  &  2       &   4  & 6 \\
 $\overline{n}_\text{pl}$ & $< 0.006 $ & 0.010$^{+0.010}_{-0.005}$    & 0.03$^{+0.01}_{-0.01}$ & 0.03$^{+0.02}_{-0.01}$  \\
 $N_\text{h}$ &  0  &  1  &  4 & 5 \\
  $F_\text{h}$ & $ <  0.006 $ & 0.006$^{+0.005}_{-0.005}$ & 0.03$^{+0.01}_{-0.01}$ & 0.03$^{+0.01}_{-0.01}$\\
\noalign{\smallskip}
\hline                                             \noalign{\smallskip}
\multicolumn{5}{c}{{\em (b) Planets with} 10\,M$_\oplus < M_\mathrm{pl} \sin i < 100$\,M$_\oplus$}\\
\noalign{\smallskip}
$N_\text{pl,det}$ &  4     &  7         &   3  & 14 \\
$\overline{n}_\text{pl}$ & 0.02$^{+0.02}_{-0.01}$ & 0.04$^{+0.02}_{-0.01}$   &     0.04$^{+0.02}_{-0.02}$   & 0.09$^{+0.03}_{-0.02}$  \\
$N_\text{h}$ &  4    &  7         &  2    & 13 \\
$F_\text{h}$ & 0.02$^{+0.02}_{-0.01}$ &  0.04$^{+0.02}_{-0.01}$  &  0.03$^{+0.02}_{-0.02}$ & 0.09$^{+0.02}_{-0.03}$\\
\noalign{\smallskip}
\hline
\noalign{\smallskip}
\multicolumn{5}{c}{{\em (c) Planets with} 1\,M$_\oplus < M_\mathrm{pl} \sin i < 10$\,M$_\oplus$} \\
\noalign{\smallskip}
  $N_\text{pl,det}$ &  18    &  15    &  0 & 33 \\
$\overline{n}_\text{pl}$ & 0.39$^{+0.10}_{-0.07}$  & 0.67$^{+0.18}_{-0.15}$   &  < 0.40    & 1.37$^{+0.24}_{-0.24}$   \\
$N_\text{h}$ &  15    & 10    & 0 & 25 \\
$F_\text{h}$ & 0.33$^{+0.08}_{-0.07}$  & 0.47$^{+0.13}_{-0.13}$   &  < 0.40   &  0.89$^{+0.08}_{-0.11}$  \\
\noalign{\smallskip}
\hline
\noalign{\smallskip}
\multicolumn{5}{c}{{\em (d) Planets with} 1\,M$_\oplus < M_\mathrm{pl} \sin i < 1000$\,M$_\oplus$} \\
\noalign{\smallskip}
  $N_\text{pl,det}$ &  22    &  24    &  7 & 53 \\
$\overline{n}_\text{pl}$ &  0.37$^{+0.09}_{-0.07}$  &  0.63$^{+0.14}_{-0.12}$   &   0.54$^{+0.23}_{-0.17}$    &  1.44$^{+0.20}_{-0.20}$ \\
$N_\text{h}$ &  19    & 18    & 6 & 43 \\
$F_\text{h}$ & 0.32$^{+0.07}_{-0.07}$  &  0.47$^{+0.13}_{-0.09}$  & 0.47$^{+0.20}_{-0.16}$   &  0.94$^{+0.04}_{-0.09}$  \\
\noalign{\smallskip}\hline
\end{tabular}

\tablefoot{
$N_\text{pl,det}$: number of detected planets,
$\overline{n}_\text{pl}$: average number of planets per star,
$N_\text{h}$: number of planet host stars,
$F_\text{h}$: frequency of stars with planets.
}
\end{table}

The analysis of Sab21 yielded occurrence rates that are larger by a factor of two for planets with
10\,M$_\oplus < M_{\rm pl} \sin i < 100$\,M$_\oplus$ and by 30\,\% for the low-mass planets with
1\,M$_\oplus < M_{\rm pl} \sin i < 10$\,M$_\oplus$ with respect to the results obtained here for the
full sample. The lower occurrence rates cannot be due to the looser requirement on orbital period
coverage (1.5 instead of two orbital periods), since, if anything, this would result in larger
occurrence rates. The smaller occurrence rates observed for the full CARMENES sample thus illustrate
the effectiveness in terms of planet discovery of the pre-selection of targets that are observed more
intensively than others. Human intervention bias in this case leads to an over-estimation of occurrence
rates. The survey sensitivity is higher for stars with planets because targets showing interesting
signals that could be of planetary nature were observed more intensively. Sab21 pointed out this effect,
and explicitly introduced the bias by rejecting all targets with fewer than 50 RVs, but it affects all
targeted surveys that change the observing strategy based on acquired knowledge. In fact, by aiming
at a specific number of observations for all of our targets, we minimised this effect. In the CARMENES
DR1, we reach this number of 50 RVs for 42\,\% of our targets, which corresponds to 112 stars. 
We are continuing the survey as part of the CARMENES Legacy+ programme. Even if the planet detection
efficiency may not be as high as in the early stages of the GTO, the statistical value of the sample
will greatly increase.

\begin{figure}[!t]
    \centering
    \includegraphics[width=\linewidth]{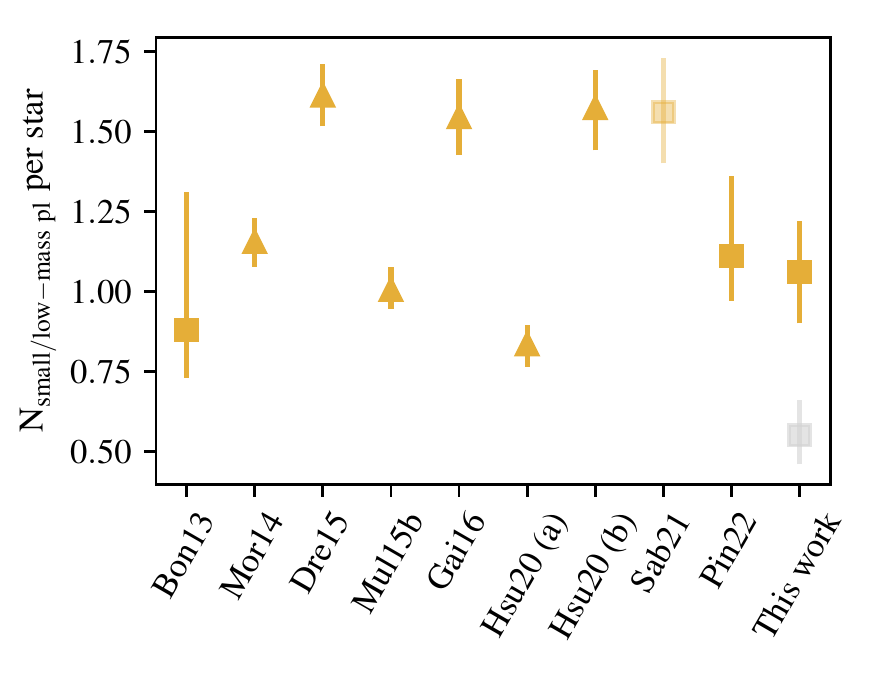}
    \caption{Comparison of low-mass/small planet occurrence rates from various surveys (adapted from Sab21).
    The parameter range is 1 to 10\,M$_\oplus$ in $M_\text{pl} \sin i$ for RV surveys and 1.3 to 3.7\,R$_\oplus$ in
    $R_\text{pl}$ for transit surveys. The error bars for transit surveys are Poisson errors based on
    the number of planet detections in the respective bins, and error bars from RV surveys are the 16\,\% and
    84\,\% levels of the occurrence rate posterior distributions. Results using transiting planets from the
    \emph{Kepler} mission are represented as triangles (\citealt{Morton2014}, \citealt{Dressing2015},
    \citealt{Mulders2015a}, \citealt{Gaidos2016},  \citealt{Hsu2020}), and rates from the HARPS, HADES, and
    CARMENES RV surveys are represented as squares
    \citep[][and this work]{Bonfils2013, Sabotta2021A&A...653A.114S, Pinamonti2022}. The grey
    square shows the occurrence rate from this work with the assumption of a log-uniform distribution in
    $M_\text{pl}\sin i$.}
    \label{fig:occurrence_comparison}
\end{figure}

We compare our low-mass planet occurrence rates around M dwarfs to those of other surveys in
Fig.~\ref{fig:occurrence_comparison}. Our updated occurrence rates are consistent with the values
obtained from the HARPS \citep{Bonfils2013} and HADES  \citep{Pinamonti2022} surveys, but our results
are based on a significantly larger statistical sample. The agreement is good despite the fact that
both estimates by \citet{Bonfils2013} and \citet{Pinamonti2022} assumed a log-uniform distribution
in planet mass, as opposed to our power-law relationship. If we also utilised a uniform distribution
for our occurrence rate calculation, we would have obtained a lower occurrence rate of
0.58$^{+0.11}_{-0.09}$ low-mass planets per star in orbits of up to 100\,d (indicated as the grey
square in Fig.~\ref{fig:occurrence_comparison}). In this parameter range, we obtained instead
$1.06\pm0.17$ planets per star assuming a power-law distribution in $M_\text{pl} \sin i$. The
difference occurs only in those regions of the period-mass plane with a strong sensitivity gradient,
that is, below 10\,M$_\oplus$. For higher planet masses, the choice of distribution does not affect our
results significantly.

The comparison to transit surveys is not as straightforward due to the lack of an exact correspondence between
the observed parameters. The expected value of $\sin i$ assuming randomly oriented orbits is $\sim$0.8
\citep[e.g.][]{Hatzes2019} and, therefore, our $M_\text{pl} \sin i$ bin of 1--10\,M$_\oplus$ on average
corresponds to a bin of 1.25--12.5\,M$_\oplus$ in true $M_\text{pl}$. In this mass regime, planets could
be rocky, water worlds, or puffy sub-Neptunes with very different densities \citep{Luque2022b}. According
to the mass-radius relation of \cite{Kanodia2019}, this mass interval corresponds on average to the
$R_\text{pl}$ interval of 1.3--3.7\,R$_\oplus$. In log-space this is only 75\,\% of the radius interval
of 1--4\,R$_\oplus$ that Sab21 used for comparison with transiting planet statistics. Thus, in
Fig.~\ref{fig:occurrence_comparison}, we plot lower occurrence rates for the transit surveys (75\,\%
of those in Sab21). In any case, all occurrence rate estimates agree within a factor of two despite
all the involved assumptions and the fact that we infer the occurrence rates from an overall detection
sensitivity of 15\,\% (considering the full period-mass plane).

The discussion above is relevant if one wants to find an absolute number of planets per star or to compare
with transiting planet surveys or surveys targeting other stellar masses. Moreover, these calculations also
serve as a valuable benchmark for planet formation theories that aim to reproduce population-level trends of
exoplanets \citep[e.g.][]{Ida2004,Bitsch2015,Miguel2020,Izidoro2021,Schlecker2021b,Schlecker2021a,Mishra2021}.
Using the results of Sab21 as input, \cite{Schlecker2022} compared a planet sample based on the HARPS and
CARMENES M-dwarf surveys to a synthetic population computed with the Bern model of planet formation
\citep{Mordasini2012,Emsenhuber2021,Burn2021} and found three main discrepancies.

The first one is the observational finding of an excess of giant planets around lower-mass stars compared
to the theoretical prediction. The simulations do not produce any giant planets around host stars with masses
below 0.5\,M$_\odot$. As was done by Sab21, we split the full CARMENES sample at a stellar mass of
0.337\,M$_\odot$ and calculated giant planet occurrence rates. The median stellar masses of the two
subsamples are 0.24\,M$_\odot$ and 0.45\,M$_\odot$. Using a strict limit for the giant planet mass of
$M_{\rm pl} > 100$\,M$_\oplus$, we obtained a rate of $0.021^{+0.018}_{-0.011}$ planets per star and
$0.045^{+0.021}_{-0.016}$ planets per star for the low-mass and the high-mass stellar subsamples,
respectively. The resulting occurrence rate ratio, $f_{\rm high-mass}/f_{\rm low-mass}=2.14,$ is marginally
consistent with the giant planet frequency as a function of stellar mass published by \cite{Ghezzi2018},
\begin{equation}
f(M_\star,{\rm [Fe/H]}) = 0.085^{+0.008}_{-0.010} M_\star^{1.05^{+0.28}_{-0.24}} 10^{1.05^{+0.21}_{-0.17} {\rm [Fe/H]}},
\end{equation}
assuming similar stellar metallicities, [Fe/H], in both samples.

The second discrepancy between model and observation concerns the shape of the planet mass distribution.
The distribution of $M_\text{pl} \sin i$ in the synthetic population is bimodal, whereas its counterpart
in the observed sample is consistent with a power-law. In fact, our planet mass distribution does not deviate
significantly from that in Sab21 (see Fig.~6 therein).

A third mismatch between the observed and model-predicted planet demographics as identified by
\cite{Schlecker2022} is the orbital period distribution around stars with masses higher than
0.4\,M$_\odot$. Short-period planets ($P_{\rm orb} < 10$\,d) are found to be significantly scarcer
in the observed population compared to the synthetic one. The drop in occurrence rates at periods of
less than 10\,d, which was previously observed for stars with different stellar masses, does not
hold for targets with masses below 0.4\,M$_\odot$, with period distributions showing a good match.
One possible explanation is a migration barrier having higher efficiency in protoplanetary disks
around early M dwarfs that is not adequately accounted for by current models. For targets with
$M < 0.337$\,M$_\odot$ we calculate $0.56^{+0.15}_{-0.14}$ and $0.63^{+0.23}_{-0.18}$ low-mass
planets per star for the intervals 1--10~d and 10--100~d, respectively.

\section{Conclusions} \label{sec:conclusions}

The CARMENES GTO survey ran from 1 January 2016 to 31 December 2020 and obtained 19\,633 spectroscopic
measurements of a sample of 362 targets. The sample was designed to be as complete as possible by including
M~dwarfs observable from the Calar Alto Observatory with no selection criteria other than brightness
limits and visual binarity restrictions. To best exploit the capabilities of the instrument, variable
brightness cuts were applied as a function of spectral type to increase the presence of late-type targets.
This effectively leads to a sample that does not deviate significantly from a volume-limited one for each
spectral type. The global completeness of the sample is 15\,\% of all known M dwarfs out to a distance of 20\,pc
and 48\,\% at 10\,pc.

The present paper accompanies the release of a large dataset, the CARMENES DR1. Raw data,
pipeline-processed data, and high-level data products are provided, including 18\,642 precise
RVs for 345 targets (removing double- and triple-line systems). After correction of a NZP
offset, the median internal precision of early and intermediate M-dwarf types is $\sim$1.2\,m\,s$^{-1}$. This
value increases to $\sim$5.4\,m\,s$^{-1}$ for late M spectral types due to their intrinsic faintness. The median
rms of the RV time series of all the targets in the sample is calculated to be $\sim$3.9\,m\,s$^{-1}$, where no
signal has been subtracted. A comparison between the internal and external precisions indicates that the RV
variance has a contribution of $\sim$3.5\,m\,s$^{-1}$ on top of the instrument error when treated as
uncorrelated random noise. This external noise component is unlikely to be of instrumental origin. It is
instead believed to arise from astrophysical effects, including Keplerian signals from planets but,
most importantly, RV variability arising from stellar activity (e.g. active region rotation and evolution).

The CARMENES time series data have been analysed in the search for RV signals of a planetary nature. So far we have
identified 33 new planets from the blind survey observations, which are complemented by 17 planets that we have
re-analysed with CARMENES data and 26 planets from transit search space missions that we have confirmed and
measured. The number of blind survey planets is in good agreement with the initial estimates
considering the properties of the stellar sample, the survey design, and the assumed planet occurrence
rates \citep{Garcia-Piquer2017A&A...604A..87G}. The new planets cover a broad region of the parameter space
in terms of stellar host mass, planetary mass, and orbital period. A remarkable result is that CARMENES has
discovered half of the RV planets known to orbit stars of masses below 0.25\,M$_\odot$. This fact illustrates
the prime `hunting ground' of CARMENES thanks to the competitive advantage of the optimised red-sensitive
design and the possibility of undertaking a massive survey with a large fraction of dedicated 4 m class
telescope time over five years.

With the CARMENES DR1 data, we have calculated new planet occurrence rates around M dwarfs to update the results
already presented by Sab21. We have employed a subsample of 238 stars that fulfil a set of specific requirements.
We still find a high long-period giant planet occurrence rate of around 3\,\%, a high number of low-mass planets
(1.06 planets per star in periods of 1\,d to 100\,d), and an overabundance of short-period planets around the
lowest-mass stars of our sample compared to stars with higher masses. For our complete period-mass parameter
space, we determine an overall occurrence rate of $\overline{n}_\text{pl} = 1.44\pm0.20$ planets
per star and a fraction of $F_\text{h} = 94^{+4}_{-9}\,\%$ stars with planets. We calculate the overall CARMENES
survey sensitivity to be 15\,\% and find planets around 43 of 238 targets (i.e. 18\,\% of the stars), which
again shows that nearly every M dwarf hosts at least one planet.

In the present description of the CARMENES GTO data, we have focused on their use for precise RV work in the
field of exoplanet detection and characterisation. Nevertheless, we have shown in a number of publications that
these data are also of high value to a variety of science cases within stellar astrophysics, such as
studying atmospheric parameters
\citep[$T_{\rm eff}$, $\log g$, and chemical abundances;][]{Passegger2018,Passegger2019,Passegger2020,Passegger2022,
Fuhrmeister2019b,Marfil2020,Marfil2021,Abia2020,Shan2021}, determining fundamental properties
\citep[$M$, $R$, and magnetic field;][]{Schweitzer2019,Shulyak2019,Reiners2022}, and analysing magnetic
activity \citep{TalOr2018,Fuhrmeister2018,Fuhrmeister2019,Fuhrmeister2020,Fuhrmeister2022,Schoefer2019,
Hintz2019,Hintz2020,Baroch2020,Lafarga2021,Jeffers2022}. CARMENES VIS channel data have also proved useful
in addressing the study of exoplanet atmospheres via transit transmission spectroscopy \citep{Yan2019,Yan2021,
CasasayasBarris2020,CasasayasBarris2021,SanchezLopez2020,Khalafinejad2021,Czesla2022} and the
Rossiter-McLaughlin effect \citep{Oshagh2020,Sedaghati2022}.

The CARMENES GTO survey is now complete. In terms of exoplanet RV detection, the survey has provided about
60 planet discoveries and confirmations, some of which are of very high scientific relevance, and, as a sample, is of
great statistical value, thus contributing to a complete census of the planetary population in the solar
neighbourhood. The initial goals of the survey have therefore been fulfilled. The CARMENES sample continues
to be observed within the CARMENES Legacy+ programme. The ultimate goal is to reach 50 measurements for all
suitable targets (i.e. excluding multiples, RV-loud stars, etc.). The CARMENES Legacy+ extension of the survey
is expected to run at least until the end of 2023 and, eventually, to lead to a second release of CARMENES
survey data with 50 measurements or more for about 300 nearby M dwarfs. Through the present release and
future additions, the CARMENES data will continue to yield new exoplanet discoveries and enable abundant
studies in other domains within stellar astrophysics and exoplanetary science.

\begin{acknowledgements}
CARMENES is an instrument at the Centro Astron\'omico Hispano en Andaluc\'ia (CAHA) at Calar Alto
(Almer\'{\i}a, Spain), operated jointly by the Junta de Andaluc\'ia and the Instituto de Astrof\'isica
de Andaluc\'ia (CSIC).

CARMENES was funded by the Max-Planck-Gesellschaft (MPG),
  the Consejo Superior de Investigaciones Cient\'{\i}ficas (CSIC),
  the Ministerio de Econom\'ia y Competitividad (MINECO) and the European Regional Development Fund (ERDF) through projects FICTS-2011-02, ICTS-2017-07-CAHA-4, and CAHA16-CE-3978,
  and the members of the CARMENES Consortium
  (Max-Planck-Institut f\"ur Astronomie,
  Instituto de Astrof\'{\i}sica de Andaluc\'{\i}a,
  Landessternwarte K\"onigstuhl,
  Institut de Ci\`encies de l'Espai,
  Institut f\"ur Astrophysik G\"ottingen,
  Universidad Complutense de Madrid,
  Th\"uringer Landessternwarte Tautenburg,
  Instituto de Astrof\'{\i}sica de Canarias,
  Hamburger Sternwarte,
  Centro de Astrobiolog\'{\i}a and
  Centro Astron\'omico Hispano-Alem\'an),
  with additional contributions by the MINECO,
  the Deutsche Forschungsgemeinschaft (DFG) through the Major Research Instrumentation Programme and Research Unit FOR2544 ``Blue Planets around Red Stars'',
  the Klaus Tschira Stiftung,
  the states of Baden-W\"urttemberg and Niedersachsen,
  and by the Junta de Andaluc\'{\i}a.

We acknowledge financial support from the Spanish Agencia Estatal de Investigaci\'on of the Ministerio de Ciencia e Innovaci\'on (AEI-MCIN) and the ERDF ``A way of making Europe'' through projects
  PID2020-117493GB-I00,     
  PID2019-109522GB-C5[1:4],     
  PID2019-110689RB-I00,      
  PID2019-107061GB-C61,     
  PID2019-107061GB-C64,     
  PGC2018-098153-B-C33,
  PID2021-125627OB-C31/AEI/10.13039/501100011033,           
and the Centre of Excellence ``Severo Ochoa'' and ``Mar\'ia de Maeztu'' awards to the Institut de Ci\`encies de l'Espai (CEX2020-001058-M), Instituto de Astrof\'isica de Canarias (CEX2019-000920-S), Instituto de Astrof\'isica de Andaluc\'ia (SEV-2017-0709), and Centro de Astrobiolog\'ia (MDM-2017-0737).

We also benefited from additional funding from:

the Secretaria d'Universitats i Recerca del Departament d'Empresa i Coneixement de la Generalitat de Catalunya and the Ag\`encia de Gesti\'o d’Ajuts Universitaris i de Recerca of the Generalitat de Catalunya, with additional funding from the European FEDER/ERDF funds, and from the Generalitat de Catalunya/CERCA programme;

the DFG through the  Major Research Instrumentation Programme and Research Unit FOR2544 ``Blue Planets around Red Stars'' (RE 2694/8-1); 

the University of La Laguna through the Margarita Salas Fellowship from the Spanish Ministerio de Universidades ref. UNI/551/2021-May-26, and under the EU Next Generation funds; 

the Gobierno de Canarias through projects
ProID2021010128 
and ProID2020010129; 

the Spanish MICINN under Ram\'on y Cajal programme RYC-2013-14875; 

the ``Fondi di Ricerca Scientifica d'Ateneo 2021'' of the University of Rome ``Tor Vergata''; 

and the programme ``Alien Earths'' supported by the National Aeronautics and Space Administration (NASA) under agreement No. 80NSSC21K0593.

This reasearch was based on data from the CARMENES data archive at CAB (CSIC-INTA), and made use of the NASA Exoplanet Archive, which is operated by the California Institute of Technology, under contract with the NASA under the Exoplanet Exploration Program.

We thank
B.~Arroyo,
A.~Fern\'andez,
E.~Gallego,
A.~Gardini,
A.~Guijarro,
R.~Hedrosa,
I.~Hermelo,
J.~Iglesias-P\'aramo, 
P.~Mart\'in,
R.\,J.~Mathar, 
M.~Moreno,
E.~Ramos,
R.~Rebolo, 
I.~Vico,
and many other individuals for their commitment to the CARMENES GTO.

\end{acknowledgements}

\bibliographystyle{aa}
\bibliography{bibtex}

\begin{thebibliography}{180}
\expandafter\ifx\csname natexlab\endcsname\relax\def\natexlab#1{#1}\fi

\bibitem[{{Abia} {et~al.}(2020){Abia}, {Tabernero}, {Korotin}, {Montes},
  {Marfil}, {Caballero}, {Straniero}, {Prantzos}, {Ribas}, {Reiners},
  {Quirrenbach}, {Amado}, {B{\'e}jar}, {Cort{\'e}s-Contreras}, {Dreizler},
  {Henning}, {Jeffers}, {Kaminski}, {K{\"u}rster}, {Lafarga},
  {L{\'o}pez-Gallifa}, {Morales}, {Nagel}, {Passegger}, {Pedraz},
  {Rodr{\'\i}guez L{\'o}pez}, {Schweitzer}, \& {Zechmeister}}]{Abia2020}
{Abia}, C., {Tabernero}, H.~M., {Korotin}, S.~A., {et~al.} 2020,
  \href{https://ui.adsabs.harvard.edu/abs/2020A&A...642A.227A}{\aap, 642, A227}

\bibitem[{{Affer} {et~al.}(2016){Affer}, {Micela}, {Damasso}, {Perger},
  {Ribas}, {Su{\'a}rez Mascare{\~n}o}, {Gonz{\'a}lez Hern{\'a}ndez}, {Rebolo},
  {Poretti}, {Maldonado}, {Leto}, {Pagano}, {Scandariato}, {Zanmar Sanchez},
  {Sozzetti}, {Bonomo}, {Malavolta}, {Morales}, {Rosich}, {Bignamini},
  {Gratton}, {Velasco}, {Cenadelli}, {Claudi}, {Cosentino}, {Desidera},
  {Giacobbe}, {Herrero}, {Lafarga}, {Lanza}, {Molinari}, \&
  {Piotto}}]{Affer2016}
{Affer}, L., {Micela}, G., {Damasso}, M., {et~al.} 2016,
  \href{https://ui.adsabs.harvard.edu/abs/2016A&A...593A.117A}{\aap, 593, A117}

\bibitem[{{Alonso-Floriano} {et~al.}(2015){Alonso-Floriano}, {Morales},
  {Caballero}, {Montes}, {Klutsch}, {Mundt}, {Cort{\'e}s-Contreras}, {Ribas},
  {Reiners}, {Amado}, {Quirrenbach}, \& {Jeffers}}]{Alonso2015}
{Alonso-Floriano}, F.~J., {Morales}, J.~C., {Caballero}, J.~A., {et~al.} 2015,
  \href{http://adsabs.harvard.edu/abs/2015A%26A...577A.128A}{\aap, 577, A128}

\bibitem[{{Amado} {et~al.}(2021){Amado}, {Bauer}, {Rodr{\'\i}guez L{\'o}pez},
  {Rodr{\'\i}guez}, {Cardona Guill{\'e}n}, {Perger}, {Caballero},
  {L{\'o}pez-Gonz{\'a}lez}, {Mu{\~n}oz Rodr{\'\i}guez}, {Pozuelos},
  {S{\'a}nchez-Rivero}, {Schlecker}, {Quirrenbach}, {Ribas}, {Reiners},
  {Almenara}, {Astudillo-Defru}, {Azzaro}, {B{\'e}jar}, {Bohemann}, {Bonfils},
  {Bouchy}, {Cifuentes}, {Cort{\'e}s-Contreras}, {Delfosse}, {Dreizler},
  {Forveille}, {Hatzes}, {Henning}, {Jeffers}, {Kaminski}, {K{\"u}rster},
  {Lafarga}, {Lodieu}, {Lovis}, {Mayor}, {Montes}, {Morales}, {Morales},
  {Murgas}, {Ortiz}, {Pall{\'e}}, {Pepe}, {Perdelwitz}, {Pollaco}, {Santos},
  {Sch{\"o}fer}, {Schweitzer}, {S{\'e}gransan}, {Shan}, {Stock}, {Tal-Or},
  {Udry}, {Zapatero Osorio}, \& {Zechmeister}}]{Amado2021}
{Amado}, P.~J., {Bauer}, F.~F., {Rodr{\'\i}guez L{\'o}pez}, C., {et~al.} 2021,
  \href{https://ui.adsabs.harvard.edu/abs/2021A&A...650A.188A}{\aap, 650, A188}

\bibitem[{{Anglada-Escud{\'e}} {et~al.}(2012){Anglada-Escud{\'e}}, {Boss},
  {Weinberger}, {Thompson}, {Butler}, {Vogt}, \&
  {Rivera}}]{2012ApJ...746...37A}
{Anglada-Escud{\'e}}, G., {Boss}, A.~P., {Weinberger}, A.~J., {et~al.} 2012,
  \href{https://ui.adsabs.harvard.edu/abs/2012ApJ...746...37A}{\apj, 746, 37}

\bibitem[{{Anglada-Escud{\'e}} \& {Butler}(2012)}]{Anglada2012}
{Anglada-Escud{\'e}}, G. \& {Butler}, R.~P. 2012,
  \href{https://ui.adsabs.harvard.edu/abs/2012ApJS..200...15A}{\apjs, 200, 15}

\bibitem[{{Astudillo-Defru} {et~al.}(2017{\natexlab{a}}){Astudillo-Defru},
  {D{\'\i}az}, {Bonfils}, {Almenara}, {Delisle}, {Bouchy}, {Delfosse},
  {Forveille}, {Lovis}, {Mayor}, {Murgas}, {Pepe}, {Santos}, {S{\'e}gransan},
  {Udry}, \& {W{\"u}nsche}}]{Astudillo2017A&A...605L..11A}
{Astudillo-Defru}, N., {D{\'\i}az}, R.~F., {Bonfils}, X., {et~al.}
  2017{\natexlab{a}},
  \href{https://ui.adsabs.harvard.edu/abs/2017A&A...605L..11A}{\aap, 605, L11}

\bibitem[{{Astudillo-Defru} {et~al.}(2017{\natexlab{b}}){Astudillo-Defru},
  {Forveille}, {Bonfils}, {S{\'e}gransan}, {Bouchy}, {Delfosse}, {Lovis},
  {Mayor}, {Murgas}, {Pepe}, {Santos}, {Udry}, \&
  {W{\"u}nsche}}]{2017A&A...602A..88A}
{Astudillo-Defru}, N., {Forveille}, T., {Bonfils}, X., {et~al.}
  2017{\natexlab{b}},
  \href{https://ui.adsabs.harvard.edu/abs/2017A&A...602A..88A}{\aap, 602, A88}

\bibitem[{{Baranne} {et~al.}(1996){Baranne}, {Queloz}, {Mayor}, {Adrianzyk},
  {Knispel}, {Kohler}, {Lacroix}, {Meunier}, {Rimbaud}, \& {Vin}}]{Baranne1996}
{Baranne}, A., {Queloz}, D., {Mayor}, M., {et~al.} 1996,
  \href{https://ui.adsabs.harvard.edu/abs/1996A&AS..119..373B}{\aaps, 119, 373}

\bibitem[{{Baroch} {et~al.}(2021){Baroch}, {Morales}, {Ribas}, {B{\'e}jar},
  {Reffert}, {Cardona Guill{\'e}n}, {Reiners}, {Caballero}, {Quirrenbach},
  {Amado}, {Anglada-Escud{\'e}}, {Colom{\'e}}, {Cort{\'e}s-Contreras},
  {Dreizler}, {Galad{\'\i}-Enr{\'\i}quez}, {Hatzes}, {Jeffers}, {Henning},
  {Herrero}, {Kaminski}, {K{\"u}rster}, {Lafarga}, {Lodieu},
  {L{\'o}pez-Gonz{\'a}lez}, {Montes}, {Pall{\'e}}, {Perger}, {Pollacco},
  {Rodr{\'\i}guez-L{\'o}pez}, {Rodr{\'\i}guez}, {Rosich}, {Sch{\"o}fer},
  {Schweitzer}, {Shan}, {Tal-Or}, \& {Zechmeister}}]{Baroch2021A&A...653A..49B}
{Baroch}, D., {Morales}, J.~C., {Ribas}, I., {et~al.} 2021,
  \href{https://ui.adsabs.harvard.edu/abs/2021A&A...653A..49B}{\aap, 653, A49}

\bibitem[{{Baroch} {et~al.}(2020){Baroch}, {Morales}, {Ribas}, {Herrero},
  {Rosich}, {Perger}, {Anglada-Escud{\'e}}, {Reiners}, {Caballero},
  {Quirrenbach}, {Amado}, {Jeffers}, {Cifuentes}, {Passegger}, {Schweitzer},
  {Lafarga}, {Bauer}, {B{\'e}jar}, {Colom{\'e}}, {Cort{\'e}s-Contreras},
  {Dreizler}, {Galad{\'\i}-Enr{\'\i}quez}, {Hatzes}, {Henning}, {Kaminski},
  {K{\"u}rster}, {Montes}, {Rodr{\'\i}guez-L{\'o}pez}, \&
  {Zechmeister}}]{Baroch2020}
{Baroch}, D., {Morales}, J.~C., {Ribas}, I., {et~al.} 2020,
  \href{https://ui.adsabs.harvard.edu/abs/2020A&A...641A..69B}{\aap, 641, A69}

\bibitem[{{Baroch} {et~al.}(2018){Baroch}, {Morales}, {Ribas}, {Tal-Or},
  {Zechmeister}, {Reiners}, {Caballero}, {Quirrenbach}, {Amado}, {Dreizler},
  {Lalitha}, {Jeffers}, {Lafarga}, {B{\'e}jar}, {Colom{\'e}},
  {Cort{\'e}s-Contreras}, {D{\'\i}ez-Alonso}, {Galad{\'\i}-Enr{\'\i}quez},
  {Guenther}, {Hagen}, {Henning}, {Herrero}, {K{\"u}rster}, {Montes}, {Nagel},
  {Passegger}, {Perger}, {Rosich}, {Schweitzer}, \& {Seifert}}]{Baroch2018}
{Baroch}, D., {Morales}, J.~C., {Ribas}, I., {et~al.} 2018,
  \href{https://ui.adsabs.harvard.edu/abs/2018A&A...619A..32B}{\aap, 619, A32}

\bibitem[{{Bauer} {et~al.}(2020){Bauer}, {Zechmeister}, {Kaminski},
  {Rodr{\'\i}guez L{\'o}pez}, {Caballero}, {Azzaro}, {Stahl}, {Kossakowski},
  {Quirrenbach}, {Becerril Jarque}, {Rodr{\'\i}guez}, {Amado}, {Seifert},
  {Reiners}, {Sch{\"a}fer}, {Ribas}, {B{\'e}jar}, {Cort{\'e}s-Contreras},
  {Dreizler}, {Hatzes}, {Henning}, {Jeffers}, {K{\"u}rster}, {Lafarga},
  {Montes}, {Morales}, {Schmitt}, {Schweitzer}, \& {Solano}}]{Bauer2020}
{Bauer}, F.~F., {Zechmeister}, M., {Kaminski}, A., {et~al.} 2020,
  \href{https://ui.adsabs.harvard.edu/abs/2020A&A...640A..50B}{\aap, 640, A50}

\bibitem[{{Bauer} {et~al.}(2015){Bauer}, {Zechmeister}, \&
  {Reiners}}]{Bauer2015A&A...581A.117B}
{Bauer}, F.~F., {Zechmeister}, M., \& {Reiners}, A. 2015,
  \href{https://ui.adsabs.harvard.edu/abs/2015A&A...581A.117B}{\aap, 581, A117}

\bibitem[{{Bayliss} {et~al.}(2018){Bayliss}, {Gillen}, {Eigm{\"u}ller},
  {McCormac}, {Alexander}, {Armstrong}, {Booth}, {Bouchy}, {Burleigh},
  {Cabrera}, {Casewell}, {Chaushev}, {Chazelas}, {Csizmadia}, {Erikson},
  {Faedi}, {Foxell}, {G{\"a}nsicke}, {Goad}, {Grange}, {G{\"u}nther},
  {Hodgkin}, {Jackman}, {Jenkins}, {Lambert}, {Louden}, {Metrailler}, {Moyano},
  {Pollacco}, {Poppenhaeger}, {Queloz}, {Raddi}, {Rauer}, {Raynard}, {Smith},
  {Soto}, {Thompson}, {Titz-Weider}, {Udry}, {Walker}, {Watson}, {West}, \&
  {Wheatley}}]{Bayliss2018}
{Bayliss}, D., {Gillen}, E., {Eigm{\"u}ller}, P., {et~al.} 2018,
  \href{https://ui.adsabs.harvard.edu/abs/2018MNRAS.475.4467B}{\mnras, 475,
  4467}

\bibitem[{{Bitsch} {et~al.}(2015){Bitsch}, {Lambrechts}, \&
  {Johansen}}]{Bitsch2015}
{Bitsch}, B., {Lambrechts}, M., \& {Johansen}, A. 2015,
  \href{https://ui.adsabs.harvard.edu/abs/2015A&A...582A.112B}{\aap, 582, A112}

\bibitem[{{Blanco-Pozo} {et~al.}(2022){Blanco-Pozo}, {Perger}, {Damasso},
  {Anglada-Escud{\'e}}, {Ribas}, {Baroch}, {Caballero}, {Cifuentes}, {Jeffers},
  {Lafarga}, {Kaminski}, {Kaur}, {Nagel}, {Perdelwitz}, {P{\' e}rez-Torres}, \&
  {et al.}}]{BlancoPozo2022}
{Blanco-Pozo}, J., {Perger}, M., {Damasso}, M., {et~al.} 2022,
  \href{https://ui.adsabs.harvard.edu/abs/2022arXiv221207332S}{\aap, submitted}

\bibitem[{{Bluhm} {et~al.}(2020){Bluhm}, {Luque}, {Espinoza}, {Pall{\'e}},
  {Caballero}, {Dreizler}, {Livingston}, {Mathur}, {Quirrenbach}, {Stock}, {Van
  Eylen}, {Nowak}, {L{\'o}pez}, {Csizmadia}, {Zapatero Osorio}, {Sch{\"o}fer},
  {Lillo-Box}, {Oshagh}, {Gonz{\'a}lez-{\'A}lvarez}, {Amado}, {Barrado},
  {B{\'e}jar}, {Cale}, {Chaturvedi}, {Cifuentes}, {Cochran}, {Collins},
  {Collins}, {Cort{\'e}s-Contreras}, {D{\'\i}ez Alonso}, {El Mufti},
  {Ercolino}, {Fridlund}, {Gaidos}, {Garc{\'\i}a}, {Georgieva},
  {Gonz{\'a}lez-Cuesta}, {Guerra}, {Hatzes}, {Henning}, {Herrero}, {Hidalgo},
  {Isopi}, {Jeffers}, {Jenkins}, {Jensen}, {K{\'a}bath}, {Kaminski}, {Kemmer},
  {Korth}, {Kossakowski}, {K{\"u}rster}, {Lafarga}, {Mallia}, {Montes},
  {Morales}, {Morales-Calder{\'o}n}, {Murgas}, {Narita}, {Passegger}, {Pedraz},
  {Persson}, {Plavchan}, {Rauer}, {Redfield}, {Reffert}, {Reiners}, {Ribas},
  {Ricker}, {Rodr{\'\i}guez-L{\'o}pez}, {Santos}, {Seager}, {Schlecker},
  {Schweitzer}, {Shan}, {Soto}, {Subjak}, {Tal-Or}, {Trifonov}, {Vanaverbeke},
  {Vanderspek}, {Wittrock}, {Zechmeister}, \& {Zohrabi}}]{Bluhm2020}
{Bluhm}, P., {Luque}, R., {Espinoza}, N., {et~al.} 2020,
  \href{https://ui.adsabs.harvard.edu/abs/2020A&A...639A.132B}{\aap, 639, A132}

\bibitem[{{Bluhm} {et~al.}(2021){Bluhm}, {Pall{\'e}}, {Molaverdikhani},
  {Kemmer}, {Hatzes}, {Kossakowski}, {Stock}, {Caballero}, {Lillo-Box},
  {B{\'e}jar}, {Soto}, {Amado}, {Brown}, {Cadieux}, {Cloutier}, {Collins},
  {Collins}, {Cort{\'e}s-Contreras}, {Doyon}, {Dreizler}, {Espinoza}, {Fukui},
  {Gonz{\'a}lez-{\'A}lvarez}, {Henning}, {Horne}, {Jeffers}, {Jenkins},
  {Jensen}, {Kaminski}, {Kielkopf}, {Kusakabe}, {K{\"u}rster},
  {Lafreni{\`e}re}, {Luque}, {Murgas}, {Montes}, {Morales}, {Narita},
  {Passegger}, {Quirrenbach}, {Sch{\"o}fer}, {Reffert}, {Reiners}, {Ribas},
  {Ricker}, {Seager}, {Schweitzer}, {Schwarz}, {Tamura}, {Trifonov},
  {Vanderspek}, {Winn}, {Zechmeister}, \& {Zapatero Osorio}}]{Bluhm2021}
{Bluhm}, P., {Pall{\'e}}, E., {Molaverdikhani}, K., {et~al.} 2021,
  \href{https://ui.adsabs.harvard.edu/abs/2021A&A...650A..78B}{\aap, 650, A78}

\bibitem[{{Bonfils} {et~al.}(2018){Bonfils}, {Astudillo-Defru}, {D{\'\i}az},
  {Almenara}, {Forveille}, {Bouchy}, {Delfosse}, {Lovis}, {Mayor}, {Murgas},
  {Pepe}, {Santos}, {S{\'e}gransan}, {Udry}, \&
  {W{\"u}nsche}}]{2018A&A...613A..25B}
{Bonfils}, X., {Astudillo-Defru}, N., {D{\'\i}az}, R., {et~al.} 2018,
  \href{https://ui.adsabs.harvard.edu/abs/2018A&A...613A..25B}{\aap, 613, A25}

\bibitem[{{Bonfils} {et~al.}(2005){Bonfils}, {Forveille}, {Delfosse}, {Udry},
  {Mayor}, {Perrier}, {Bouchy}, {Pepe}, {Queloz}, \& {Bertaux}}]{Bonfils2005}
{Bonfils}, X., {Forveille}, T., {Delfosse}, X., {et~al.} 2005,
  \href{https://ui.adsabs.harvard.edu/abs/2005A&A...443L..15B}{\aap, 443, L15}

\bibitem[{{Bonfils} {et~al.}(2013){Bonfils}, {Lo Curto}, {Correia}, {Laskar},
  {Udry}, {Delfosse}, {Forveille}, {Astudillo-Defru}, {Benz}, {Bouchy},
  {Gillon}, {H{\'e}brard}, {Lovis}, {Mayor}, {Moutou}, {Naef}, {Neves}, {Pepe},
  {Perrier}, {Queloz}, {Santos}, \& {S{\'e}gransan}}]{Bonfils2013}
{Bonfils}, X., {Lo Curto}, G., {Correia}, A.~C.~M., {et~al.} 2013,
  \href{https://ui.adsabs.harvard.edu/abs/2013A&A...556A.110B}{\aap, 556, A110}

\bibitem[{{Burn} {et~al.}(2021){Burn}, {Schlecker}, {Mordasini}, {Emsenhuber},
  {Alibert}, {Henning}, {Klahr}, \& {Benz}}]{Burn2021}
{Burn}, R., {Schlecker}, M., {Mordasini}, C., {et~al.} 2021,
  \href{https://ui.adsabs.harvard.edu/abs/2021A&A...656A..72B}{\aap, 656, A72}

\bibitem[{{Burt} {et~al.}(2014){Burt}, {Vogt}, {Butler}, {Hanson}, {Meschiari},
  {Rivera}, {Henry}, \& {Laughlin}}]{2014ApJ...789..114B}
{Burt}, J., {Vogt}, S.~S., {Butler}, R.~P., {et~al.} 2014,
  \href{https://ui.adsabs.harvard.edu/abs/2014ApJ...789..114B}{\apj, 789, 114}

\bibitem[{{Butler} {et~al.}(2009){Butler}, {Howard}, {Vogt}, \&
  {Wright}}]{Butler2009}
{Butler}, R.~P., {Howard}, A.~W., {Vogt}, S.~S., \& {Wright}, J.~T. 2009,
  \href{https://ui.adsabs.harvard.edu/abs/2009ApJ...691.1738B}{\apj, 691, 1738}

\bibitem[{{Butler} {et~al.}(2006){Butler}, {Johnson}, {Marcy}, {Wright},
  {Vogt}, \& {Fischer}}]{2006PASP..118.1685B}
{Butler}, R.~P., {Johnson}, J.~A., {Marcy}, G.~W., {et~al.} 2006,
  \href{https://ui.adsabs.harvard.edu/abs/2006PASP..118.1685B}{\pasp, 118,
  1685}

\bibitem[{{Butler} {et~al.}(2017){Butler}, {Vogt}, {Laughlin}, {Burt},
  {Rivera}, {Tuomi}, {Teske}, {Arriagada}, {Diaz}, {Holden}, \&
  {Keiser}}]{Butler2017AJ....153..208B}
{Butler}, R.~P., {Vogt}, S.~S., {Laughlin}, G., {et~al.} 2017,
  \href{https://ui.adsabs.harvard.edu/abs/2017AJ....153..208B}{\aj, 153, 208}

\bibitem[{{Caballero} {et~al.}(2016){Caballero}, {Gu{\`a}rdia}, {L{\'o}pez del
  Fresno}, {Zechmeister}, {de Juan}, {Alonso-Floriano}, {Amado}, {Colom{\'e}},
  {Cort{\'e}s-Contreras}, {Garc{\'{\i}}a-Piquer}, {Gesa}, {de Guindos},
  {Hagen}, {Helmling}, {Hern{\'a}ndez Casta{\~n}o}, {K{\"u}rster},
  {L{\'o}pez-Santiago}, {Montes}, {Morales Mu{\~n}oz}, {Pavlov}, {Quirrenbach},
  {Reiners}, {Ribas}, {Seifert}, \& {Solano}}]{Caballero2016}
{Caballero}, J.~A., {Gu{\`a}rdia}, J., {L{\'o}pez del Fresno}, M., {et~al.}
  2016, in \procspie, Vol. 9910, Observatory Operations: Strategies, Processes,
  and Systems VI, 99100E

\bibitem[{{Cadieux} {et~al.}(2022){Cadieux}, {Doyon}, {Plotnykov},
  {H{\'e}brard}, {Jahandar}, {Artigau}, {Valencia}, {Cook}, {Martioli},
  {Vandal}, {Donati}, {Cloutier}, {Narita}, {Fukui}, {Hirano}, {Bouchy},
  {Cowan}, {Gonzales}, {Ciardi}, {Stassun}, {Arnold}, {Benneke}, {Boisse},
  {Bonfils}, {Carmona}, {Cort{\'e}s-Zuleta}, {Delfosse}, {Forveille},
  {Fouqu{\'e}}, {da Silva}, {Jenkins}, {Kiefer}, {K{\'o}sp{\'a}l},
  {Lafreni{\`e}re}, {Martins}, {Moutou}, {do Nascimento}, {Ould-Elhkim},
  {Pelletier}, {Twicken}, {Bouma}, {Cartwright}, {Darveau-Bernier}, {Grankin},
  {Ikoma}, {Kagetani}, {Kawauchi}, {Kodama}, {Kotani}, {Latham}, {Menou},
  {Ricker}, {Seager}, {Tamura}, {Vanderspek}, \&
  {Watanabe}}]{2022AJ....164...96C}
{Cadieux}, C., {Doyon}, R., {Plotnykov}, M., {et~al.} 2022,
  \href{https://ui.adsabs.harvard.edu/abs/2022AJ....164...96C}{\aj, 164, 96}

\bibitem[{{Cale} {et~al.}(2021){Cale}, {Reefe}, {Plavchan}, {Tanner}, {Gaidos},
  {Gagn{\'e}}, {Gao}, {Kane}, {B{\'e}jar}, {Lodieu}, {Anglada-Escud{\'e}},
  {Ribas}, {Pall{\'e}}, {Quirrenbach}, {Amado}, {Reiners}, {Caballero}, {Rosa
  Zapatero Osorio}, {Dreizler}, {Howard}, {Fulton}, {Xuesong Wang}, {Collins},
  {El Mufti}, {Wittrock}, {Gilbert}, {Barclay}, {Klein}, {Martioli},
  {Wittenmyer}, {Wright}, {Addison}, {Hirano}, {Tamura}, {Kotani}, {Narita},
  {Vermilion}, {Lee}, {Geneser}, {Teske}, {Quinn}, {Latham}, {Esquerdo},
  {Calkins}, {Berlind}, {Zohrabi}, {Stibbards}, {Kotnana}, {Jenkins},
  {Twicken}, {Henze}, {Kidwell}, {Burke}, {Villase{\~n}or}, \&
  {Boyd}}]{Cale2021}
{Cale}, B.~L., {Reefe}, M., {Plavchan}, P., {et~al.} 2021,
  \href{https://ui.adsabs.harvard.edu/abs/2021AJ....162..295C}{\aj, 162, 295}

\bibitem[{{Carleo} {et~al.}(2020){Carleo}, {Malavolta}, {Lanza}, {Damasso},
  {Desidera}, {Borsa}, {Mallonn}, {Pinamonti}, {Gratton}, {Alei}, {Benatti},
  {Mancini}, {Maldonado}, {Biazzo}, {Esposito}, {Frustagli},
  {Gonz{\'a}lez-{\'A}lvarez}, {Micela}, {Scandariato}, {Sozzetti}, {Affer},
  {Bignamini}, {Bonomo}, {Claudi}, {Cosentino}, {Covino}, {Fiorenzano},
  {Giacobbe}, {Harutyunyan}, {Leto}, {Maggio}, {Molinari}, {Nascimbeni},
  {Pagano}, {Pedani}, {Piotto}, {Poretti}, {Rainer}, {Redfield}, {Baffa},
  {Baruffolo}, {Buchschacher}, {Billotti}, {Cecconi}, {Falcini}, {Fantinel},
  {Fini}, {Galli}, {Ghedina}, {Ghinassi}, {Giani}, {Gonzalez}, {Gonzalez},
  {Guerra}, {Hernandez Diaz}, {Hernandez}, {Iuzzolino}, {Lodi}, {Oliva},
  {Origlia}, {Perez Ventura}, {Puglisi}, {Riverol}, {Riverol}, {San Juan},
  {Sanna}, {Scuderi}, {Seemann}, {Sozzi}, \& {Tozzi}}]{Carleo2020}
{Carleo}, I., {Malavolta}, L., {Lanza}, A.~F., {et~al.} 2020,
  \href{https://ui.adsabs.harvard.edu/abs/2020A&A...638A...5C}{\aap, 638, A5}

\bibitem[{{Casasayas-Barris} {et~al.}(2021){Casasayas-Barris}, {Orell-Miquel},
  {Stangret}, {Nortmann}, {Yan}, {Oshagh}, {Palle}, {Sanz-Forcada},
  {L{\'o}pez-Puertas}, {Nagel}, {Luque}, {Morello}, {Snellen}, {Zechmeister},
  {Quirrenbach}, {Caballero}, {Ribas}, {Reiners}, {Amado}, {Bergond}, {Czesla},
  {Henning}, {Khalafinejad}, {Molaverdikhani}, {Montes}, {Perger},
  {S{\'a}nchez-L{\'o}pez}, \& {Sedaghati}}]{CasasayasBarris2021}
{Casasayas-Barris}, N., {Orell-Miquel}, J., {Stangret}, M., {et~al.} 2021,
  \href{https://ui.adsabs.harvard.edu/abs/2021A&A...654A.163C}{\aap, 654, A163}

\bibitem[{{Casasayas-Barris} {et~al.}(2020){Casasayas-Barris}, {Pall{\'e}},
  {Yan}, {Chen}, {Luque}, {Stangret}, {Nagel}, {Zechmeister}, {Oshagh},
  {Sanz-Forcada}, {Nortmann}, {Alonso-Floriano}, {Amado}, {Caballero},
  {Czesla}, {Khalafinejad}, {L{\'o}pez-Puertas}, {L{\'o}pez-Santiago},
  {Molaverdikhani}, {Montes}, {Quirrenbach}, {Reiners}, {Ribas},
  {S{\'a}nchez-L{\'o}pez}, \& {Zapatero Osorio}}]{CasasayasBarris2020}
{Casasayas-Barris}, N., {Pall{\'e}}, E., {Yan}, F., {et~al.} 2020,
  \href{https://ui.adsabs.harvard.edu/abs/2020A&A...635A.206C}{\aap, 635, A206}

\bibitem[{{Chaturvedi} {et~al.}(2022){Chaturvedi}, {Bluhm}, {Nagel}, {Hatzes},
  {Morello}, {Brady}, {Korth}, {Molaverdikhani}, {Kossakowski}, {Caballero},
  {Guenther}, {Pall{\'e}}, {Espinoza}, {Seifahrt}, {Lodieu}, {Cifuentes},
  {Furlan}, {Amado}, {Barclay}, {Bean}, {B{\'e}jar}, {Bergond}, {Boyle},
  {Ciardi}, {Collins}, {Collins}, {Esparza-Borges}, {Fukui}, {Gnilka}, {Goeke},
  {Guerra}, {Henning}, {Herrero}, {Howell}, {Jeffers}, {Jenkins}, {Jensen},
  {Kasper}, {Kodama}, {Latham}, {L{\'o}pez-Gonz{\'a}lez}, {Luque}, {Montes},
  {Morales}, {Mori}, {Murgas}, {Narita}, {Nowak}, {Parviainen}, {Passegger},
  {Quirrenbach}, {Reffert}, {Reiners}, {Ribas}, {Ricker}, {Rodriguez},
  {Rodr{\'\i}guez-L{\'o}pez}, {Schlecker}, {Schwarz}, {Schweitzer}, {Seager},
  {Stef{\'a}nsson}, {Stockdale}, {Tal-Or}, {Twicken}, {Vanaverbeke}, {Wang},
  {Watanabe}, {Winn}, \& {Zechmeister}}]{Chaturvedi2022}
{Chaturvedi}, P., {Bluhm}, P., {Nagel}, E., {et~al.} 2022,
  \href{https://ui.adsabs.harvard.edu/abs/2022A&A...666A.155C}{\aap, 666, A155}

\bibitem[{{Cifuentes} {et~al.}(2020){Cifuentes}, {Caballero},
  {Cort{\'e}s-Contreras}, {Montes}, {Abell{\'a}n}, {Dorda}, {Holgado},
  {Zapatero Osorio}, {Morales}, {Amado}, {Passegger}, {Quirrenbach}, {Reiners},
  {Ribas}, {Sanz-Forcada}, {Schweitzer}, {Seifert}, \&
  {Solano}}]{Cifuentes2020A&A...642A.115C}
{Cifuentes}, C., {Caballero}, J.~A., {Cort{\'e}s-Contreras}, M., {et~al.} 2020,
  \href{https://ui.adsabs.harvard.edu/abs/2020A&A...642A.115C}{\aap, 642, A115}

\bibitem[{{Cloutier} {et~al.}(2017){Cloutier}, {Astudillo-Defru}, {Doyon},
  {Bonfils}, {Almenara}, {Benneke}, {Bouchy}, {Delfosse}, {Ehrenreich},
  {Forveille}, {Lovis}, {Mayor}, {Menou}, {Murgas}, {Pepe}, {Rowe}, {Santos},
  {Udry}, \& {W{\"u}nsche}}]{Cloutier2017A&A...608A..35C}
{Cloutier}, R., {Astudillo-Defru}, N., {Doyon}, R., {et~al.} 2017,
  \href{https://ui.adsabs.harvard.edu/abs/2017A&A...608A..35C}{\aap, 608, A35}

\bibitem[{{Cort{\'e}s-Contreras} {et~al.}(2017){Cort{\'e}s-Contreras},
  {B{\'e}jar}, {Caballero}, {Gauza}, {Montes}, {Alonso-Floriano}, {Jeffers},
  {Morales}, {Reiners}, {Ribas}, {Sch{\"o}fer}, {Quirrenbach}, {Amado},
  {Mundt}, \& {Seifert}}]{Cortes2017}
{Cort{\'e}s-Contreras}, M., {B{\'e}jar}, V.~J.~S., {Caballero}, J.~A., {et~al.}
  2017, \href{https://ui.adsabs.harvard.edu/abs/2017A&A...597A..47C}{\aap, 597,
  A47}

\bibitem[{{Czesla} {et~al.}(2022){Czesla}, {Lamp{\'o}n}, {Sanz-Forcada},
  {Garc{\'\i}a Mu{\~n}oz}, {L{\'o}pez-Puertas}, {Nortmann}, {Yan}, {Nagel},
  {Yan}, {Schmitt}, {Aceituno}, {Amado}, {Caballero}, {Casasayas-Barris},
  {Henning}, {Khalafinejad}, {Molaverdikhani}, {Montes}, {Pall{\'e}},
  {Reiners}, {Schneider}, {Ribas}, {Quirrenbach}, {Zapatero Osorio}, \&
  {Zechmeister}}]{Czesla2022}
{Czesla}, S., {Lamp{\'o}n}, M., {Sanz-Forcada}, J., {et~al.} 2022,
  \href{https://ui.adsabs.harvard.edu/abs/2022A&A...657A...6C}{\aap, 657, A6}

\bibitem[{{Damasso} {et~al.}(2022){Damasso}, {Perger}, {Almenara}, {Nardiello},
  {P{\'e}rez-Torres}, {Sozzetti}, {Hara}, {Quirrenbach}, {Bonfils}, {Zapatero
  Osorio}, {Astudillo-Defru}, {Gonz{\'a}lez Hern{\'a}ndez}, {Su{\'a}rez
  Mascareno}, {Amado}, {Forveille}, {Lillo-Box}, {Alibert}, {Caballero},
  {Cifuentes}, {Delfosse}, {Figueira}, {Galad{\'\i}-Enr{\'\i}quez}, {Hatzes},
  {Henning}, {Kaminski}, {Mayor}, {Murgas}, {Montes}, {Pinamonti}, {Reiners},
  {Ribas}, {B{\'e}jar}, {Schweitzer}, \& {Zechmeister}}]{Damasso2022}
{Damasso}, M., {Perger}, M., {Almenara}, J.~M., {et~al.} 2022,
  \href{https://ui.adsabs.harvard.edu/abs/2022A&A...666A.187D}{\aap, 666, A187}

\bibitem[{{David} {et~al.}(2016){David}, {Hillenbrand}, {Petigura},
  {Carpenter}, {Crossfield}, {Hinkley}, {Ciardi}, {Howard}, {Isaacson}, {Cody},
  {Schlieder}, {Beichman}, \& {Barenfeld}}]{2016Natur.534..658D}
{David}, T.~J., {Hillenbrand}, L.~A., {Petigura}, E.~A., {et~al.} 2016,
  \href{https://ui.adsabs.harvard.edu/abs/2016Natur.534..658D}{\nat, 534, 658}

\bibitem[{{Delfosse} {et~al.}(1999){Delfosse}, {Forveille}, {Beuzit}, {Udry},
  {Mayor}, \& {Perrier}}]{Delfosse1999}
{Delfosse}, X., {Forveille}, T., {Beuzit}, J.~L., {et~al.} 1999,
  \href{https://ui.adsabs.harvard.edu/abs/1999A&A...344..897D}{\aap, 344, 897}

\bibitem[{{Demory} {et~al.}(2020){Demory}, {Pozuelos}, {G{\'o}mez Maqueo Chew},
  {Sabin}, {Petrucci}, {Schroffenegger}, {Grimm}, {Sestovic}, {Gillon},
  {McCormac}, {Barkaoui}, {Benz}, {Bieryla}, {Bouchy}, {Burdanov}, {Collins},
  {de Wit}, {Dressing}, {Garcia}, {Giacalone}, {Guerra}, {Haldemann}, {Heng},
  {Jehin}, {Jofr{\'e}}, {Kane}, {Lillo-Box}, {Maign{\'e}}, {Mordasini},
  {Morris}, {Niraula}, {Queloz}, {Rackham}, {Savel}, {Soubkiou}, {Srdoc},
  {Stassun}, {Triaud}, {Zambelli}, {Ricker}, {Latham}, {Seager}, {Winn},
  {Jenkins}, {Calvario-Vel{\'a}squez}, {Franco Herrera}, {Colorado}, {Cadena
  Zepeda}, {Figueroa}, {Watson}, {Lugo-Ibarra}, {Carigi}, {Guisa}, {Herrera},
  {Sierra D{\'\i}az}, {Su{\'a}rez}, {Barrado}, {Batalha}, {Benkhaldoun},
  {Chontos}, {Dai}, {Essack}, {Ghachoui}, {Huang}, {Huber}, {Isaacson},
  {Lissauer}, {Morales-Calder{\'o}n}, {Robertson}, {Roy}, {Twicken},
  {Vanderburg}, \& {Weiss}}]{2020A&A...642A..49D}
{Demory}, B.~O., {Pozuelos}, F.~J., {G{\'o}mez Maqueo Chew}, Y., {et~al.} 2020,
  \href{https://ui.adsabs.harvard.edu/abs/2020A&A...642A..49D}{\aap, 642, A49}

\bibitem[{{D{\'\i}az} {et~al.}(2019){D{\'\i}az}, {Delfosse}, {Hobson},
  {Boisse}, {Astudillo-Defru}, {Bonfils}, {Henry}, {Arnold}, {Bouchy},
  {Bourrier}, {Brugger}, {Dalal}, {Deleuil}, {Demangeon}, {Dolon}, {Dumusque},
  {Forveille}, {Hara}, {H{\'e}brard}, {Kiefer}, {Lopez}, {Mignon}, {Moreau},
  {Mousis}, {Moutou}, {Pepe}, {Perruchot}, {Richaud}, {Santerne}, {Santos},
  {Sottile}, {Stalport}, {S{\'e}gransan}, {Udry}, {Unger}, \&
  {Wilson}}]{Diaz2019A&A...625A..17D}
{D{\'\i}az}, R.~F., {Delfosse}, X., {Hobson}, M.~J., {et~al.} 2019,
  \href{https://ui.adsabs.harvard.edu/abs/2019A&A...625A..17D}{\aap, 625, A17}

\bibitem[{{D{\'\i}ez Alonso} {et~al.}(2019){D{\'\i}ez Alonso}, {Caballero},
  {Montes}, {de Cos Juez}, {Dreizler}, {Dubois}, {Jeffers}, {Lalitha}, {Naves},
  {Reiners}, {Ribas}, {Vanaverbeke}, {Amado}, {B{\'e}jar},
  {Cort{\'e}s-Contreras}, {Herrero}, {Hidalgo}, {K{\"u}rster}, {Logie},
  {Quirrenbach}, {Rau}, {Seifert}, {Sch{\"o}fer}, \& {Tal-Or}}]{DiezAlonso2019}
{D{\'\i}ez Alonso}, E., {Caballero}, J.~A., {Montes}, D., {et~al.} 2019,
  \href{https://ui.adsabs.harvard.edu/abs/2019A&A...621A.126D}{\aap, 621, A126}

\bibitem[{{Dreizler} {et~al.}(2020){Dreizler}, {Crossfield}, {Kossakowski},
  {Plavchan}, {Jeffers}, {Kemmer}, {Luque}, {Espinoza}, {Pall{\'e}}, {Stassun},
  {Matthews}, {Cale}, {Caballero}, {Schlecker}, {Lillo-Box}, {Zechmeister},
  {Lalitha}, {Reiners}, {Soubkiou}, {Bitsch}, {Zapatero Osorio}, {Chaturvedi},
  {Hatzes}, {Ricker}, {Vanderspek}, {Latham}, {Seager}, {Winn}, {Jenkins},
  {Aceituno}, {Amado}, {Barkaoui}, {Barbieri}, {Batalha}, {Bauer}, {Benneke},
  {Benkhaldoun}, {Beichman}, {Berberian}, {Burt}, {Butler}, {Caldwell},
  {Chintada}, {Chontos}, {Christiansen}, {Ciardi}, {Cifuentes}, {Collins},
  {Collins}, {Combs}, {Cort{\'e}s-Contreras}, {Crane}, {Daylan}, {Dragomir},
  {Esparza-Borges}, {Evans}, {Feng}, {Flowers}, {Fukui}, {Fulton}, {Furlan},
  {Gaidos}, {Geneser}, {Giacalone}, {Gillon}, {Gonzales}, {Gorjian}, {Hellier},
  {Hidalgo}, {Howard}, {Howell}, {Huber}, {Isaacson}, {Jehin}, {Jensen},
  {Kaminski}, {Kane}, {Kawauchi}, {Kielkopf}, {Klahr}, {Kosiarek}, {Kreidberg},
  {K{\"u}rster}, {Lafarga}, {Livingston}, {Louie}, {Mann}, {Madrigal-Aguado},
  {Matson}, {Mocnik}, {Morales}, {Muirhead}, {Murgas}, {Nandakumar}, {Narita},
  {Nowak}, {Oshagh}, {Parviainen}, {Passegger}, {Pollacco}, {Pozuelos},
  {Quirrenbach}, {Reefe}, {Ribas}, {Robertson}, {Rodr{\'\i}guez-L{\'o}pez},
  {Rose}, {Roy}, {Schweitzer}, {Schlieder}, {Shectman}, {Tanner},
  {{\c{S}}enavc{\i}}, {Teske}, {Twicken}, {Villasenor}, {Wang}, {Weiss},
  {Wittrock}, {Y{\i}lmaz}, \& {Zohrabi}}]{Dreizler2020}
{Dreizler}, S., {Crossfield}, I.~J.~M., {Kossakowski}, D., {et~al.} 2020,
  \href{https://ui.adsabs.harvard.edu/abs/2020A&A...644A.127D}{\aap, 644, A127}

\bibitem[{{Dressing} \& {Charbonneau}(2015)}]{Dressing2015}
{Dressing}, C.~D. \& {Charbonneau}, D. 2015,
  \href{https://ui.adsabs.harvard.edu/abs/2015ApJ...807...45D}{\apj, 807, 45}

\bibitem[{{Emsenhuber} {et~al.}(2021){Emsenhuber}, {Mordasini}, {Burn},
  {Alibert}, {Benz}, \& {Asphaug}}]{Emsenhuber2021}
{Emsenhuber}, A., {Mordasini}, C., {Burn}, R., {et~al.} 2021,
  \href{https://ui.adsabs.harvard.edu/abs/2021A&A...656A..69E}{\aap, 656, A69}

\bibitem[{{Endl} {et~al.}(2003){Endl}, {Cochran}, {Tull}, \&
  {MacQueen}}]{Endl2003}
{Endl}, M., {Cochran}, W.~D., {Tull}, R.~G., \& {MacQueen}, P.~J. 2003,
  \href{https://ui.adsabs.harvard.edu/abs/2003AJ....126.3099E}{\aj, 126, 3099}

\bibitem[{{Endl} {et~al.}(2008){Endl}, {Cochran}, {Wittenmyer}, \&
  {Boss}}]{Endl2008}
{Endl}, M., {Cochran}, W.~D., {Wittenmyer}, R.~A., \& {Boss}, A.~P. 2008,
  \href{https://ui.adsabs.harvard.edu/abs/2008ApJ...673.1165E}{\apj, 673, 1165}

\bibitem[{{Espinoza} {et~al.}(2022){Espinoza}, {Pall{\'e}}, {Kemmer}, {Luque},
  {Caballero}, {Cifuentes}, {Herrero}, {S{\'a}nchez B{\'e}jar}, {Stock},
  {Molaverdikhani}, {Morello}, {Kossakowski}, {Schlecker}, {Amado}, {Bluhm},
  {Cort{\'e}s-Contreras}, {Henning}, {Kreidberg}, {K{\"u}rster}, {Lafarga},
  {Lodieu}, {Morales}, {Oshagh}, {Passegger}, {Pavlov}, {Quirrenbach},
  {Reffert}, {Reiners}, {Ribas}, {Rodr{\'\i}guez}, {L{\'o}pez}, {Schweitzer},
  {Trifonov}, {Chaturvedi}, {Dreizler}, {Jeffers}, {Kaminski},
  {L{\'o}pez-Gonz{\'a}lez}, {Lillo-Box}, {Montes}, {Nowak}, {Pedraz},
  {Vanaverbeke}, {Zapatero Osorio}, {Zechmeister}, {Collins}, {Girardin},
  {Guerra}, {Naves}, {Crossfield}, {Matthews}, {Howell}, {Ciardi}, {Gonzales},
  {Matson}, {Beichman}, {Schlieder}, {Barclay}, {Vezie}, {Villase{\~n}or},
  {Daylan}, {Mireies}, {Dragomir}, {Twicken}, {Jenkins}, {Winn}, {Latham},
  {Ricker}, \& {Seager}}]{Espinoza22}
{Espinoza}, N., {Pall{\'e}}, E., {Kemmer}, J., {et~al.} 2022,
  \href{https://ui.adsabs.harvard.edu/abs/2022AJ....163..133E}{\aj, 163, 133}

\bibitem[{{Feng} {et~al.}(2020){Feng}, {Shectman}, {Clement}, {Vogt}, {Tuomi},
  {Teske}, {Burt}, {Crane}, {Holden}, {Wang}, {Thompson}, {D{\'\i}az}, \&
  {Butler}}]{Feng2020}
{Feng}, F., {Shectman}, S.~A., {Clement}, M.~S., {et~al.} 2020,
  \href{https://ui.adsabs.harvard.edu/abs/2020ApJS..250...29F}{\apjs, 250, 29}

\bibitem[{{Forveille} {et~al.}(2009){Forveille}, {Bonfils}, {Delfosse},
  {Gillon}, {Udry}, {Bouchy}, {Lovis}, {Mayor}, {Pepe}, {Perrier}, {Queloz},
  {Santos}, \& {Bertaux}}]{Forveille2009}
{Forveille}, T., {Bonfils}, X., {Delfosse}, X., {et~al.} 2009,
  \href{https://ui.adsabs.harvard.edu/abs/2009A&A...493..645F}{\aap, 493, 645}

\bibitem[{{Fuhrmeister} {et~al.}(2019{\natexlab{a}}){Fuhrmeister}, {Czesla},
  {Hildebrandt}, {Nagel}, {Schmitt}, {Hintz}, {Johnson}, {Sanz-Forcada},
  {Sch{\"o}fer}, {Jeffers}, {Caballero}, {Zechmeister}, {Reiners}, {Ribas},
  {Amado}, {Quirrenbach}, {Bauer}, {B{\'e}jar}, {Cort{\'e}s-Contreras},
  {D{\'\i}ez-Alonso}, {Dreizler}, {Galad{\'\i}-Enr{\'\i}quez}, {Guenther},
  {Kaminski}, {K{\"u}rster}, {Lafarga}, \& {Montes}}]{Fuhrmeister2019b}
{Fuhrmeister}, B., {Czesla}, S., {Hildebrandt}, L., {et~al.}
  2019{\natexlab{a}},
  \href{https://ui.adsabs.harvard.edu/abs/2019A&A...632A..24F}{\aap, 632, A24}

\bibitem[{{Fuhrmeister} {et~al.}(2020){Fuhrmeister}, {Czesla}, {Hildebrandt},
  {Nagel}, {Schmitt}, {Jeffers}, {Caballero}, {Hintz}, {Johnson},
  {Sch{\"o}fer}, {Zechmeister}, {Reiners}, {Ribas}, {Amado}, {Quirrenbach},
  {Nortmann}, {Bauer}, {B{\'e}jar}, {Cort{\'e}s-Contreras}, {Dreizler},
  {Galad{\'\i}-Enr{\'\i}quez}, {Hatzes}, {Kaminski}, {K{\"u}rster}, {Lafarga},
  \& {Montes}}]{Fuhrmeister2020}
{Fuhrmeister}, B., {Czesla}, S., {Hildebrandt}, L., {et~al.} 2020,
  \href{https://ui.adsabs.harvard.edu/abs/2020A&A...640A..52F}{\aap, 640, A52}

\bibitem[{{Fuhrmeister} {et~al.}(2022){Fuhrmeister}, {Czesla}, {Nagel},
  {Reiners}, {Schmitt}, {Jeffers}, {Caballero}, {Shulyak}, {Johnson},
  {Zechmeister}, {Montes}, {L{\'o}pez-Gallifa}, {Ribas}, {Quirrenbach},
  {Amado}, {Galad{\'\i}-Enr{\'\i}quez}, {Hatzes}, {K{\"u}rster}, {Danielski},
  {B{\'e}jar}, {Kaminski}, {Morales}, \& {Zapatero Osorio}}]{Fuhrmeister2022}
{Fuhrmeister}, B., {Czesla}, S., {Nagel}, E., {et~al.} 2022,
  \href{https://ui.adsabs.harvard.edu/abs/2022A&A...657A.125F}{\aap, 657, A125}

\bibitem[{{Fuhrmeister} {et~al.}(2018){Fuhrmeister}, {Czesla}, {Schmitt},
  {Jeffers}, {Caballero}, {Zechmeister}, {Reiners}, {Ribas}, {Amado},
  {Quirrenbach}, {B{\'e}jar}, {Galad{\'\i}-Enr{\'\i}quez}, {Guenther},
  {K{\"u}rster}, {Montes}, \& {Seifert}}]{Fuhrmeister2018}
{Fuhrmeister}, B., {Czesla}, S., {Schmitt}, J.~H.~M.~M., {et~al.} 2018,
  \href{https://ui.adsabs.harvard.edu/abs/2018A&A...615A..14F}{\aap, 615, A14}

\bibitem[{{Fuhrmeister} {et~al.}(2019{\natexlab{b}}){Fuhrmeister}, {Czesla},
  {Schmitt}, {Johnson}, {Sch{\"o}fer}, {Jeffers}, {Caballero}, {Zechmeister},
  {Reiners}, {Ribas}, {Amado}, {Quirrenbach}, {Bauer}, {B{\'e}jar},
  {Cort{\'e}s-Contreras}, {D{\'\i}ez Alonso}, {Dreizler},
  {Galad{\'\i}-Enr{\'\i}quez}, {Guenther}, {Kaminski}, {K{\"u}rster},
  {Lafarga}, \& {Montes}}]{Fuhrmeister2019}
{Fuhrmeister}, B., {Czesla}, S., {Schmitt}, J.~H.~M.~M., {et~al.}
  2019{\natexlab{b}},
  \href{https://ui.adsabs.harvard.edu/abs/2019A&A...623A..24F}{\aap, 623, A24}

\bibitem[{{Gaia Collaboration} {et~al.}(2021){Gaia Collaboration}, {Smart},
  {Sarro}, {Rybizki}, {Reyl{\'e}}, {Robin}, {Hambly}, {Abbas}, {Barstow}, {de
  Bruijne}, {Bucciarelli}, {Carrasco}, {Cooper}, {Hodgkin}, {Masana},
  {Michalik}, {Sahlmann}, {Sozzetti}, {Brown}, {Vallenari}, {Prusti},
  {Babusiaux}, {Biermann}, {Creevey}, {Evans}, {Eyer}, {Hutton}, {Jansen},
  {Jordi}, {Klioner}, {Lammers}, {Lindegren}, {Luri}, {Mignard}, {Panem},
  {Pourbaix}, {Randich}, {Sartoretti}, {Soubiran}, {Walton}, {Arenou},
  {Bailer-Jones}, {Bastian}, {Cropper}, {Drimmel}, {Katz}, {Lattanzi}, {van
  Leeuwen}, {Bakker}, {Casta{\~n}eda}, {De Angeli}, {Ducourant}, {Fabricius},
  {Fouesneau}, {Fr{\'e}mat}, {Guerra}, {Guerrier}, {Guiraud}, {Jean-Antoine
  Piccolo}, {Messineo}, {Mowlavi}, {Nicolas}, {Nienartowicz}, {Pailler},
  {Panuzzo}, {Riclet}, {Roux}, {Seabroke}, {Sordo}, {Tanga}, {Th{\'e}venin},
  {Gracia-Abril}, {Portell}, {Teyssier}, {Altmann}, {Andrae}, {Bellas-Velidis},
  {Benson}, {Berthier}, {Blomme}, {Brugaletta}, {Burgess}, {Busso}, {Carry},
  {Cellino}, {Cheek}, {Clementini}, {Damerdji}, {Davidson}, {Delchambre},
  {Dell'Oro}, {Fern{\'a}ndez-Hern{\'a}ndez}, {Galluccio}, {Garc{\'\i}a-Lario},
  {Garcia-Reinaldos}, {Gonz{\'a}lez-N{\'u}{\~n}ez}, {Gosset}, {Haigron},
  {Halbwachs}, {Harrison}, {Hatzidimitriou}, {Heiter}, {Hern{\'a}ndez},
  {Hestroffer}, {Holl}, {Jan{\ss}en}, {Jevardat de Fombelle}, {Jordan},
  {Krone-Martins}, {Lanzafame}, {L{\"o}ffler}, {Lorca}, {Manteiga}, {Marchal},
  {Marrese}, {Moitinho}, {Mora}, {Muinonen}, {Osborne}, {Pancino}, {Pauwels},
  {Recio-Blanco}, {Richards}, {Riello}, {Rimoldini}, {Roegiers}, {Siopis},
  {Smith}, {Ulla}, {Utrilla}, {van Leeuwen}, {van Reeven}, {Abreu Aramburu},
  {Accart}, {Aerts}, {Aguado}, {Ajaj}, {Altavilla}, {{\'A}lvarez}, {{\'A}lvarez
  Cid-Fuentes}, {Alves}, {Anderson}, {Anglada Varela}, {Antoja}, {Audard},
  {Baines}, {Baker}, {Balaguer-N{\'u}{\~n}ez}, {Balbinot}, {Balog}, {Barache},
  {Barbato}, {Barros}, {Bartolom{\'e}}, {Bassilana}, {Bauchet},
  {Baudesson-Stella}, {Becciani}, {Bellazzini}, {Bernet}, {Bertone}, {Bianchi},
  {Blanco-Cuaresma}, {Boch}, {Bombrun}, {Bossini}, {Bouquillon}, {Bragaglia},
  {Bramante}, {Breedt}, {Bressan}, {Brouillet}, {Burlacu}, {Busonero},
  {Butkevich}, {Buzzi}, {Caffau}, {Cancelliere}, {C{\'a}novas},
  {Cantat-Gaudin}, {Carballo}, {Carlucci}, {Carnerero}, {Casamiquela},
  {Castellani}, {Castro-Ginard}, {Castro Sampol}, {Chaoul}, {Charlot},
  {Chemin}, {Chiavassa}, {Cioni}, {Comoretto}, {Cornez}, {Cowell}, {Crifo},
  {Crosta}, {Crowley}, {Dafonte}, {Dapergolas}, {David}, {David}, {de Laverny},
  {De Luise}, {De March}, {De Ridder}, {de Souza}, {de Teodoro}, {de Torres},
  {del Peloso}, {del Pozo}, {Delgado}, {Delgado}, {Delisle}, {Di Matteo},
  {Diakite}, {Diener}, {Distefano}, {Dolding}, {Eappachen}, {Edvardsson},
  {Enke}, {Esquej}, {Fabre}, {Fabrizio}, {Faigler}, {Fedorets}, {Fernique},
  {Fienga}, {Figueras}, {Fouron}, {Fragkoudi}, {Fraile}, {Franke}, {Gai},
  {Garabato}, {Garcia-Gutierrez}, {Garc{\'\i}a-Torres}, {Garofalo}, {Gavras},
  {Gerlach}, {Geyer}, {Giacobbe}, {Gilmore}, {Girona}, {Giuffrida}, {Gomel},
  {Gomez}, {Gonzalez-Santamaria}, {Gonz{\'a}lez-Vidal}, {Granvik},
  {Guti{\'e}rrez-S{\'a}nchez}, {Guy}, {Hauser}, {Haywood}, {Helmi}, {Hidalgo},
  {Hilger}, {H{\l}adczuk}, {Hobbs}, {Holland}, {Huckle}, {Jasniewicz},
  {Jonker}, {Juaristi Campillo}, {Julbe}, {Karbevska}, {Kervella}, {Khanna},
  {Kochoska}, {Kontizas}, {Kordopatis}, {Korn}, {Kostrzewa-Rutkowska},
  {Kruszy{\'n}ska}, {Lambert}, {Lanza}, {Lasne}, {Le Campion}, {Le Fustec},
  {Lebreton}, {Lebzelter}, {Leccia}, {Leclerc}, {Lecoeur-Taibi}, {Liao},
  {Licata}, {Lindstr{\o}m}, {Lister}, {Livanou}, {Lobel}, {Madrero Pardo},
  {Managau}, {Mann}, {Marchant}, {Marconi}, {Marcos Santos}, {Marinoni},
  {Marocco}, {Marshall}, {Martin Polo}, {Mart{\'\i}n-Fleitas}, {Masip},
  {Massari}, {Mastrobuono-Battisti}, {Mazeh}, {McMillan}, {Messina}, {Millar},
  {Mints}, {Molina}, {Molinaro}, {Moln{\'a}r}, {Montegriffo}, {Mor},
  {Morbidelli}, {Morel}, {Morris}, {Mulone}, {Munoz}, {Muraveva}, {Murphy},
  {Musella}, {Noval}, {Ord{\'e}novic}, {Orr{\`u}}, {Osinde}, {Pagani},
  {Pagano}, {Palaversa}, {Palicio}, {Panahi}, {Pawlak}, {Pe{\~n}alosa
  Esteller}, {Penttil{\"a}}, {Piersimoni}, {Pineau}, {Plachy}, {Plum},
  {Poggio}, {Poretti}, {Poujoulet}, {Pr{\v{s}}a}, {Pulone}, {Racero},
  {Ragaini}, {Rainer}, {Raiteri}, {Rambaux}, {Ramos}, {Ramos-Lerate}, {Re
  Fiorentin}, {Regibo}, {Ripepi}, {Riva}, {Rixon}, {Robichon}, {Robin},
  {Roelens}, {Rohrbasser}, {Romero-G{\'o}mez}, {Rowell}, {Royer}, {Rybicki},
  {Sadowski}, {Sagrist{\`a} Sell{\'e}s}, {Salgado}, {Salguero}, {Samaras},
  {Sanchez Gimenez}, {Sanna}, {Santove{\~n}a}, {Sarasso}, {Schultheis},
  {Sciacca}, {Segol}, {Segovia}, {S{\'e}gransan}, {Semeux}, {Shahaf},
  {Siddiqui}, {Siebert}, {Siltala}, {Slezak}, {Solano}, {Solitro}, {Souami},
  {Souchay}, {Spagna}, {Spoto}, {Steele}, {Steidelm{\"u}ller}, {Stephenson},
  {S{\"u}veges}, {Szabados}, {Szegedi-Elek}, {Taris}, {Tauran}, {Taylor},
  {Teixeira}, {Thuillot}, {Tonello}, {Torra}, {Torra}, {Turon}, {Unger},
  {Vaillant}, {van Dillen}, {Vanel}, {Vecchiato}, {Viala}, {Vicente},
  {Voutsinas}, {Weiler}, {Wevers}, {Wyrzykowski}, {Yoldas}, {Yvard}, {Zhao},
  {Zorec}, {Zucker}, {Zurbach}, \& {Zwitter}}]{Gaia2021A&A...649A...6G}
{Gaia Collaboration}, {Smart}, R.~L., {Sarro}, L.~M., {et~al.} 2021,
  \href{https://ui.adsabs.harvard.edu/abs/2021A&A...649A...6G}{\aap, 649, A6}

\bibitem[{{Gaidos} {et~al.}(2016){Gaidos}, {Mann}, {Kraus}, \&
  {Ireland}}]{Gaidos2016}
{Gaidos}, E., {Mann}, A.~W., {Kraus}, A.~L., \& {Ireland}, M. 2016,
  \href{https://ui.adsabs.harvard.edu/abs/2016MNRAS.457.2877G}{\mnras, 457,
  2877}

\bibitem[{{Garcia-Piquer} {et~al.}(2017){Garcia-Piquer}, {Morales}, {Ribas},
  {Colom{\'e}}, {Gu{\`a}rdia}, {Perger}, {Caballero}, {Cort{\'e}s-Contreras},
  {Jeffers}, {Reiners}, {Amado}, {Quirrenbach}, \&
  {Seifert}}]{Garcia-Piquer2017A&A...604A..87G}
{Garcia-Piquer}, A., {Morales}, J.~C., {Ribas}, I., {et~al.} 2017,
  \href{https://ui.adsabs.harvard.edu/abs/2017A&A...604A..87G}{\aap, 604, A87}

\bibitem[{{Ghezzi} {et~al.}(2018){Ghezzi}, {Montet}, \& {Johnson}}]{Ghezzi2018}
{Ghezzi}, L., {Montet}, B.~T., \& {Johnson}, J.~A. 2018,
  \href{https://ui.adsabs.harvard.edu/abs/2018ApJ...860..109G}{\apj, 860, 109}

\bibitem[{{Gillon} {et~al.}(2016){Gillon}, {Jehin}, {Lederer}, {Delrez}, {de
  Wit}, {Burdanov}, {Van Grootel}, {Burgasser}, {Triaud}, {Opitom}, {Demory},
  {Sahu}, {Bardalez Gagliuffi}, {Magain}, \& {Queloz}}]{2016Natur.533..221G}
{Gillon}, M., {Jehin}, E., {Lederer}, S.~M., {et~al.} 2016,
  \href{https://ui.adsabs.harvard.edu/abs/2016Natur.533..221G}{\nat, 533, 221}

\bibitem[{{Gonz{\'a}lez-{\'A}lvarez} {et~al.}(2021){Gonz{\'a}lez-{\'A}lvarez},
  {Petralia}, {Micela}, {Maldonado}, {Affer}, {Maggio}, {Covino}, {Damasso},
  {Lanza}, {Perger}, {Pinamonti}, {Poretti}, {Scandariato}, {Sozzetti},
  {Bignamini}, {Giacobbe}, {Leto}, {Pagano}, {Zanmar S{\'a}nchez},
  {Gonz{\'a}lez Hern{\'a}ndez}, {Rebolo}, {Ribas}, {Su{\'a}rez Mascare{\~n}o},
  \& {Toledo-Padr{\'o}n}}]{2021A&A...649A.157G}
{Gonz{\'a}lez-{\'A}lvarez}, E., {Petralia}, A., {Micela}, G., {et~al.} 2021,
  \href{https://ui.adsabs.harvard.edu/abs/2021A&A...649A.157G}{\aap, 649, A157}

\bibitem[{{Gonz{\'a}lez-{\'A}lvarez} {et~al.}(2020){Gonz{\'a}lez-{\'A}lvarez},
  {Zapatero Osorio}, {Caballero}, {Sanz-Forcada}, {B{\'e}jar},
  {Gonz{\'a}lez-Cuesta}, {Dreizler}, {Bauer}, {Rodr{\'\i}guez}, {Tal-Or},
  {Zechmeister}, {Montes}, {L{\'o}pez-Gonz{\'a}lez}, {Ribas}, {Reiners},
  {Quirrenbach}, {Amado}, {Anglada-Escud{\'e}}, {Azzaro},
  {Cort{\'e}s-Contreras}, {Hatzes}, {Henning}, {Jeffers}, {Kaminski},
  {K{\"u}rster}, {Lafarga}, {Morales}, {Pall{\'e}}, {Perger}, \&
  {Schmitt}}]{Gonzalez2020}
{Gonz{\'a}lez-{\'A}lvarez}, E., {Zapatero Osorio}, M.~R., {Caballero}, J.~A.,
  {et~al.} 2020,
  \href{https://ui.adsabs.harvard.edu/abs/2020A&A...637A..93G}{\aap, 637, A93}

\bibitem[{{Gonz{\'a}lez-{\'A}lvarez} {et~al.}(2022){Gonz{\'a}lez-{\'A}lvarez},
  {Zapatero Osorio}, {Sanz-Forcada}, {Caballero}, {Reffert}, {B{\'e}jar},
  {Hatzes}, {Herrero}, {Jeffers}, {Kemmer}, {L{\'o}pez-Gonz{\'a}lez}, {Luque},
  {Molaverdikhani}, {Morello}, {Nagel}, {Quirrenbach}, {Rodr{\'\i}guez},
  {Rodr{\'\i}guez-L{\'o}pez}, {Schlecker}, {Schweitzer}, {Stock}, {Passegger},
  {Trifonov}, {Amado}, {Baker}, {Boyd}, {Cadieux}, {Charbonneau}, {Collins},
  {Doyon}, {Dreizler}, {Espinoza}, {F{\H{u}}r{\'e}sz}, {Furlan}, {Hesse},
  {Howell}, {Jenkins}, {Kidwell}, {Latham}, {McLeod}, {Montes}, {Morales},
  {O'Dwyer}, {Pall{\'e}}, {Pedraz}, {Reiners}, {Ribas}, {Quinn}, {Schnaible},
  {Seager}, {Skinner}, {Smith}, {Schwarz}, {Shporer}, {Vanderspek}, \&
  {Winn}}]{Gonzalez2022a}
{Gonz{\'a}lez-{\'A}lvarez}, E., {Zapatero Osorio}, M.~R., {Sanz-Forcada}, J.,
  {et~al.} 2022,
  \href{https://ui.adsabs.harvard.edu/abs/2022A&A...658A.138G}{\aap, 658, A138}

\bibitem[{{Hatzes}(2019)}]{Hatzes2019}
{Hatzes}, A.~P. 2019, {The Doppler Method for the Detection of Exoplanets} (IOP
  Publishing)

\bibitem[{{Hintz} {et~al.}(2019){Hintz}, {Fuhrmeister}, {Czesla}, {Schmitt},
  {Johnson}, {Schweitzer}, {Caballero}, {Zechmeister}, {Jeffers}, {Reiners},
  {Ribas}, {Amado}, {Quirrenbach}, {Anglada-Escud{\'e}}, {Bauer}, {B{\'e}jar},
  {Cort{\'e}s-Contreras}, {Dreizler}, {Galad{\'\i}-Enr{\'\i}quez}, {Guenther},
  {Hauschildt}, {Kaminski}, {K{\"u}rster}, {Lafarga}, {L{\'o}pez del Fresno},
  {Montes}, {Morales}, {Passegger}, \& {Seifert}}]{Hintz2019}
{Hintz}, D., {Fuhrmeister}, B., {Czesla}, S., {et~al.} 2019,
  \href{https://ui.adsabs.harvard.edu/abs/2019A&A...623A.136H}{\aap, 623, A136}

\bibitem[{{Hintz} {et~al.}(2020){Hintz}, {Fuhrmeister}, {Czesla}, {Schmitt},
  {Schweitzer}, {Nagel}, {Johnson}, {Caballero}, {Zechmeister}, {Jeffers},
  {Reiners}, {Ribas}, {Amado}, {Quirrenbach}, {Anglada-Escud{\'e}}, {Bauer},
  {B{\'e}jar}, {Cort{\'e}s-Contreras}, {Dreizler}, {Galad{\'\i}-Enr{\'\i}quez},
  {Guenther}, {Hauschildt}, {Kaminski}, {K{\"u}rster}, {Lafarga}, {L{\'o}pez
  del Fresno}, {Montes}, \& {Morales}}]{Hintz2020}
{Hintz}, D., {Fuhrmeister}, B., {Czesla}, S., {et~al.} 2020,
  \href{https://ui.adsabs.harvard.edu/abs/2020A&A...638A.115H}{\aap, 638, A115}

\bibitem[{{Hirano} {et~al.}(2018){Hirano}, {Dai}, {Livingston}, {Fujii},
  {Cochran}, {Endl}, {Gandolfi}, {Redfield}, {Winn}, {Guenther},
  {Prieto-Arranz}, {Albrecht}, {Barragan}, {Cabrera}, {Cauley}, {Csizmadia},
  {Deeg}, {Eigm{\"u}ller}, {Erikson}, {Fridlund}, {Fukui}, {Grziwa}, {Hatzes},
  {Korth}, {Narita}, {Nespral}, {Niraula}, {Nowak}, {P{\"a}tzold}, {Palle},
  {Persson}, {Rauer}, {Ribas}, {Smith}, \& {Van Eylen}}]{2018AJ....155..124H}
{Hirano}, T., {Dai}, F., {Livingston}, J.~H., {et~al.} 2018,
  \href{https://ui.adsabs.harvard.edu/abs/2018AJ....155..124H}{\aj, 155, 124}

\bibitem[{{Hobson} {et~al.}(2019){Hobson}, {Delfosse}, {Astudillo-Defru},
  {Boisse}, {D{\'\i}az}, {Bouchy}, {Bonfils}, {Forveille}, {Arnold},
  {Borgniet}, {Bourrier}, {Brugger}, {Cabrera Salazar}, {Courcol}, {Dalal},
  {Deleuil}, {Demangeon}, {Dumusque}, {Hara}, {H{\'e}brard}, {Kiefer}, {Lopez},
  {Mignon}, {Montagnier}, {Mousis}, {Moutou}, {Pepe}, {Rey}, {Santerne},
  {Santos}, {Stalport}, {S{\'e}gransan}, {Udry}, \&
  {Wilson}}]{2019A&A...625A..18H}
{Hobson}, M.~J., {Delfosse}, X., {Astudillo-Defru}, N., {et~al.} 2019,
  \href{https://ui.adsabs.harvard.edu/abs/2019A&A...625A..18H}{\aap, 625, A18}

\bibitem[{{Hobson} {et~al.}(2018){Hobson}, {D{\'\i}az}, {Delfosse},
  {Astudillo-Defru}, {Boisse}, {Bouchy}, {Bonfils}, {Forveille}, {Hara},
  {Arnold}, {Borgniet}, {Bourrier}, {Brugger}, {Cabrera}, {Courcol}, {Dalal},
  {Deleuil}, {Demangeon}, {Dumusque}, {Ehrenreich}, {H{\'e}brard}, {Kiefer},
  {Lopez}, {Mignon}, {Montagnier}, {Mousis}, {Moutou}, {Pepe}, {Rey},
  {Santerne}, {Santos}, {Stalport}, {S{\'e}gransan}, {Udry}, \&
  {Wilson}}]{Hobson2018A&A...618A.103H}
{Hobson}, M.~J., {D{\'\i}az}, R.~F., {Delfosse}, X., {et~al.} 2018,
  \href{https://ui.adsabs.harvard.edu/abs/2018A&A...618A.103H}{\aap, 618, A103}

\bibitem[{{Howard} {et~al.}(2010){Howard}, {Johnson}, {Marcy}, {Fischer},
  {Wright}, {Bernat}, {Henry}, {Peek}, {Isaacson}, {Apps}, {Endl}, {Cochran},
  {Valenti}, {Anderson}, \& {Piskunov}}]{2010ApJ...721.1467H}
{Howard}, A.~W., {Johnson}, J.~A., {Marcy}, G.~W., {et~al.} 2010,
  \href{https://ui.adsabs.harvard.edu/abs/2010ApJ...721.1467H}{\apj, 721, 1467}

\bibitem[{{Howell} {et~al.}(2014){Howell}, {Sobeck}, {Haas}, {Still},
  {Barclay}, {Mullally}, {Troeltzsch}, {Aigrain}, {Bryson}, {Caldwell},
  {Chaplin}, {Cochran}, {Huber}, {Marcy}, {Miglio}, {Najita}, {Smith},
  {Twicken}, \& {Fortney}}]{Howell2014}
{Howell}, S.~B., {Sobeck}, C., {Haas}, M., {et~al.} 2014,
  \href{https://ui.adsabs.harvard.edu/abs/2014PASP..126..398H}{\pasp, 126, 398}

\bibitem[{{Hoyer} {et~al.}(2021){Hoyer}, {Gandolfi}, {Armstrong}, {Deleuil},
  {Acu{\~n}a}, {de Medeiros}, {Goffo}, {Lillo-Box}, {Delgado Mena}, {Lopez},
  {Santerne}, {Sousa}, {Fridlund}, {Adibekyan}, {Collins}, {Serrano},
  {Cort{\'e}s-Zuleta}, {Howell}, {Deeg}, {Aguichine}, {Barrag{\'a}n}, {Bryant},
  {Canto Martins}, {Collins}, {Cooke}, {D{\'\i}az}, {Esposito}, {Furlan},
  {Hojjatpanah}, {Jackman}, {Jenkins}, {Jensen}, {Latham}, {Le{\~a}o},
  {Matson}, {Nielsen}, {Osborn}, {Otegi}, {Rodler}, {Sabotta}, {Scott},
  {Seager}, {Stockdale}, {Str{\o}m}, {Vanderspek}, {Van Eylen}, {Wheatley},
  {Winn}, {Almenara}, {Barrado}, {Barros}, {Bayliss}, {Bouchy}, {Boyd},
  {Cabrera}, {Cochran}, {Demangeon}, {Doty}, {Dumusque}, {Figueira}, {Fong},
  {Grziwa}, {Hatzes}, {Kab{\'a}th}, {Knudstrup}, {Korth}, {Livingston},
  {Luque}, {Mousis}, {Mullally}, {Osborn}, {Pall{\'e}}, {Persson}, {Redfield},
  {Santos}, {Smith}, {{\v{S}}ubjak}, {Twicken}, {Udry}, \&
  {Yahalomi}}]{Hoyer2021}
{Hoyer}, S., {Gandolfi}, D., {Armstrong}, D.~J., {et~al.} 2021,
  \href{https://ui.adsabs.harvard.edu/abs/2021MNRAS.505.3361H}{\mnras, 505,
  3361}

\bibitem[{{Hsu} {et~al.}(2020){Hsu}, {Ford}, \& {Terrien}}]{Hsu2020}
{Hsu}, D.~C., {Ford}, E.~B., \& {Terrien}, R. 2020,
  \href{https://ui.adsabs.harvard.edu/abs/2020MNRAS.498.2249H}{\mnras, 498,
  2249}

\bibitem[{{Ida} \& {Lin}(2004)}]{Ida2004}
{Ida}, S. \& {Lin}, D.~N.~C. 2004,
  \href{https://ui.adsabs.harvard.edu/abs/2004ApJ...604..388I}{\apj, 604, 388}

\bibitem[{{Izidoro} {et~al.}(2021){Izidoro}, {Bitsch}, {Raymond}, {Johansen},
  {Morbidelli}, {Lambrechts}, \& {Jacobson}}]{Izidoro2021}
{Izidoro}, A., {Bitsch}, B., {Raymond}, S.~N., {et~al.} 2021,
  \href{https://ui.adsabs.harvard.edu/abs/2021A&A...650A.152I}{\aap, 650, A152}

\bibitem[{{Jeffers} {et~al.}(2022){Jeffers}, {Barnes}, {Sch{\"o}fer},
  {Quirrenbach}, {Zechmeister}, {Amado}, {Caballero}, {Fern{\'a}ndez},
  {Rodr{\'\i}guez}, {Ribas}, {Reiners}, {Cardona Guill{\'e}n}, {Cifuentes},
  {Czesla}, {Hatzes}, {K{\"u}rster}, {Montes}, {Morales}, {Pedraz}, \&
  {Sadegi}}]{Jeffers2022}
{Jeffers}, S.~V., {Barnes}, J.~R., {Sch{\"o}fer}, P., {et~al.} 2022,
  \href{https://ui.adsabs.harvard.edu/abs/2022A&A...663A..27J}{\aap, 663, A27}

\bibitem[{{Jeffers} {et~al.}(2018){Jeffers}, {Sch{\"o}fer}, {Lamert},
  {Reiners}, {Montes}, {Caballero}, {Cort{\'e}s-Contreras}, {Marvin},
  {Passegger}, {Zechmeister}, {Quirrenbach}, {Alonso-Floriano}, {Amado},
  {Bauer}, {Casal}, {Alonso}, {Herrero}, {Morales}, {Mundt}, {Ribas}, \&
  {Sarmiento}}]{Jeffers2018}
{Jeffers}, S.~V., {Sch{\"o}fer}, P., {Lamert}, A., {et~al.} 2018,
  \href{http://adsabs.harvard.edu/abs/2018A\%26A...614A..76J}{\aap, 614, A76}

\bibitem[{{Johnson} {et~al.}(2010){Johnson}, {Howard}, {Marcy}, {Bowler},
  {Henry}, {Fischer}, {Apps}, {Isaacson}, \& {Wright}}]{Johnson2010}
{Johnson}, J.~A., {Howard}, A.~W., {Marcy}, G.~W., {et~al.} 2010,
  \href{https://ui.adsabs.harvard.edu/abs/2010PASP..122..149J}{\pasp, 122, 149}

\bibitem[{{Kaminski} {et~al.}(2018){Kaminski}, {Trifonov}, {Caballero},
  {Quirrenbach}, {Ribas}, {Reiners}, {Amado}, {Zechmeister}, {Dreizler},
  {Perger}, {Tal-Or}, {Bonfils}, {Mayor}, {Astudillo-Defru}, {Bauer},
  {B{\'e}jar}, {Cifuentes}, {Colom{\'e}}, {Cort{\'e}s-Contreras}, {Delfosse},
  {D{\'\i}ez-Alonso}, {Forveille}, {Guenther}, {Hatzes}, {Henning}, {Jeffers},
  {K{\"u}rster}, {Lafarga}, {Luque}, {Mandel}, {Montes}, {Morales},
  {Passegger}, {Pedraz}, {Reffert}, {Sadegi}, {Schweitzer}, {Seifert}, {Stahl},
  \& {Udry}}]{Kaminski2018}
{Kaminski}, A., {Trifonov}, T., {Caballero}, J.~A., {et~al.} 2018,
  \href{https://ui.adsabs.harvard.edu/abs/2018A&A...618A.115K}{\aap, 618, A115}

\bibitem[{{Kanodia} {et~al.}(2019){Kanodia}, {Wolfgang}, {Stef\'ansson},
  {Ning}, \& {Mahadevan}}]{Kanodia2019}
{Kanodia}, S., {Wolfgang}, A., {Stef\'ansson}, G.~K., {Ning}, B., \&
  {Mahadevan}, S. 2019,
  \href{https://ui.adsabs.harvard.edu/abs/2019ApJ...882...38K}{\apj, 882, 38}

\bibitem[{{Kanodia} \& {Wright}(2018)}]{Kanodia2018RNAAS...2....4K}
{Kanodia}, S. \& {Wright}, J. 2018,
  \href{https://ui.adsabs.harvard.edu/abs/2018RNAAS...2....4K}{Research Notes
  of the American Astronomical Society, 2, 4}

\bibitem[{{Kemmer} {et~al.}(2022){Kemmer}, {Dreizler}, {Kossakowski}, {Stock},
  {Quirrenbach}, {Caballero}, {Amado}, {Collins}, {Espinoza}, {Herrero},
  {Jenkins}, {Latham}, {Lillo-Box}, {Narita}, {Pall{\'e}}, {Reiners}, {Ribas},
  {Ricker}, {Rodr{\'\i}guez}, {Seager}, {Vanderspek}, {Wells}, {Winn},
  {Aceituno}, {B{\'e}jar}, {Barclay}, {Bluhm}, {Chaturvedi}, {Cifuentes},
  {Collins}, {Cort{\'e}s-Contreras}, {Demory}, {Fausnaugh}, {Fukui}, {G{\'o}mez
  Maqueo Chew}, {Galad{\'\i}-Enr{\'\i}quez}, {Gan}, {Gillon}, {Golovin},
  {Hatzes}, {Henning}, {Huang}, {Jeffers}, {Kaminski}, {Kunimoto},
  {K{\"u}rster}, {L{\'o}pez-Gonz{\'a}lez}, {Lafarga}, {Luque}, {McCormac},
  {Molaverdikhani}, {Montes}, {Morales}, {Passegger}, {Reffert}, {Sabin},
  {Sch{\"o}fer}, {Schanche}, {Schlecker}, {Schroffenegger}, {Schwarz},
  {Schweitzer}, {Sota}, {Tenenbaum}, {Trifonov}, {Vanaverbeke}, \&
  {Zechmeister}}]{Kemmer2022}
{Kemmer}, J., {Dreizler}, S., {Kossakowski}, D., {et~al.} 2022,
  \href{https://ui.adsabs.harvard.edu/abs/2022A&A...659A..17K}{\aap, 659, A17}

\bibitem[{{Kemmer} {et~al.}(2020){Kemmer}, {Stock}, {Kossakowski}, {Kaminski},
  {Molaverdikhani}, {Schlecker}, {Caballero}, {Amado}, {Astudillo-Defru},
  {Bonfils}, {Ciardi}, {Collins}, {Espinoza}, {Fukui}, {Hirano}, {Jenkins},
  {Latham}, {Matthews}, {Narita}, {Pall{\'e}}, {Parviainen}, {Quirrenbach},
  {Reiners}, {Ribas}, {Ricker}, {Schlieder}, {Seager}, {Vanderspek}, {Winn},
  {Almenara}, {B{\'e}jar}, {Bluhm}, {Bouchy}, {Boyd}, {Christiansen},
  {Cifuentes}, {Cloutier}, {Collins}, {Cort{\'e}s-Contreras}, {Crossfield},
  {Crouzet}, {de Leon}, {Della-Rose}, {Delfosse}, {Dreizler}, {Esparza-Borges},
  {Essack}, {Forveille}, {Figueira}, {Galad{\'\i}-Enr{\'\i}quez}, {Gan},
  {Glidden}, {Gonzales}, {Guerra}, {Harakawa}, {Hatzes}, {Henning}, {Herrero},
  {Hodapp}, {Hori}, {Howell}, {Ikoma}, {Isogai}, {Jeffers}, {K{\"u}rster},
  {Kawauchi}, {Kimura}, {Klagyivik}, {Kotani}, {Kurokawa}, {Kusakabe},
  {Kuzuhara}, {Lafarga}, {Livingston}, {Luque}, {Matson}, {Morales}, {Mori},
  {Muirhead}, {Murgas}, {Nishikawa}, {Nishiumi}, {Omiya}, {Reffert},
  {Rodr{\'\i}guez L{\'o}pez}, {Santos}, {Sch{\"o}fer}, {Schwarz}, {Shiao},
  {Tamura}, {Terada}, {Twicken}, {Ueda}, {Vievard}, {Watanabe}, \&
  {Zechmeister}}]{Kemmer2020}
{Kemmer}, J., {Stock}, S., {Kossakowski}, D., {et~al.} 2020,
  \href{https://ui.adsabs.harvard.edu/abs/2020A&A...642A.236K}{\aap, 642, A236}

\bibitem[{{Khalafinejad} {et~al.}(2021){Khalafinejad}, {Molaverdikhani},
  {Blecic}, {Mallonn}, {Nortmann}, {Caballero}, {Rahmati}, {Kaminski},
  {Sadegi}, {Nagel}, {Carone}, {Amado}, {Azzaro}, {Bauer}, {Casasayas-Barris},
  {Czesla}, {von Essen}, {Fossati}, {G{\"u}del}, {Henning},
  {L{\'o}pez-Puertas}, {Lendl}, {L{\"u}ftinger}, {Montes}, {Oshagh},
  {Pall{\'e}}, {Quirrenbach}, {Reffert}, {Reiners}, {Ribas}, {Stock}, {Yan},
  {Zapatero Osorio}, \& {Zechmeister}}]{Khalafinejad2021}
{Khalafinejad}, S., {Molaverdikhani}, K., {Blecic}, J., {et~al.} 2021,
  \href{https://ui.adsabs.harvard.edu/abs/2021A&A...656A.142K}{\aap, 656, A142}

\bibitem[{{Kopparapu} {et~al.}(2013){Kopparapu}, {Ramirez}, {Kasting}, {Eymet},
  {Robinson}, {Mahadevan}, {Terrien}, {Domagal-Goldman}, {Meadows}, \&
  {Deshpande}}]{Kopparapu2013}
{Kopparapu}, R.~K., {Ramirez}, R., {Kasting}, J.~F., {et~al.} 2013,
  \href{https://ui.adsabs.harvard.edu/abs/2013ApJ...765..131K}{\apj, 765, 131}

\bibitem[{{Kossakowski} {et~al.}(2021){Kossakowski}, {Kemmer}, {Bluhm},
  {Stock}, {Caballero}, {B{\'e}jar}, {Guill{\'e}n}, {Lodieu}, {Collins},
  {Oshagh}, {Schlecker}, {Espinoza}, {Pall{\'e}}, {Henning}, {Kreidberg},
  {K{\"u}rster}, {Amado}, {Anderson}, {Morales}, {Cartwright}, {Charbonneau},
  {Chaturvedi}, {Cifuentes}, {Conti}, {Cort{\'e}s-Contreras}, {Dreizler},
  {Galad{\'\i}-Enr{\'\i}quez}, {Guerra}, {Hart}, {Hellier}, {Henze}, {Herrero},
  {Jeffers}, {Jenkins}, {Jensen}, {Kaminski}, {Kielkopf}, {Kunimoto},
  {Lafarga}, {Latham}, {Lillo-Box}, {Luque}, {Molaverdikhani}, {Montes},
  {Morello}, {Morgan}, {Nowak}, {Pavlov}, {Perger}, {Quintana}, {Quirrenbach},
  {Reffert}, {Reiners}, {Ricker}, {Ribas}, {L{\'o}pez}, {Osorio}, {Seager},
  {Sch{\"o}fer}, {Schweitzer}, {Trifonov}, {Vanaverbeke}, {Vanderspek}, {West},
  {Winn}, \& {Zechmeister}}]{Kossakowski2021}
{Kossakowski}, D., {Kemmer}, J., {Bluhm}, P., {et~al.} 2021,
  \href{https://ui.adsabs.harvard.edu/abs/2021A&A...656A.124K}{\aap, 656, A124}

\bibitem[{{Kossakowski} {et~al.}(2022){Kossakowski}, {K{\"u}rster}, {Henning},
  {Trifonov}, {Caballero}, {Lafarga}, {Bauer}, {Stock}, {Kemmer}, {Jeffers},
  {Amado}, {P{\'e}rez-Torres}, {B{\'e}jar}, {Cort{\'e}s-Contreras}, {Ribas},
  {Reiners}, {Quirrenbach}, {Aceituno}, {Baroch}, {Cifuentes}, {Dreizler},
  {Hatzes}, {Kaminski}, {Montes}, {Morales}, {Pavlov}, {Pena}, {Perdelwitz},
  {Reffert}, {Revilla}, {Lopez}, {Rosich}, {Sadegi}, {Sanz-Forcada},
  {Sch{\"o}fer}, {Schweitzer}, \& {Zechmeister}}]{Kossakowski2022}
{Kossakowski}, D., {K{\"u}rster}, M., {Henning}, T., {et~al.} 2022,
  \href{https://ui.adsabs.harvard.edu/abs/2022A&A...666A.143K}{\aap, 666, A143}

\bibitem[{{Kossakowski} {et~al.}(2023){Kossakowski}, {K{\"u}rster}, {Trifonov},
  {Henning}, {Kemmer}, {Caballero}, {Burn}, {Sabotta}, {Crouse}, {Fauchez},
  {Nagel}, {Kaminski}, {Herrero}, {Rodr{\'\i}guez}, {Gonz{\'a}lez-{\'A}lvarez},
  {Quirrenbach}, {Amado}, {Ribas}, {Reiners}, {Aceituno}, {B{\'e}jar},
  {Baroch}, {Bastelberger}, {Chaturvedi}, {Cifuentes}, {Dreizler}, {Jeffers},
  {Kopparapu}, {Lafarga}, {L{\'o}pez-Gonz{\'a}lez}, {Mart{\'\i}n-Ruiz},
  {Montes}, {Morales}, {Pall{\'e}}, {Pavlov}, {Pedraz}, {Perdelwitz},
  {P{\'e}rez-Torres}, {Perger}, {Reffert}, {Rodr{\'\i}guez L{\'o}pez},
  {Schlecker}, {Sch{\"o}fer}, {Schweitzer}, {Shan}, {Shields}, {Stock}, {Wolf},
  {Zapatero Osorio}, \& {Zechmeister}}]{Kossakowski2022b}
{Kossakowski}, D., {K{\"u}rster}, M., {Trifonov}, T., {et~al.} 2023,
  \href{https://ui.adsabs.harvard.edu/abs/2023A&A...670A..84K}{\aap, 670, A84}

\bibitem[{{K{\"u}rster} {et~al.}(2003){K{\"u}rster}, {Endl}, {Rouesnel}, {Els},
  {Kaufer}, {Brillant}, {Hatzes}, {Saar}, \& {Cochran}}]{Kuerster2003}
{K{\"u}rster}, M., {Endl}, M., {Rouesnel}, F., {et~al.} 2003,
  \href{https://ui.adsabs.harvard.edu/abs/2003A&A...403.1077K}{\aap, 403, 1077}

\bibitem[{{Lafarga} {et~al.}(2020){Lafarga}, {Ribas}, {Lovis}, {Perger},
  {Zechmeister}, {Bauer}, {K{\"u}rster}, {Cort{\'e}s-Contreras}, {Morales},
  {Herrero}, {Rosich}, {Baroch}, {Reiners}, {Caballero}, {Quirrenbach},
  {Amado}, {Alacid}, {B{\'e}jar}, {Dreizler}, {Hatzes}, {Henning}, {Jeffers},
  {Kaminski}, {Montes}, {Pedraz}, {Rodr{\'\i}guez-L{\'o}pez}, \&
  {Schmitt}}]{Lafarga2020A&A...636A..36L}
{Lafarga}, M., {Ribas}, I., {Lovis}, C., {et~al.} 2020,
  \href{https://ui.adsabs.harvard.edu/abs/2020A&A...636A..36L}{\aap, 636, A36}

\bibitem[{{Lafarga} {et~al.}(2021){Lafarga}, {Ribas}, {Reiners}, {Quirrenbach},
  {Amado}, {Caballero}, {Azzaro}, {B{\'e}jar}, {Cort{\'e}s-Contreras},
  {Dreizler}, {Hatzes}, {Henning}, {Jeffers}, {Kaminski}, {K{\"u}rster},
  {Montes}, {Morales}, {Oshagh}, {Rodr{\'\i}guez-L{\'o}pez}, {Sch{\"o}fer},
  {Schweitzer}, \& {Zechmeister}}]{Lafarga2021}
{Lafarga}, M., {Ribas}, I., {Reiners}, A., {et~al.} 2021,
  \href{https://ui.adsabs.harvard.edu/abs/2021A&A...652A..28L}{\aap, 652, A28}

\bibitem[{{Lalitha} {et~al.}(2019){Lalitha}, {Baroch}, {Morales}, {Passegger},
  {Bauer}, {Cardona Guill{\'e}n}, {Dreizler}, {Oshagh}, {Reiners}, {Ribas},
  {Caballero}, {Quirrenbach}, {Amado}, {B{\'e}jar}, {Colom{\'e}},
  {Cort{\'e}s-Contreras}, {Galad{\'\i}-Enr{\'\i}quez}, {Gonz{\'a}lez-Cuesta},
  {Guenther}, {Hagen}, {Henning}, {Herrero}, {Husser}, {Jeffers}, {Kaminski},
  {K{\"u}rster}, {Lafarga}, {Lodieu}, {L{\'o}pez-Gonz{\'a}lez}, {Montes},
  {Perger}, {Rosich}, {Rodr{\'\i}guez}, {Rodr{\'\i}guez-L{\'o}pez}, {Schmitt},
  {Tal-Or}, \& {Zechmeister}}]{Lalitha2019}
{Lalitha}, S., {Baroch}, D., {Morales}, J.~C., {et~al.} 2019,
  \href{https://ui.adsabs.harvard.edu/abs/2019A&A...627A.116L}{\aap, 627, A116}

\bibitem[{{Luque} {et~al.}(2022){Luque}, {Fulton}, {Kunimoto}, {Amado},
  {Gorrini}, {Dreizler}, {Hellier}, {Henry}, {Molaverdikhani}, {Morello},
  {Pe{\~n}a-Mo{\~n}ino}, {P{\'e}rez-Torres}, {Pozuelos}, {Shan},
  {Anglada-Escud{\'e}}, {B{\'e}jar}, {Bergond}, {Boyle}, {Caballero},
  {Charbonneau}, {Ciardi}, {Dufoer}, {Espinoza}, {Everett}, {Fischer},
  {Hatzes}, {Henning}, {Hesse}, {Howard}, {Howell}, {Isaacson}, {Jeffers},
  {Jenkins}, {Kane}, {Kemmer}, {Khalafinejad}, {Kidwell}, {Kossakowski},
  {Latham}, {Lillo-Box}, {Lissauer}, {Montes}, {Orell-Miquel}, {Pall{\'e}},
  {Pollacco}, {Quirrenbach}, {Reffert}, {Reiners}, {Ribas}, {Ricker}, {Rogers},
  {Sanz-Forcada}, {Schlecker}, {Schweitzer}, {Seager}, {Shporer}, {Stassun},
  {Stock}, {Tal-Or}, {Ting}, {Trifonov}, {Vanaverbeke}, {Vanderspek},
  {Villase{\~n}or}, {Winn}, {Winters}, \& {Zapatero Osorio}}]{Luque2022}
{Luque}, R., {Fulton}, B.~J., {Kunimoto}, M., {et~al.} 2022,
  \href{https://ui.adsabs.harvard.edu/abs/2022A&A...664A.199L}{\aap, 664, A199}

\bibitem[{{Luque} {et~al.}(2018){Luque}, {Nowak}, {Pall{\'e}}, {Kossakowski},
  {Trifonov}, {Zechmeister}, {B{\'e}jar}, {Cardona Guill{\'e}n}, {Tal-Or},
  {Hidalgo}, {Ribas}, {Reiners}, {Caballero}, {Amado}, {Quirrenbach},
  {Aceituno}, {Cort{\'e}s-Contreras}, {D{\'\i}ez-Alonso}, {Dreizler},
  {Guenther}, {Henning}, {Jeffers}, {Kaminski}, {K{\"u}rster}, {Lafarga},
  {Montes}, {Morales}, {Passegger}, {Schmitt}, \& {Schweitzer}}]{Luque2018}
{Luque}, R., {Nowak}, G., {Pall{\'e}}, E., {et~al.} 2018,
  \href{https://ui.adsabs.harvard.edu/abs/2018A&A...620A.171L}{\aap, 620, A171}

\bibitem[{{Luque} \& {Pall{\'e}}(2022)}]{Luque2022b}
{Luque}, R. \& {Pall{\'e}}, E. 2022,
  \href{https://ui.adsabs.harvard.edu/abs/2022Sci...377.1211L}{Science, 377,
  1211}

\bibitem[{{Luque} {et~al.}(2019){Luque}, {Pall{\'e}}, {Kossakowski},
  {Dreizler}, {Kemmer}, {Espinoza}, {Burt}, {Anglada-Escud{\'e}}, {B{\'e}jar},
  {Caballero}, {Collins}, {Collins}, {Cort{\'e}s-Contreras},
  {D{\'\i}ez-Alonso}, {Feng}, {Hatzes}, {Hellier}, {Henning}, {Jeffers},
  {Kaltenegger}, {K{\"u}rster}, {Madden}, {Molaverdikhani}, {Montes}, {Narita},
  {Nowak}, {Ofir}, {Oshagh}, {Parviainen}, {Quirrenbach}, {Reffert}, {Reiners},
  {Rodr{\'\i}guez-L{\'o}pez}, {Schlecker}, {Stock}, {Trifonov}, {Winn},
  {Zapatero Osorio}, {Zechmeister}, {Amado}, {Anderson}, {Batalha}, {Bauer},
  {Bluhm}, {Burke}, {Butler}, {Caldwell}, {Chen}, {Crane}, {Dragomir},
  {Dressing}, {Dynes}, {Jenkins}, {Kaminski}, {Klahr}, {Kotani}, {Lafarga},
  {Latham}, {Lewin}, {McDermott}, {Monta{\~n}{\'e}s-Rodr{\'\i}guez}, {Morales},
  {Murgas}, {Nagel}, {Pedraz}, {Ribas}, {Ricker}, {Rowden}, {Seager},
  {Shectman}, {Tamura}, {Teske}, {Twicken}, {Vanderspeck}, {Wang}, \&
  {Wohler}}]{Luque2019}
{Luque}, R., {Pall{\'e}}, E., {Kossakowski}, D., {et~al.} 2019,
  \href{https://ui.adsabs.harvard.edu/abs/2019A&A...628A..39L}{\aap, 628, A39}

\bibitem[{{Mahadevan} {et~al.}(2021){Mahadevan}, {Stef{\'a}nsson}, {Robertson},
  {Terrien}, {Ninan}, {Holcomb}, {Halverson}, {Cochran}, {Kanodia}, {Ramsey},
  {Wolszczan}, {Endl}, {Bender}, {Diddams}, {Fredrick}, {Hearty}, {Monson},
  {Metcalf}, {Roy}, \& {Schwab}}]{Mahadevan2021ApJ...919L...9M}
{Mahadevan}, S., {Stef{\'a}nsson}, G., {Robertson}, P., {et~al.} 2021,
  \href{https://ui.adsabs.harvard.edu/abs/2021ApJ...919L...9M}{\apjl, 919, L9}

\bibitem[{{Marfil} {et~al.}(2021){Marfil}, {Tabernero}, {Montes}, {Caballero},
  {L{\'a}zaro}, {Gonz{\'a}lez Hern{\'a}ndez}, {Nagel}, {Passegger},
  {Schweitzer}, {Ribas}, {Reiners}, {Quirrenbach}, {Amado}, {Cifuentes},
  {Cort{\'e}s-Contreras}, {Dreizler}, {Duque-Arribas},
  {Galad{\'\i}-Enr{\'\i}quez}, {Henning}, {Jeffers}, {Kaminski}, {K{\"u}rster},
  {Lafarga}, {L{\'o}pez-Gallifa}, {Morales}, {Shan}, \&
  {Zechmeister}}]{Marfil2021}
{Marfil}, E., {Tabernero}, H.~M., {Montes}, D., {et~al.} 2021,
  \href{https://ui.adsabs.harvard.edu/abs/2021A&A...656A.162M}{\aap, 656, A162}

\bibitem[{{Marfil} {et~al.}(2020){Marfil}, {Tabernero}, {Montes}, {Caballero},
  {Soto}, {Gonz{\'a}lez Hern{\'a}ndez}, {Kaminski}, {Nagel}, {Jeffers},
  {Reiners}, {Ribas}, {Quirrenbach}, \& {Amado}}]{Marfil2020}
{Marfil}, E., {Tabernero}, H.~M., {Montes}, D., {et~al.} 2020,
  \href{https://ui.adsabs.harvard.edu/abs/2020MNRAS.492.5470M}{\mnras, 492,
  5470}

\bibitem[{{Miguel} {et~al.}(2020){Miguel}, {Cridland}, {Ormel}, {Fortney}, \&
  {Ida}}]{Miguel2020}
{Miguel}, Y., {Cridland}, A., {Ormel}, C.~W., {Fortney}, J.~J., \& {Ida}, S.
  2020, \href{https://ui.adsabs.harvard.edu/abs/2020MNRAS.491.1998M}{\mnras,
  491, 1998}

\bibitem[{{Mishra} {et~al.}(2021){Mishra}, {Alibert}, {Leleu}, {Emsenhuber},
  {Mordasini}, {Burn}, {Udry}, \& {Benz}}]{Mishra2021}
{Mishra}, L., {Alibert}, Y., {Leleu}, A., {et~al.} 2021,
  \href{https://ui.adsabs.harvard.edu/abs/2021A&A...656A..74M}{\aap, 656, A74}

\bibitem[{{Morales} {et~al.}(2019){Morales}, {Mustill}, {Ribas}, {Davies},
  {Reiners}, {Bauer}, {Kossakowski}, {Herrero}, {Rodr{\'\i}guez},
  {L{\'o}pez-Gonz{\'a}lez}, {Rodr{\'\i}guez-L{\'o}pez}, {B{\'e}jar},
  {Gonz{\'a}lez-Cuesta}, {Luque}, {Pall{\'e}}, {Perger}, {Baroch}, {Johansen},
  {Klahr}, {Mordasini}, {Anglada-Escud{\'e}}, {Caballero},
  {Cort{\'e}s-Contreras}, {Dreizler}, {Lafarga}, {Nagel}, {Passegger},
  {Reffert}, {Rosich}, {Schweitzer}, {Tal-Or}, {Trifonov}, {Zechmeister},
  {Quirrenbach}, {Amado}, {Guenther}, {Hagen}, {Henning}, {Jeffers},
  {Kaminski}, {K{\"u}rster}, {Montes}, {Seifert}, {Abell{\'a}n}, {Abril},
  {Aceituno}, {Aceituno}, {Alonso-Floriano}, {Ammler-von Eiff}, {Antona},
  {Arroyo-Torres}, {Azzaro}, {Barrado}, {Becerril-Jarque}, {Ben{\'\i}tez},
  {Berdi{\~n}as}, {Bergond}, {Brinkm{\"o}ller}, {del Burgo}, {Burn},
  {Calvo-Ortega}, {Cano}, {C{\'a}rdenas}, {Cardona Guill{\'e}n}, {Carro},
  {Casal}, {Casanova}, {Casasayas-Barris}, {Chaturvedi}, {Cifuentes}, {Claret},
  {Colom{\'e}}, {Czesla}, {D{\'\i}ez-Alonso}, {Dorda}, {Emsenhuber},
  {Fern{\'a}ndez}, {Fern{\'a}ndez-Mart{\'\i}n}, {Ferro}, {Fuhrmeister},
  {Galad{\'\i}-Enr{\'\i}quez}, {Gallardo Cava}, {Garc{\'\i}a Vargas},
  {Garcia-Piquer}, {Gesa}, {Gonz{\'a}lez-{\'A}lvarez}, {Gonz{\'a}lez
  Hern{\'a}ndez}, {Gonz{\'a}lez-Peinado}, {Gu{\`a}rdia}, {Guijarro}, {de
  Guindos}, {Hatzes}, {Hauschildt}, {Hedrosa}, {Hermelo}, {Hern{\'a}ndez
  Arabi}, {Hern{\'a}ndez Otero}, {Hintz}, {Holgado}, {Huber}, {Huke},
  {Johnson}, {de Juan}, {Kehr}, {Kemmer}, {Kim}, {Kl{\"u}ter}, {Klutsch},
  {Labarga}, {Labiche}, {Lalitha}, {Lamp{\'o}n}, {Lara}, {Launhardt},
  {L{\'a}zaro}, {Lizon}, {Llamas}, {Lodieu}, {L{\'o}pez del Fresno}, {L{\'o}pez
  Salas}, {L{\'o}pez-Santiago}, {Mag{\'a}n Madinabeitia}, {Mall}, {Mancini},
  {Mandel}, {Marfil}, {Mar{\'\i}n Molina}, {Mart{\'\i}n},
  {Mart{\'\i}n-Fern{\'a}ndez}, {Mart{\'\i}n-Ruiz},
  {Mart{\'\i}nez-Rodr{\'\i}guez}, {Marvin}, {Mirabet}, {Moya}, {Naranjo},
  {Nelson}, {Nortmann}, {Nowak}, {Ofir}, {Pascual}, {Pavlov}, {Pedraz},
  {P{\'e}rez Medialdea}, {P{\'e}rez-Calpena}, {Perryman}, {Rabaza}, {Ram{\'o}n
  Ballesta}, {Rebolo}, {Redondo}, {Rix}, {Rodler}, {Rodr{\'\i}guez Trinidad},
  {Sabotta}, {Sadegi}, {Salz}, {S{\'a}nchez-Blanco}, {S{\'a}nchez Carrasco},
  {S{\'a}nchez-L{\'o}pez}, {Sanz-Forcada}, {Sarkis}, {Sarmiento},
  {Sch{\"a}fer}, {Schlecker}, {Schmitt}, {Sch{\"o}fer}, {Solano}, {Sota},
  {Stahl}, {Stock}, {Stuber}, {St{\"u}rmer}, {Su{\'a}rez}, {Tabernero},
  {Tulloch}, {Veredas}, {Vico-Linares}, {Vilardell}, {Wagner}, {Winkler},
  {Wolthoff}, {Yan}, \& {Zapatero Osorio}}]{Morales2019}
{Morales}, J.~C., {Mustill}, A.~J., {Ribas}, I., {et~al.} 2019,
  \href{https://ui.adsabs.harvard.edu/abs/2019Sci...365.1441M}{Science, 365,
  1441}

\bibitem[{{Mordasini} {et~al.}(2012){Mordasini}, {Alibert}, {Klahr}, \&
  {Henning}}]{Mordasini2012}
{Mordasini}, C., {Alibert}, Y., {Klahr}, H., \& {Henning}, T. 2012,
  \href{https://ui.adsabs.harvard.edu/abs/2012A&A...547A.111M}{\aap, 547, A111}

\bibitem[{{Morton} \& {Swift}(2014)}]{Morton2014}
{Morton}, T.~D. \& {Swift}, J. 2014,
  \href{https://ui.adsabs.harvard.edu/abs/2014ApJ...791...10M}{\apj, 791, 10}

\bibitem[{{Mulders} {et~al.}(2015){Mulders}, {Pascucci}, \&
  {Apai}}]{Mulders2015a}
{Mulders}, G.~D., {Pascucci}, I., \& {Apai}, D. 2015,
  \href{https://ui.adsabs.harvard.edu/abs/2015ApJ...814..130M}{\apj, 814, 130}

\bibitem[{Nagel(2019)}]{Nagel2019PhD}
Nagel, E. 2019, PhD thesis, Universit{\"a}t Hamburg,
  \url{https://ediss.sub.uni-hamburg.de/handle/ediss/6280}

\bibitem[{{Nagel} {et~al.}(2019){Nagel}, {Czesla}, {Schmitt}, {Dreizler},
  {Anglada-Escud{\'e}}, {Rodr{\'\i}guez}, {Ribas}, {Reiners}, {Quirrenbach},
  {Amado}, {Caballero}, {Aceituno}, {B{\'e}jar}, {Cort{\'e}s-Contreras},
  {Gonz{\'a}lez-Cuesta}, {Guenther}, {Henning}, {Jeffers}, {Kaminski},
  {K{\"u}rster}, {Lafarga}, {L{\'o}pez-Gonz{\'a}lez}, {Montes}, {Morales},
  {Passegger}, {Rodr{\'\i}guez-L{\'o}pez}, {Schweitzer}, \&
  {Zechmeister}}]{Nagel2019}
{Nagel}, E., {Czesla}, S., {Schmitt}, J.~H.~M.~M., {et~al.} 2019,
  \href{https://ui.adsabs.harvard.edu/abs/2019A&A...622A.153N}{\aap, 622, A153}

\bibitem[{{Nowak} {et~al.}(2020){Nowak}, {Luque}, {Parviainen}, {Pall{\'e}},
  {Molaverdikhani}, {B{\'e}jar}, {Lillo-Box}, {Rodr{\'\i}guez-L{\'o}pez},
  {Caballero}, {Zechmeister}, {Passegger}, {Cifuentes}, {Schweitzer}, {Narita},
  {Cale}, {Espinoza}, {Murgas}, {Hidalgo}, {Zapatero Osorio}, {Pozuelos},
  {Aceituno}, {Amado}, {Barkaoui}, {Barrado}, {Bauer}, {Benkhaldoun},
  {Caldwell}, {Casasayas Barris}, {Chaturvedi}, {Chen}, {Collins}, {Collins},
  {Cort{\'e}s-Contreras}, {Crossfield}, {de Le{\'o}n}, {D{\'\i}ez Alonso},
  {Dreizler}, {El Mufti}, {Esparza-Borges}, {Essack}, {Fukui}, {Gaidos},
  {Gillon}, {Gonzales}, {Guerra}, {Hatzes}, {Henning}, {Herrero}, {Hesse},
  {Hirano}, {Howell}, {Jeffers}, {Jehin}, {Jenkins}, {Kaminski}, {Kemmer},
  {Kielkopf}, {Kossakowski}, {Kotani}, {K{\"u}rster}, {Lafarga}, {Latham},
  {Law}, {Lissauer}, {Lodieu}, {Madrigal-Aguado}, {Mann}, {Massey}, {Matson},
  {Matthews}, {Monta{\~n}{\'e}s-Rodr{\'\i}guez}, {Montes}, {Morales}, {Mori},
  {Nagel}, {Oshagh}, {Pedraz}, {Plavchan}, {Pollacco}, {Quirrenbach},
  {Reffert}, {Reiners}, {Ribas}, {Ricker}, {Rose}, {Schlecker}, {Schlieder},
  {Seager}, {Stangret}, {Stock}, {Tamura}, {Tanner}, {Teske}, {Trifonov},
  {Twicken}, {Vanderspek}, {Watanabe}, {Wittrock}, {Ziegler}, \&
  {Zohrabi}}]{Nowak2020}
{Nowak}, G., {Luque}, R., {Parviainen}, H., {et~al.} 2020,
  \href{https://ui.adsabs.harvard.edu/abs/2020A&A...642A.173N}{\aap, 642, A173}

\bibitem[{{Nutzman} \& {Charbonneau}(2008)}]{Nutzmann2008}
{Nutzman}, P. \& {Charbonneau}, D. 2008,
  \href{https://ui.adsabs.harvard.edu/abs/2008PASP..120..317N}{\pasp, 120, 317}

\bibitem[{{Oshagh} {et~al.}(2020){Oshagh}, {Bauer}, {Lafarga},
  {Molaverdikhani}, {Amado}, {Nortmann}, {Reiners}, {Guzm{\'a}n-Mesa},
  {Pall{\'e}}, {Nagel}, {Caballero}, {Casasayas-Barris}, {Claret}, {Czesla},
  {Galad{\'\i}}, {Henning}, {Khalafinejad}, {L{\'o}pez-Puertas}, {Montes},
  {Quirrenbach}, {Ribas}, {Stangret}, {Yan}, {Zapatero Osorio}, \&
  {Zechmeister}}]{Oshagh2020}
{Oshagh}, M., {Bauer}, F.~F., {Lafarga}, M., {et~al.} 2020,
  \href{https://ui.adsabs.harvard.edu/abs/2020A&A...643A..64O}{\aap, 643, A64}

\bibitem[{{Pall\'e} \& {et al.}(2022)}]{Palle2022}
{Pall\'e}, E. \& {et al.} 2022,
  \href{https://ui.adsabs.harvard.edu/abs/2022arXiv220410261L}{\aap, submitted}

\bibitem[{{Passegger} {et~al.}(2022){Passegger}, {Bello-Garc{\'\i}a},
  {Ordieres-Mer{\'e}}, {Antoniadis-Karnavas}, {Marfil}, {Duque-Arribas},
  {Amado}, {Delgado-Mena}, {Montes}, {Rojas-Ayala}, {Schweitzer}, {Tabernero},
  {B{\'e}jar}, {Caballero}, {Hatzes}, {Henning}, {Pedraz}, {Quirrenbach},
  {Reiners}, \& {Ribas}}]{Passegger2022}
{Passegger}, V.~M., {Bello-Garc{\'\i}a}, A., {Ordieres-Mer{\'e}}, J., {et~al.}
  2022, \href{https://ui.adsabs.harvard.edu/abs/2022A&A...658A.194P}{\aap, 658,
  A194}

\bibitem[{{Passegger} {et~al.}(2020){Passegger}, {Bello-Garc{\'\i}a},
  {Ordieres-Mer{\'e}}, {Caballero}, {Schweitzer}, {Gonz{\'a}lez-Marcos},
  {Ribas}, {Reiners}, {Quirrenbach}, {Amado}, {Azzaro}, {Bauer}, {B{\'e}jar},
  {Cort{\'e}s-Contreras}, {Dreizler}, {Hatzes}, {Henning}, {Jeffers},
  {Kaminski}, {K{\"u}rster}, {Lafarga}, {Marfil}, {Montes}, {Morales}, {Nagel},
  {Sarro}, {Solano}, {Tabernero}, \& {Zechmeister}}]{Passegger2020}
{Passegger}, V.~M., {Bello-Garc{\'\i}a}, A., {Ordieres-Mer{\'e}}, J., {et~al.}
  2020, \href{https://ui.adsabs.harvard.edu/abs/2020A&A...642A..22P}{\aap, 642,
  A22}

\bibitem[{{Passegger} {et~al.}(2018){Passegger}, {Reiners}, {Jeffers},
  {Wende-von Berg}, {Sch{\"o}fer}, {Caballero}, {Schweitzer}, {Amado},
  {B{\'e}jar}, {Cort{\'e}s-Contreras}, {Hatzes}, {K{\"u}rster}, {Montes},
  {Pedraz}, {Quirrenbach}, {Ribas}, \& {Seifert}}]{Passegger2018}
{Passegger}, V.~M., {Reiners}, A., {Jeffers}, S.~V., {et~al.} 2018,
  \href{https://ui.adsabs.harvard.edu/abs/2018A&A...615A...6P}{\aap, 615, A6}

\bibitem[{{Passegger} {et~al.}(2019){Passegger}, {Schweitzer}, {Shulyak},
  {Nagel}, {Hauschildt}, {Reiners}, {Amado}, {Caballero},
  {Cort{\'e}s-Contreras}, {Dom{\'\i}nguez-Fern{\'a}ndez}, {Quirrenbach},
  {Ribas}, {Azzaro}, {Anglada-Escud{\'e}}, {Bauer}, {B{\'e}jar}, {Dreizler},
  {Guenther}, {Henning}, {Jeffers}, {Kaminski}, {K{\"u}rster}, {Lafarga},
  {Mart{\'\i}n}, {Montes}, {Morales}, {Schmitt}, \&
  {Zechmeister}}]{Passegger2019}
{Passegger}, V.~M., {Schweitzer}, A., {Shulyak}, D., {et~al.} 2019,
  \href{https://ui.adsabs.harvard.edu/abs/2019A&A...627A.161P}{\aap, 627, A161}

\bibitem[{{Perdelwitz} {et~al.}(2021){Perdelwitz}, {Mittag}, {Tal-Or},
  {Schmitt}, {Caballero}, {Jeffers}, {Reiners}, {Schweitzer}, {Trifonov},
  {Ribas}, {Quirrenbach}, {Amado}, {Seifert}, {Cifuentes},
  {Cort{\'e}s-Contreras}, {Montes}, {Revilla}, \&
  {Skrzypinski}}]{Perdelwitz2021A&A...652A.116P}
{Perdelwitz}, V., {Mittag}, M., {Tal-Or}, L., {et~al.} 2021,
  \href{https://ui.adsabs.harvard.edu/abs/2021A&A...652A.116P}{\aap, 652, A116}

\bibitem[{{Perger} {et~al.}(2017){Perger}, {Garc{\'\i}a-Piquer}, {Ribas},
  {Morales}, {Affer}, {Micela}, {Damasso}, {Su{\'a}rez-Mascare{\~n}o},
  {Gonz{\'a}lez-Hern{\'a}ndez}, {Rebolo}, {Herrero}, {Rosich}, {Lafarga},
  {Bignamini}, {Sozzetti}, {Claudi}, {Cosentino}, {Molinari}, {Maldonado},
  {Maggio}, {Lanza}, {Poretti}, {Pagano}, {Desidera}, {Gratton}, {Piotto},
  {Bonomo}, {Martinez Fiorenzano}, {Giacobbe}, {Malavolta}, {Nascimbeni},
  {Rainer}, \& {Scandariato}}]{Perger2017}
{Perger}, M., {Garc{\'\i}a-Piquer}, A., {Ribas}, I., {et~al.} 2017,
  \href{https://ui.adsabs.harvard.edu/abs/2017A&A...598A..26P}{\aap, 598, A26}

\bibitem[{{Perger} {et~al.}(2021){Perger}, {Ribas}, {Anglada-Escud{\'e}},
  {Morales}, {Amado}, {Caballero}, {Quirrenbach}, {Reiners}, {B{\'e}jar},
  {Dreizler}, {Galad{\'\i}-Enr{\'\i}quez}, {Hatzes}, {Henning}, {Jeffers},
  {Kaminski}, {K{\"u}rster}, {Lafarga}, {Montes}, {Pall{\'e}},
  {Rodr{\'\i}guez-L{\'o}pez}, {Schweitzer}, {Zapatero Osorio}, \&
  {Zechmeister}}]{Perger2021A&A...649L..12P}
{Perger}, M., {Ribas}, I., {Anglada-Escud{\'e}}, G., {et~al.} 2021,
  \href{https://ui.adsabs.harvard.edu/abs/2021A&A...649L..12P}{\aap, 649, L12}

\bibitem[{{Perger} {et~al.}(2019){Perger}, {Scandariato}, {Ribas}, {Morales},
  {Affer}, {Azzaro}, {Amado}, {Anglada-Escud{\'e}}, {Baroch}, {Barrado},
  {Bauer}, {B{\'e}jar}, {Caballero}, {Cort{\'e}s-Contreras}, {Damasso},
  {Dreizler}, {Gonz{\'a}lez-Cuesta}, {Gonz{\'a}lez Hern{\'a}ndez}, {Guenther},
  {Henning}, {Herrero}, {Jeffers}, {Kaminski}, {K{\"u}rster}, {Lafarga},
  {Leto}, {L{\'o}pez-Gonz{\'a}lez}, {Maldonado}, {Micela}, {Montes},
  {Pinamonti}, {Quirrenbach}, {Rebolo}, {Reiners}, {Rodr{\'\i}guez},
  {Rodr{\'\i}guez-L{\'o}pez}, {Schmitt}, {Sozzetti}, {Su{\'a}rez
  Mascare{\~n}o}, {Toledo-Padr{\'o}n}, {Zanmar S{\'a}nchez}, {Zapatero Osorio},
  \& {Zechmeister}}]{Perger2019}
{Perger}, M., {Scandariato}, G., {Ribas}, I., {et~al.} 2019,
  \href{https://ui.adsabs.harvard.edu/abs/2019A&A...624A.123P}{\aap, 624, A123}

\bibitem[{{Pinamonti} {et~al.}(2019){Pinamonti}, {Sozzetti}, {Giacobbe},
  {Damasso}, {Scandariato}, {Perger}, {Gonz{\'a}lez Hern{\'a}ndez}, {Lanza},
  {Maldonado}, {Micela}, {Su{\'a}rez Mascare{\~n}o}, {Toledo-Padr{\'o}n},
  {Affer}, {Benatti}, {Bignamini}, {Bonomo}, {Claudi}, {Cosentino}, {Desidera},
  {Maggio}, {Martinez Fiorenzano}, {Pagano}, {Piotto}, {Rainer}, {Rebolo}, \&
  {Ribas}}]{2019A&A...625A.126P}
{Pinamonti}, M., {Sozzetti}, A., {Giacobbe}, P., {et~al.} 2019,
  \href{https://ui.adsabs.harvard.edu/abs/2019A&A...625A.126P}{\aap, 625, A126}

\bibitem[{{Pinamonti} {et~al.}(2022){Pinamonti}, {Sozzetti}, {Maldonado},
  {Affer}, {Micela}, {Bonomo}, {Lanza}, {Perger}, {Ribas}, {Gonz{\'a}lez
  Hern{\'a}ndez}, {Bignamini}, {Claudi}, {Covino}, {Damasso}, {Desidera},
  {Giacobbe}, {Gonz{\'a}lez-{\'A}lvarez}, {Herrero}, {Leto}, {Maggio},
  {Molinari}, {Morales}, {Pagano}, {Petralia}, {Piotto}, {Poretti}, {Rebolo},
  {Scandariato}, {Su{\'a}rez Mascare{\~n}o}, {Toledo-Padr{\'o}n}, \& {Zanmar
  S{\'a}nchez}}]{Pinamonti2022}
{Pinamonti}, M., {Sozzetti}, A., {Maldonado}, J., {et~al.} 2022,
  \href{https://ui.adsabs.harvard.edu/abs/2022A&A...664A..65P}{\aap, 664, A65}

\bibitem[{{Piskunov} \& {Valenti}(2002)}]{Piskunov2002A&A...385.1095P}
{Piskunov}, N.~E. \& {Valenti}, J.~A. 2002,
  \href{https://ui.adsabs.harvard.edu/abs/2002A&A...385.1095P}{\aap, 385, 1095}

\bibitem[{{Quirrenbach} {et~al.}(2014){Quirrenbach}, {Amado}, {Caballero},
  {Mundt}, {Reiners}, {Ribas}, {Seifert}, {Abril}, {Aceituno},
  {Alonso-Floriano}, {Ammler-von Eiff}, {Antona Jim{\'e}nez},
  {Anwand-Heerwart}, {Azzaro}, {Bauer}, {Barrado}, {Becerril}, {B{\'e}jar},
  {Ben{\'{\i}}tez}, {Berdi{\~n}as}, {C{\'a}rdenas}, {Casal}, {Claret},
  {Colom{\'e}}, {Cort{\'e}s-Contreras}, {Czesla}, {Doellinger}, {Dreizler},
  {Feiz}, {Fern{\'a}ndez}, {Galad{\'{\i}}}, {G{\'a}lvez-Ortiz},
  {Garc{\'{\i}}a-Piquer}, {Garc{\'{\i}}a-Vargas}, {Garrido}, {Gesa}, {G{\'o}mez
  Galera}, {Gonz{\'a}lez {\'A}lvarez}, {Gonz{\'a}lez Hern{\'a}ndez},
  {Gr{\"o}zinger}, {Gu{\`a}rdia}, {Guenther}, {de Guindos},
  {Guti{\'e}rrez-Soto}, {Hagen}, {Hatzes}, {Hauschildt}, {Helmling}, {Henning},
  {Hermann}, {Hern{\'a}ndez Casta{\~n}o}, {Herrero}, {Hidalgo}, {Holgado},
  {Huber}, {Huber}, {Jeffers}, {Joergens}, {de Juan}, {Kehr}, {Klein},
  {K{\"u}rster}, {Lamert}, {Lalitha}, {Laun}, {Lemke}, {Lenzen}, {L{\'o}pez del
  Fresno}, {L{\'o}pez Mart{\'{\i}}}, {L{\'o}pez-Santiago}, {Mall}, {Mandel},
  {Mart{\'{\i}}n}, {Mart{\'{\i}}n-Ruiz}, {Mart{\'{\i}}nez-Rodr{\'{\i}}guez},
  {Marvin}, {Mathar}, {Mirabet}, {Montes}, {Morales Mu{\~n}oz}, {Moya},
  {Naranjo}, {Ofir}, {Oreiro}, {Pall{\'e}}, {Panduro}, {Passegger},
  {P{\'e}rez-Calpena}, {P{\'e}rez Medialdea}, {Perger}, {Pluto}, {Ram{\'o}n},
  {Rebolo}, {Redondo}, {Reffert}, {Reinhardt}, {Rhode}, {Rix}, {Rodler},
  {Rodr{\'{\i}}guez}, {Rodr{\'{\i}}guez-L{\'o}pez},
  {Rodr{\'{\i}}guez-P{\'e}rez}, {Rohloff}, {Rosich}, {S{\'a}nchez-Blanco},
  {S{\'a}nchez Carrasco}, {Sanz-Forcada}, {Sarmiento}, {Sch{\"a}fer},
  {Schiller}, {Schmidt}, {Schmitt}, {Solano}, {Stahl}, {Storz}, {St{\"u}rmer},
  {Su{\'a}rez}, {Ulbrich}, {Veredas}, {Wagner}, {Winkler}, {Zapatero Osorio},
  {Zechmeister}, {Abell{\'a}n de Paco}, {Anglada-Escud{\'e}}, {del Burgo},
  {Klutsch}, {Lizon}, {L{\'o}pez-Morales}, {Morales}, {Perryman}, {Tulloch}, \&
  {Xu}}]{Quirrenbach2014}
{Quirrenbach}, A., {Amado}, P.~J., {Caballero}, J.~A., {et~al.} 2014, in
  \procspie, Vol. 9147, Ground-based and Airborne Instrumentation for Astronomy
  V, 91471F

\bibitem[{{Quirrenbach} {et~al.}(2016){Quirrenbach}, {Amado}, {Caballero},
  {Mundt}, {Reiners}, {Ribas}, {Seifert}, {Abril}, {Aceituno},
  {Alonso-Floriano}, {Anwand-Heerwart}, {Azzaro}, {Bauer}, {Barrado},
  {Becerril}, {Bejar}, {Benitez}, {Berdinas}, {Brinkm{\"o}ller}, {Cardenas},
  {Casal}, {Claret}, {Colom{\'e}}, {Cortes-Contreras}, {Czesla}, {Doellinger},
  {Dreizler}, {Feiz}, {Fernandez}, {Ferro}, {Fuhrmeister}, {Galadi},
  {Gallardo}, {G{\'a}lvez-Ortiz}, {Garcia-Piquer}, {Garrido}, {Gesa},
  {G{\'o}mez Galera}, {Gonz{\'a}lez Hern{\'a}ndez}, {Gonzalez Peinado},
  {Gr{\"o}zinger}, {Gu{\`a}rdia}, {Guenther}, {de Guindos}, {Hagen}, {Hatzes},
  {Hauschildt}, {Helmling}, {Henning}, {Hermann}, {Hern{\'a}ndez Arabi},
  {Hern{\'a}ndez Casta{\~n}o}, {Hern{\'a}ndez Hernando}, {Herrero}, {Huber},
  {Huber}, {Huke}, {Jeffers}, {de Juan}, {Kaminski}, {Kehr}, {Kim}, {Klein},
  {Kl{\"u}ter}, {K{\"u}rster}, {Lafarga}, {Lara}, {Lamert}, {Laun},
  {Launhardt}, {Lemke}, {Lenzen}, {Llamas}, {Lopez del Fresno},
  {L{\'o}pez-Puertas}, {L{\'o}pez-Santiago}, {Lopez Salas}, {Magan
  Madinabeitia}, {Mall}, {Mandel}, {Mancini}, {Marin Molina}, {Maroto
  Fern{\'a}ndez}, {Mart{\'{\i}}n}, {Mart{\'{\i}}n-Ruiz}, {Marvin}, {Mathar},
  {Mirabet}, {Montes}, {Morales}, {Morales Mu{\~n}oz}, {Nagel}, {Naranjo},
  {Nowak}, {Palle}, {Panduro}, {Passegger}, {Pavlov}, {Pedraz}, {Perez},
  {P{\'e}rez-Medialdea}, {Perger}, {Pluto}, {Ram{\'o}n}, {Rebolo}, {Redondo},
  {Reffert}, {Reinhart}, {Rhode}, {Rix}, {Rodler}, {Rodr{\'{\i}}guez},
  {Rodr{\'{\i}}guez L{\'o}pez}, {Rohloff}, {Rosich}, {Sanchez Carrasco},
  {Sanz-Forcada}, {Sarkis}, {Sarmiento}, {Sch{\"a}fer}, {Schiller}, {Schmidt},
  {Schmitt}, {Sch{\"o}fer}, {Schweitzer}, {Shulyak}, {Solano}, {Stahl},
  {Storz}, {Tabernero}, {Tala}, {Tal-Or}, {Ulbrich}, {Veredas}, {Vico Linares},
  {Vilardell}, {Wagner}, {Winkler}, {Zapatero Osorio}, {Zechmeister},
  {Ammler-von Eiff}, {Anglada-Escud{\'e}}, {del Burgo}, {Garcia-Vargas},
  {Klutsch}, {Lizon}, {Lopez-Morales}, {Ofir}, {P{\'e}rez-Calpena}, {Perryman},
  {S{\'a}nchez-Blanco}, {Strachan}, {St{\"u}rmer}, {Su{\'a}rez}, {Trifonov},
  {Tulloch}, \& {Xu}}]{Quirrenbach2016}
{Quirrenbach}, A., {Amado}, P.~J., {Caballero}, J.~A., {et~al.} 2016, in
  \procspie, Vol. 9908, Ground-based and Airborne Instrumentation for Astronomy
  VI, 990812

\bibitem[{{Quirrenbach} {et~al.}(2022){Quirrenbach}, {Passegger}, {Trifonov},
  {Amado}, {Caballero}, {Reiners}, {Ribas}, {Aceituno}, {B{\'e}jar},
  {Chaturvedi}, {Gonz{\'a}lez-Cuesta}, {Henning}, {Herrero}, {Kaminski},
  {K{\"u}rster}, {Lalitha}, {Lodieu}, {L{\'o}pez-Gonz{\'a}lez}, {Montes},
  {Pall{\'e}}, {Perger}, {Pollacco}, {Reffert}, {Rodr{\'\i}guez}, {L{\'o}pez},
  {Shan}, {Tal-Or}, {Osorio}, \& {Zechmeister}}]{Quirrenbach2022}
{Quirrenbach}, A., {Passegger}, V.~M., {Trifonov}, T., {et~al.} 2022,
  \href{https://ui.adsabs.harvard.edu/abs/2022A&A...663A..48Q}{\aap, 663, A48}

\bibitem[{{Reiners} {et~al.}(2012){Reiners}, {Joshi}, \&
  {Goldman}}]{Reiners2012}
{Reiners}, A., {Joshi}, N., \& {Goldman}, B. 2012,
  \href{https://ui.adsabs.harvard.edu/abs/2012AJ....143...93R}{\aj, 143, 93}

\bibitem[{{Reiners} {et~al.}(2018{\natexlab{a}}){Reiners}, {Ribas},
  {Zechmeister}, {Caballero}, {Trifonov}, {Dreizler}, {Morales}, {Tal-Or},
  {Lafarga}, {Quirrenbach}, {Amado}, {Kaminski}, {Jeffers}, {Aceituno},
  {B{\'e}jar}, {Gu{\`a}rdia}, {Guenther}, {Hagen}, {Montes}, {Passegger},
  {Seifert}, {Schweitzer}, {Cort{\'e}s-Contreras}, {Abril}, {Alonso-Floriano},
  {Ammler-von Eiff}, {Antona}, {Anglada-Escud{\'e}}, {Anwand-Heerwart},
  {Arroyo-Torres}, {Azzaro}, {Baroch}, {Barrado}, {Bauer}, {Becerril},
  {Ben{\'\i}tez}, {Berdi{\~n}as}, {Bergond}, {Bl{\"u}mcke}, {Brinkm{\"o}ller},
  {del Burgo}, {Cano}, {C{\'a}rdenas V{\'a}zquez}, {Casal}, {Cifuentes},
  {Claret}, {Colom{\'e}}, {Czesla}, {D{\'\i}ez-Alonso}, {Feiz},
  {Fern{\'a}ndez}, {Ferro}, {Fuhrmeister}, {Galad{\'\i}-Enr{\'\i}quez},
  {Garcia-Piquer}, {Garc{\'\i}a Vargas}, {Gesa}, {G{\'o}mez Galera},
  {Gonz{\'a}lez Hern{\'a}ndez}, {Gonz{\'a}lez-Peinado}, {Gr{\"o}zinger},
  {Grohnert}, {Guijarro}, {de Guindos}, {Guti{\'e}rrez-Soto}, {Hatzes},
  {Hauschildt}, {Hedrosa}, {Helmling}, {Henning}, {Hermelo}, {Hern{\'a}ndez
  Arab{\'\i}}, {Hern{\'a}ndez Casta{\~n}o}, {Hern{\'a}ndez Hernando},
  {Herrero}, {Huber}, {Huke}, {Johnson}, {de Juan}, {Kim}, {Klein},
  {Kl{\"u}ter}, {Klutsch}, {K{\"u}rster}, {Labarga}, {Lamert}, {Lamp{\'o}n},
  {Lara}, {Laun}, {Lemke}, {Lenzen}, {Launhardt}, {L{\'o}pez del Fresno},
  {L{\'o}pez-Gonz{\'a}lez}, {L{\'o}pez-Puertas}, {L{\'o}pez Salas},
  {L{\'o}pez-Santiago}, {Luque}, {Mag{\'a}n Madinabeitia}, {Mall}, {Mancini},
  {Mandel}, {Marfil}, {Mar{\'\i}n Molina}, {Maroto Fern{\'a}ndez},
  {Mart{\'\i}n}, {Mart{\'\i}n-Ruiz}, {Marvin}, {Mathar}, {Mirabet},
  {Moreno-Raya}, {Moya}, {Mundt}, {Nagel}, {Naranjo}, {Nortmann}, {Nowak},
  {Ofir}, {Oreiro}, {Pall{\'e}}, {Panduro}, {Pascual}, {Pavlov}, {Pedraz},
  {P{\'e}rez-Calpena}, {P{\'e}rez Medialdea}, {Perger}, {Perryman}, {Pluto},
  {Rabaza}, {Ram{\'o}n}, {Rebolo}, {Redondo}, {Reffert}, {Reinhart}, {Rhode},
  {Rix}, {Rodler}, {Rodr{\'\i}guez}, {Rodr{\'\i}guez-L{\'o}pez},
  {Rodr{\'\i}guez Trinidad}, {Rohloff}, {Rosich}, {Sadegi},
  {S{\'a}nchez-Blanco}, {S{\'a}nchez Carrasco}, {S{\'a}nchez-L{\'o}pez},
  {Sanz-Forcada}, {Sarkis}, {Sarmiento}, {Sch{\"a}fer}, {Schmitt}, {Schiller},
  {Sch{\"o}fer}, {Solano}, {Stahl}, {Strachan}, {St{\"u}rmer}, {Su{\'a}rez},
  {Tabernero}, {Tala}, {Tulloch}, {Ulbrich}, {Veredas}, {Vico Linares},
  {Vilardell}, {Wagner}, {Winkler}, {Wolthoff}, {Xu}, {Yan}, \& {Zapatero
  Osorio}}]{Reiners2018a}
{Reiners}, A., {Ribas}, I., {Zechmeister}, M., {et~al.} 2018{\natexlab{a}},
  \href{https://ui.adsabs.harvard.edu/abs/2018A&A...609L...5R}{\aap, 609, L5}

\bibitem[{{Reiners} {et~al.}(2022){Reiners}, {Shulyak}, {K{\"a}pyl{\"a}},
  {Ribas}, {Nagel}, {Zechmeister}, {Caballero}, {Shan}, {Fuhrmeister},
  {Quirrenbach}, {Amado}, {Montes}, {Jeffers}, {Azzaro}, {B{\'e}jar},
  {Chaturvedi}, {Henning}, {K{\"u}rster}, \& {Pall{\'e}}}]{Reiners2022}
{Reiners}, A., {Shulyak}, D., {K{\"a}pyl{\"a}}, P.~J., {et~al.} 2022,
  \href{https://ui.adsabs.harvard.edu/abs/2022A&A...662A..41R}{\aap, 662, A41}

\bibitem[{{Reiners} \& {Zechmeister}(2020)}]{Reiners2020}
{Reiners}, A. \& {Zechmeister}, M. 2020,
  \href{https://ui.adsabs.harvard.edu/abs/2020ApJS..247...11R}{\apjs, 247, 11}

\bibitem[{{Reiners} {et~al.}(2018{\natexlab{b}}){Reiners}, {Zechmeister},
  {Caballero}, {Ribas}, {Morales}, {Jeffers}, {Sch{\"o}fer}, {Tal-Or},
  {Quirrenbach}, {Amado}, {Kaminski}, {Seifert}, {Abril}, {Aceituno},
  {Alonso-Floriano}, {Ammler-von Eiff}, {Antona}, {Anglada-Escud{\'e}},
  {Anwand-Heerwart}, {Arroyo-Torres}, {Azzaro}, {Baroch}, {Barrado}, {Bauer},
  {Becerril}, {B{\'e}jar}, {Ben{\'{\i}}tez}, {Berdi{\~n}as}, {Bergond},
  {Bl{\"u}mcke}, {Brinkm{\"o}ller}, {del Burgo}, {Cano}, {C{\'a}rdenas
  V{\'a}zquez}, {Casal}, {Cifuentes}, {Claret}, {Colom{\'e}},
  {Cort{\'e}s-Contreras}, {Czesla}, {D{\'{\i}}ez-Alonso}, {Dreizler}, {Feiz},
  {Fern{\'a}ndez}, {Ferro}, {Fuhrmeister}, {Galad{\'{\i}}-Enr{\'{\i}}quez},
  {Garcia-Piquer}, {Garc{\'{\i}}a Vargas}, {Gesa}, {G{\'o}mez Galera},
  {Gonz{\'a}lez Hern{\'a}ndez}, {Gonz{\'a}lez-Peinado}, {Gr{\"o}zinger},
  {Grohnert}, {Gu{\`a}rdia}, {Guenther}, {Guijarro}, {de Guindos},
  {Guti{\'e}rrez-Soto}, {Hagen}, {Hatzes}, {Hauschildt}, {Hedrosa}, {Helmling},
  {Henning}, {Hermelo}, {Hern{\'a}ndez Arab{\'{\i}}}, {Hern{\'a}ndez
  Casta{\~n}o}, {Hern{\'a}ndez Hernando}, {Herrero}, {Huber}, {Huke},
  {Johnson}, {de Juan}, {Kim}, {Klein}, {Kl{\"u}ter}, {Klutsch}, {K{\"u}rster},
  {Lafarga}, {Lamert}, {Lamp{\'o}n}, {Lara}, {Laun}, {Lemke}, {Lenzen},
  {Launhardt}, {L{\'o}pez del Fresno}, {L{\'o}pez-Gonz{\'a}lez},
  {L{\'o}pez-Puertas}, {L{\'o}pez Salas}, {L{\'o}pez-Santiago}, {Luque},
  {Mag{\'a}n Madinabeitia}, {Mall}, {Mancini}, {Mandel}, {Marfil},
  {Mar{\'{\i}}n Molina}, {Maroto Fern{\'a}ndez}, {Mart{\'{\i}}n},
  {Mart{\'{\i}}n-Ruiz}, {Marvin}, {Mathar}, {Mirabet}, {Montes}, {Moreno-Raya},
  {Moya}, {Mundt}, {Nagel}, {Naranjo}, {Nortmann}, {Nowak}, {Ofir}, {Oreiro},
  {Pall{\'e}}, {Panduro}, {Pascual}, {Passegger}, {Pavlov}, {Pedraz},
  {P{\'e}rez-Calpena}, {P{\'e}rez Medialdea}, {Perger}, {Perryman}, {Pluto},
  {Rabaza}, {Ram{\'o}n}, {Rebolo}, {Redondo}, {Reffert}, {Reinhart}, {Rhode},
  {Rix}, {Rodler}, {Rodr{\'{\i}}guez}, {Rodr{\'{\i}}guez-L{\'o}pez},
  {Rodr{\'{\i}}guez Trinidad}, {Rohloff}, {Rosich}, {Sadegi},
  {S{\'a}nchez-Blanco}, {S{\'a}nchez Carrasco}, {S{\'a}nchez-L{\'o}pez},
  {Sanz-Forcada}, {Sarkis}, {Sarmiento}, {Sch{\"a}fer}, {Schmitt}, {Schiller},
  {Schweitzer}, {Solano}, {Stahl}, {Strachan}, {St{\"u}rmer}, {Su{\'a}rez},
  {Tabernero}, {Tala}, {Trifonov}, {Tulloch}, {Ulbrich}, {Veredas}, {Vico
  Linares}, {Vilardell}, {Wagner}, {Winkler}, {Wolthoff}, {Xu}, {Yan}, \&
  {Zapatero Osorio}}]{Reiners2018}
{Reiners}, A., {Zechmeister}, M., {Caballero}, J.~A., {et~al.}
  2018{\natexlab{b}},
  \href{http://adsabs.harvard.edu/abs/2018A%26A...612A..49R}{\aap, 612, A49}

\bibitem[{{Reyl{\'e}} {et~al.}(2021){Reyl{\'e}}, {Jardine}, {Fouqu{\'e}},
  {Caballero}, {Smart}, \& {Sozzetti}}]{Reyle2021A&A...650A.201R}
{Reyl{\'e}}, C., {Jardine}, K., {Fouqu{\'e}}, P., {et~al.} 2021,
  \href{https://ui.adsabs.harvard.edu/abs/2021A&A...650A.201R}{\aap, 650, A201}

\bibitem[{{Ribas} {et~al.}(2018){Ribas}, {Tuomi}, {Reiners}, {Butler},
  {Morales}, {Perger}, {Dreizler}, {Rodr{\'\i}guez-L{\'o}pez}, {Gonz{\'a}lez
  Hern{\'a}ndez}, {Rosich}, {Feng}, {Trifonov}, {Vogt}, {Caballero}, {Hatzes},
  {Herrero}, {Jeffers}, {Lafarga}, {Murgas}, {Nelson}, {Rodr{\'\i}guez},
  {Strachan}, {Tal-Or}, {Teske}, {Toledo-Padr{\'o}n}, {Zechmeister},
  {Quirrenbach}, {Amado}, {Azzaro}, {B{\'e}jar}, {Barnes}, {Berdi{\~n}as},
  {Burt}, {Coleman}, {Cort{\'e}s-Contreras}, {Crane}, {Engle}, {Guinan},
  {Haswell}, {Henning}, {Holden}, {Jenkins}, {Jones}, {Kaminski}, {Kiraga},
  {K{\"u}rster}, {Lee}, {L{\'o}pez-Gonz{\'a}lez}, {Montes}, {Morin}, {Ofir},
  {Pall{\'e}}, {Rebolo}, {Reffert}, {Schweitzer}, {Seifert}, {Shectman},
  {Staab}, {Street}, {Su{\'a}rez Mascare{\~n}o}, {Tsapras}, {Wang}, \&
  {Anglada-Escud{\'e}}}]{Ribas2018}
{Ribas}, I., {Tuomi}, M., {Reiners}, A., {et~al.} 2018,
  \href{https://ui.adsabs.harvard.edu/abs/2018Natur.563..365R}{\nat, 563, 365}

\bibitem[{{Ricker} {et~al.}(2015){Ricker}, {Winn}, {Vanderspek}, {Latham},
  {Bakos}, {Bean}, {Berta-Thompson}, {Brown}, {Buchhave}, {Butler}, {Butler},
  {Chaplin}, {Charbonneau}, {Christensen-Dalsgaard}, {Clampin}, {Deming},
  {Doty}, {De Lee}, {Dressing}, {Dunham}, {Endl}, {Fressin}, {Ge}, {Henning},
  {Holman}, {Howard}, {Ida}, {Jenkins}, {Jernigan}, {Johnson}, {Kaltenegger},
  {Kawai}, {Kjeldsen}, {Laughlin}, {Levine}, {Lin}, {Lissauer}, {MacQueen},
  {Marcy}, {McCullough}, {Morton}, {Narita}, {Paegert}, {Palle}, {Pepe},
  {Pepper}, {Quirrenbach}, {Rinehart}, {Sasselov}, {Sato}, {Seager},
  {Sozzetti}, {Stassun}, {Sullivan}, {Szentgyorgyi}, {Torres}, {Udry}, \&
  {Villasenor}}]{Ricker2015}
{Ricker}, G.~R., {Winn}, J.~N., {Vanderspek}, R., {et~al.} 2015,
  \href{https://ui.adsabs.harvard.edu/abs/2015JATIS...1a4003R}{Journal of
  Astronomical Telescopes, Instruments, and Systems, 1, 014003}

\bibitem[{{Robertson} {et~al.}(2020){Robertson}, {Stef\'ansson}, {Mahadevan},
  {Endl}, {Cochran}, {Beard}, {Bender}, {Diddams}, {Duong}, {Ford}, {Fredrick},
  {Halverson}, {Hearty}, {Holcomb}, {Juan}, {Kanodia}, {Lubin}, {Metcalf},
  {Monson}, {Ninan}, {Palafoutas}, {Ramsey}, {Roy}, {Schwab}, {Terrien}, \&
  {Wright}}]{Robertson2020}
{Robertson}, P., {Stef\'ansson}, G., {Mahadevan}, S., {et~al.} 2020,
  \href{https://ui.adsabs.harvard.edu/abs/2020ApJ...897..125R}{\apj, 897, 125}

\bibitem[{{Sabotta} {et~al.}(2021){Sabotta}, {Schlecker}, {Chaturvedi},
  {Guenther}, {Mu{\~n}oz Rodr{\'\i}guez}, {Mu{\~n}oz S{\'a}nchez}, {Caballero},
  {Shan}, {Reffert}, {Ribas}, {Reiners}, {Hatzes}, {Amado}, {Klahr}, {Morales},
  {Quirrenbach}, {Henning}, {Dreizler}, {Pall{\'e}}, {Perger}, {Azzaro},
  {Jeffers}, {Kaminski}, {K{\"u}rster}, {Lafarga}, {Montes}, {Passegger}, \&
  {Zechmeister}}]{Sabotta2021A&A...653A.114S}
{Sabotta}, S., {Schlecker}, M., {Chaturvedi}, P., {et~al.} 2021,
  \href{https://ui.adsabs.harvard.edu/abs/2021A&A...653A.114S}{\aap, 653, A114}

\bibitem[{{S{\'a}nchez-L{\'o}pez} {et~al.}(2020){S{\'a}nchez-L{\'o}pez},
  {L{\'o}pez-Puertas}, {Snellen}, {Nagel}, {Bauer}, {Pall{\'e}}, {Tal-Or},
  {Amado}, {Caballero}, {Czesla}, {Nortmann}, {Reiners}, {Ribas},
  {Quirrenbach}, {Aceituno}, {B{\'e}jar}, {Casasayas-Barris}, {Henning},
  {Molaverdikhani}, {Montes}, {Stangret}, {Zapatero Osorio}, \&
  {Zechmeister}}]{SanchezLopez2020}
{S{\'a}nchez-L{\'o}pez}, A., {L{\'o}pez-Puertas}, M., {Snellen}, I.~A.~G.,
  {et~al.} 2020,
  \href{https://ui.adsabs.harvard.edu/abs/2020A&A...643A..24S}{\aap, 643, A24}

\bibitem[{{Sarkis} {et~al.}(2018){Sarkis}, {Henning}, {K{\"u}rster},
  {Trifonov}, {Zechmeister}, {Tal-Or}, {Anglada-Escud{\'e}}, {Hatzes},
  {Lafarga}, {Dreizler}, {Ribas}, {Caballero}, {Reiners}, {Mallonn}, {Morales},
  {Kaminski}, {Aceituno}, {Amado}, {B{\'e}jar}, {Hagen}, {Jeffers},
  {Quirrenbach}, {Launhardt}, {Marvin}, \&
  {Montes}}]{Sarkis2018AJ....155..257S}
{Sarkis}, P., {Henning}, T., {K{\"u}rster}, M., {et~al.} 2018,
  \href{https://ui.adsabs.harvard.edu/abs/2018AJ....155..257S}{\aj, 155, 257}

\bibitem[{{Schlecker} {et~al.}(2022){Schlecker}, {Burn}, {Sabotta}, {Seifert},
  {Henning}, {Emsenhuber}, {Mordasini}, {Reffert}, {Shan}, \&
  {Klahr}}]{Schlecker2022}
{Schlecker}, M., {Burn}, R., {Sabotta}, S., {et~al.} 2022,
  \href{https://ui.adsabs.harvard.edu/abs/2022A&A...664A.180S}{\aap, 664, A180}

\bibitem[{{Schlecker} {et~al.}(2021{\natexlab{a}}){Schlecker}, {Mordasini},
  {Emsenhuber}, {Klahr}, {Henning}, {Burn}, {Alibert}, \&
  {Benz}}]{Schlecker2021b}
{Schlecker}, M., {Mordasini}, C., {Emsenhuber}, A., {et~al.}
  2021{\natexlab{a}},
  \href{https://ui.adsabs.harvard.edu/abs/2021A&A...656A..71S}{\aap, 656, A71}

\bibitem[{{Schlecker} {et~al.}(2021{\natexlab{b}}){Schlecker}, {Pham}, {Burn},
  {Alibert}, {Mordasini}, {Emsenhuber}, {Klahr}, {Henning}, \&
  {Mishra}}]{Schlecker2021a}
{Schlecker}, M., {Pham}, D., {Burn}, R., {et~al.} 2021{\natexlab{b}},
  \href{https://ui.adsabs.harvard.edu/abs/2021A&A...656A..73S}{\aap, 656, A73}

\bibitem[{{Schneider} {et~al.}(2011){Schneider}, {Dedieu}, {Le Sidaner},
  {Savalle}, \& {Zolotukhin}}]{Schneider2011}
{Schneider}, J., {Dedieu}, C., {Le Sidaner}, P., {Savalle}, R., \&
  {Zolotukhin}, I. 2011,
  \href{https://ui.adsabs.harvard.edu/abs/2011A&A...532A..79S}{\aap, 532, A79}

\bibitem[{{Sch{\"o}fer} {et~al.}(2019){Sch{\"o}fer}, {Jeffers}, {Reiners},
  {Shulyak}, {Fuhrmeister}, {Johnson}, {Zechmeister}, {Ribas}, {Quirrenbach},
  {Amado}, {Caballero}, {Anglada-Escud{\'e}}, {Bauer}, {B{\'e}jar},
  {Cort{\'e}s-Contreras}, {Dreizler}, {Guenther}, {Kaminski}, {K{\"u}rster},
  {Lafarga}, {Montes}, {Morales}, {Pedraz}, \& {Tal-Or}}]{Schoefer2019}
{Sch{\"o}fer}, P., {Jeffers}, S.~V., {Reiners}, A., {et~al.} 2019,
  \href{https://ui.adsabs.harvard.edu/abs/2019A&A...623A..44S}{\aap, 623, A44}

\bibitem[{{Schweitzer} {et~al.}(2019){Schweitzer}, {Passegger}, {Cifuentes},
  {B{\'e}jar}, {Cort{\'e}s-Contreras}, {Caballero}, {del Burgo}, {Czesla},
  {K{\"u}rster}, {Montes}, {Zapatero Osorio}, {Ribas}, {Reiners},
  {Quirrenbach}, {Amado}, {Aceituno}, {Anglada-Escud{\'e}}, {Bauer},
  {Dreizler}, {Jeffers}, {Guenther}, {Henning}, {Kaminski}, {Lafarga},
  {Marfil}, {Morales}, {Schmitt}, {Seifert}, {Solano}, {Tabernero}, \&
  {Zechmeister}}]{Schweitzer2019}
{Schweitzer}, A., {Passegger}, V.~M., {Cifuentes}, C., {et~al.} 2019,
  \href{https://ui.adsabs.harvard.edu/abs/2019A&A...625A..68S}{\aap, 625, A68}

\bibitem[{{Sedaghati} {et~al.}(2022){Sedaghati}, {S{\'a}nchez-L{\'o}pez},
  {Czesla}, {L{\'o}pez-Puertas}, {Amado}, {Palle}, {Molaverdikhani},
  {Caballero}, {Nortmann}, {Quirrenbach}, {Reiners}, \&
  {Ribas}}]{Sedaghati2022}
{Sedaghati}, E., {S{\'a}nchez-L{\'o}pez}, A., {Czesla}, S., {et~al.} 2022,
  \href{https://ui.adsabs.harvard.edu/abs/2022A&A...659A..44S}{\aap, 659, A44}

\bibitem[{{Seifahrt} {et~al.}(2018){Seifahrt}, {St{\"u}rmer}, {Bean}, \&
  {Schwab}}]{Seifahrt2018}
{Seifahrt}, A., {St{\"u}rmer}, J., {Bean}, J.~L., \& {Schwab}, C. 2018, in
  Society of Photo-Optical Instrumentation Engineers (SPIE) Conference Series,
  Vol. 10702, Ground-based and Airborne Instrumentation for Astronomy VII, ed.
  C.~J. {Evans}, L.~{Simard}, \& H.~{Takami}, 107026D

\bibitem[{{Seifert} {et~al.}(2012){Seifert}, {S{\'a}nchez Carrasco}, {Xu},
  {C{\'a}rdenas}, {S{\'a}nchez-Blanco}, {Becerril}, {Feiz}, {Ram{\'o}n},
  {Dreizler}, {Rohde}, {Quirrenbach}, {Amado}, {Ribas}, {Reiners}, {Mandel}, \&
  {Caballero}}]{Seifert2012}
{Seifert}, W., {S{\'a}nchez Carrasco}, M.~A., {Xu}, W., {et~al.} 2012, in
  Society of Photo-Optical Instrumentation Engineers (SPIE) Conference Series,
  Vol. 8446, Ground-based and Airborne Instrumentation for Astronomy IV, ed.
  I.~S. {McLean}, S.~K. {Ramsay}, \& H.~{Takami}, 844633

\bibitem[{{Shan} {et~al.}(2021){Shan}, {Reiners}, {Fabbian}, {Marfil},
  {Montes}, {Tabernero}, {Ribas}, {Caballero}, {Quirrenbach}, {Amado},
  {Aceituno}, {B{\'e}jar}, {Cort{\'e}s-Contreras}, {Dreizler}, {Hatzes},
  {Henning}, {Jeffers}, {Kaminski}, {K{\"u}rster}, {Lafarga}, {Morales},
  {Nagel}, {Pall{\'e}}, {Passegger}, {Rodriguez-L{\'o}pez}, {Schweitzer}, \&
  {Zechmeister}}]{Shan2021}
{Shan}, Y., {Reiners}, A., {Fabbian}, D., {et~al.} 2021,
  \href{https://ui.adsabs.harvard.edu/abs/2021A&A...654A.118S}{\aap, 654, A118}

\bibitem[{{Shulyak} {et~al.}(2019){Shulyak}, {Reiners}, {Nagel}, {Tal-Or},
  {Caballero}, {Zechmeister}, {B{\'e}jar}, {Cort{\'e}s-Contreras}, {Martin},
  {Kaminski}, {Ribas}, {Quirrenbach}, {Amado}, {Anglada-Escud{\'e}}, {Bauer},
  {Dreizler}, {Guenther}, {Henning}, {Jeffers}, {K{\"u}rster}, {Lafarga},
  {Montes}, {Morales}, \& {Pedraz}}]{Shulyak2019}
{Shulyak}, D., {Reiners}, A., {Nagel}, E., {et~al.} 2019,
  \href{https://ui.adsabs.harvard.edu/abs/2019A&A...626A..86S}{\aap, 626, A86}

\bibitem[{{Soto} {et~al.}(2021){Soto}, {Anglada-Escud{\'e}}, {Dreizler},
  {Molaverdikhani}, {Kemmer}, {Rodr{\'\i}guez-L{\'o}pez}, {Lillo-Box},
  {Pall{\'e}}, {Espinoza}, {Caballero}, {Quirrenbach}, {Ribas}, {Reiners},
  {Narita}, {Hirano}, {Amado}, {B{\'e}jar}, {Bluhm}, {Burke}, {Caldwell},
  {Charbonneau}, {Cloutier}, {Collins}, {Cort{\'e}s-Contreras}, {Girardin},
  {Guerra}, {Harakawa}, {Hatzes}, {Irwin}, {Jenkins}, {Jensen}, {Kawauchi},
  {Kotani}, {Kudo}, {Kunimoto}, {Kuzuhara}, {Latham}, {Montes}, {Morales},
  {Mori}, {Nelson}, {Omiya}, {Pedraz}, {Passegger}, {Rackham}, {Rudat},
  {Schlieder}, {Sch{\"o}fer}, {Schweitzer}, {Selezneva}, {Stockdale}, {Tamura},
  {Trifonov}, {Vanderspek}, \& {Watanabe}}]{Soto2021}
{Soto}, M.~G., {Anglada-Escud{\'e}}, G., {Dreizler}, S., {et~al.} 2021,
  \href{https://ui.adsabs.harvard.edu/abs/2021A&A...649A.144S}{\aap, 649, A144}

\bibitem[{{Stef\'ansson} {et~al.}(2020){Stef\'ansson}, {Ca{\~n}as},
  {Wisniewski}, {Robertson}, {Mahadevan}, {Maney}, {Kanodia}, {Beard},
  {Bender}, {Brunt}, {Clemens}, {Cochran}, {Diddams}, {Endl}, {Ford},
  {Fredrick}, {Halverson}, {Hearty}, {Hebb}, {Huehnerhoff}, {Jennings},
  {Kaplan}, {Levi}, {Lubar}, {Metcalf}, {Monson}, {Morris}, {Ninan}, {Nitroy},
  {Ramsey}, {Roy}, {Schwab}, {Sigurdsson}, {Terrien}, \&
  {Wright}}]{Stefansson2020}
{Stef\'ansson}, G., {Ca{\~n}as}, C., {Wisniewski}, J., {et~al.} 2020,
  \href{https://ui.adsabs.harvard.edu/abs/2020AJ....159..100S}{\aj, 159, 100}

\bibitem[{{Stock} {et~al.}(2020{\natexlab{a}}){Stock}, {Kemmer}, {Reffert},
  {Trifonov}, {Kaminski}, {Dreizler}, {Quirrenbach}, {Caballero}, {Reiners},
  {Jeffers}, {Anglada-Escud{\'e}}, {Ribas}, {Amado}, {Barrado}, {Barnes},
  {Bauer}, {Berdi{\~n}as}, {B{\'e}jar}, {Coleman}, {Cort{\'e}s-Contreras},
  {D{\'\i}ez-Alonso}, {Dom{\'\i}nguez-Fern{\'a}ndez}, {Espinoza}, {Haswell},
  {Hatzes}, {Henning}, {Jenkins}, {Jones}, {Kossakowski}, {K{\"u}rster},
  {Lafarga}, {Lee}, {L{\'o}pez Gonz{\'a}lez}, {Montes}, {Morales}, {Morales},
  {Pall{\'e}}, {Pedraz}, {Rodr{\'\i}guez}, {Rodr{\'\i}guez-L{\'o}pez}, \&
  {Zechmeister}}]{Stock2020a}
{Stock}, S., {Kemmer}, J., {Reffert}, S., {et~al.} 2020{\natexlab{a}},
  \href{https://ui.adsabs.harvard.edu/abs/2020A&A...636A.119S}{\aap, 636, A119}

\bibitem[{{Stock} {et~al.}(2020{\natexlab{b}}){Stock}, {Nagel}, {Kemmer},
  {Passegger}, {Reffert}, {Quirrenbach}, {Caballero}, {Czesla}, {B{\'e}jar},
  {Cardona}, {D{\'\i}ez-Alonso}, {Herrero}, {Lalitha}, {Schlecker}, {Tal-Or},
  {Rodr{\'\i}guez}, {Rodr{\'\i}guez-L{\'o}pez}, {Ribas}, {Reiners}, {Amado},
  {Bauer}, {Bluhm}, {Cort{\'e}s-Contreras}, {Gonz{\'a}lez-Cuesta}, {Dreizler},
  {Hatzes}, {Henning}, {Jeffers}, {Kaminski}, {K{\"u}rster}, {Lafarga},
  {L{\'o}pez-Gonz{\'a}lez}, {Montes}, {Morales}, {Pedraz}, {Sch{\"o}fer},
  {Schweitzer}, {Trifonov}, {Zapatero Osorio}, \& {Zechmeister}}]{Stock2020b}
{Stock}, S., {Nagel}, E., {Kemmer}, J., {et~al.} 2020{\natexlab{b}},
  \href{https://ui.adsabs.harvard.edu/abs/2020A&A...643A.112S}{\aap, 643, A112}

\bibitem[{{St{\"u}rmer} {et~al.}(2014){St{\"u}rmer}, {Stahl}, {Schwab},
  {Seifert}, {Quirrenbach}, {Amado}, {Ribas}, {Reiners}, \&
  {Caballero}}]{Sturmer2014}
{St{\"u}rmer}, J., {Stahl}, O., {Schwab}, C., {et~al.} 2014, in Society of
  Photo-Optical Instrumentation Engineers (SPIE) Conference Series, Vol. 9151,
  Advances in Optical and Mechanical Technologies for Telescopes and
  Instrumentation, ed. R.~{Navarro}, C.~R. {Cunningham}, \& A.~A. {Barto},
  915152

\bibitem[{{Su{\'a}rez Mascare{\~n}o} {et~al.}(2022){Su{\'a}rez Mascare{\~n}o},
  {Gonz{\'a}lez-{\'A}lvarez}, {Zapatero Osorio}, {Lillo-Box}, {Faria},
  {Passegger}, {Gonz{\'a}lez Hern{\'a}ndez}, {Figueira}, {Sozzetti}, {Rebolo},
  {Pepe}, {Santos}, {Cristiani}, {Lovis}, {Silva}, {Ribas}, {Amado},
  {Caballero}, {Quirrenbach}, {Reiners}, {Zechmeister}, {Adibekyan}, {Alibert},
  {B{\'e}jar}, {Benatti}, {D'Odorico}, {Damasso}, {Delisle}, {Dreizler},
  {Ehrenreich}, {Hatzes}, {Hara}, {Henning}, {Kaminski},
  {L{\'o}pez-Gonz{\'a}lez}, {Martins}, {Micela}, {Montes}, {Pall{\'e}},
  {Pedraz}, {Rodr{\'\i}guez}, {Rodr{\'\i}guez-L{\'o}pez}, {Tal-Or}, {Sousa}, \&
  {Udry}}]{Suarez2022}
{Su{\'a}rez Mascare{\~n}o}, A., {Gonz{\'a}lez-{\'A}lvarez}, E., {Zapatero
  Osorio}, M.~R., {et~al.} 2022,
  \href{https://ui.adsabs.harvard.edu/abs/2022arXiv221207332S}{arXiv e-prints,
  arXiv:2212.07332}

\bibitem[{{Su{\'a}rez Mascare{\~n}o} {et~al.}(2017){Su{\'a}rez Mascare{\~n}o},
  {Gonz{\'a}lez Hern{\'a}ndez}, {Rebolo}, {Velasco}, {Toledo-Padr{\'o}n},
  {Affer}, {Perger}, {Micela}, {Ribas}, {Maldonado}, {Leto}, {Zanmar Sanchez},
  {Scandariato}, {Damasso}, {Sozzetti}, {Esposito}, {Covino}, {Maggio},
  {Lanza}, {Desidera}, {Rosich}, {Bignamini}, {Claudi}, {Benatti}, {Borsa},
  {Pedani}, {Molinari}, {Morales}, {Herrero}, \&
  {Lafarga}}]{2017A&A...605A..92S}
{Su{\'a}rez Mascare{\~n}o}, A., {Gonz{\'a}lez Hern{\'a}ndez}, J.~I., {Rebolo},
  R., {et~al.} 2017,
  \href{https://ui.adsabs.harvard.edu/abs/2017A&A...605A..92S}{\aap, 605, A92}

\bibitem[{{Su{\'a}rez Mascare{\~n}o} {et~al.}(2018){Su{\'a}rez Mascare{\~n}o},
  {Rebolo}, {Gonz{\'a}lez Hern{\'a}ndez}, {Toledo-Padr{\'o}n}, {Perger},
  {Ribas}, {Affer}, {Micela}, {Damasso}, {Maldonado}, {Gonz{\'a}lez-Alvarez},
  {Leto}, {Pagano}, {Scandariato}, {Sozzetti}, {Lanza}, {Malavolta}, {Claudi},
  {Cosentino}, {Desidera}, {Giacobbe}, {Maggio}, {Rainer}, {Esposito},
  {Benatti}, {Pedani}, {Morales}, {Herrero}, {Lafarga}, {Rosich}, \&
  {Pinamonti}}]{Suarez2018}
{Su{\'a}rez Mascare{\~n}o}, A., {Rebolo}, R., {Gonz{\'a}lez Hern{\'a}ndez},
  J.~I., {et~al.} 2018,
  \href{https://ui.adsabs.harvard.edu/abs/2018A&A...612A..89S}{\aap, 612, A89}

\bibitem[{{Tal-Or} {et~al.}(2019){Tal-Or}, {Trifonov}, {Zucker}, {Mazeh}, \&
  {Zechmeister}}]{TalOr2019MNRAS.484L...8T}
{Tal-Or}, L., {Trifonov}, T., {Zucker}, S., {Mazeh}, T., \& {Zechmeister}, M.
  2019, \href{https://ui.adsabs.harvard.edu/abs/2019MNRAS.484L...8T}{\mnras,
  484, L8}

\bibitem[{{Tal-Or} {et~al.}(2018){Tal-Or}, {Zechmeister}, {Reiners}, {Jeffers},
  {Sch{\"o}fer}, {Quirrenbach}, {Amado}, {Ribas}, {Caballero}, {Aceituno},
  {Bauer}, {B{\'e}jar}, {Czesla}, {Dreizler}, {Fuhrmeister}, {Hatzes},
  {Johnson}, {K{\"u}rster}, {Lafarga}, {Montes}, {Morales}, {Reffert},
  {Sadegi}, {Seifert}, \& {Shulyak}}]{TalOr2018}
{Tal-Or}, L., {Zechmeister}, M., {Reiners}, A., {et~al.} 2018,
  \href{https://ui.adsabs.harvard.edu/abs/2018A&A...614A.122T}{\aap, 614, A122}

\bibitem[{{Toledo-Padr{\'o}n} {et~al.}(2021){Toledo-Padr{\'o}n}, {Su{\'a}rez
  Mascare{\~n}o}, {Gonz{\'a}lez Hern{\'a}ndez}, {Rebolo}, {Pinamonti},
  {Perger}, {Scandariato}, {Damasso}, {Sozzetti}, {Maldonado}, {Desidera},
  {Ribas}, {Micela}, {Affer}, {Gonz{\'a}lez-Alvarez}, {Leto}, {Pagano}, {Zanmar
  S{\'a}nchez}, {Giacobbe}, {Herrero}, {Morales}, {Amado}, {Caballero},
  {Quirrenbach}, {Reiners}, \& {Zechmeister}}]{Toledo2021}
{Toledo-Padr{\'o}n}, B., {Su{\'a}rez Mascare{\~n}o}, A., {Gonz{\'a}lez
  Hern{\'a}ndez}, J.~I., {et~al.} 2021,
  \href{https://ui.adsabs.harvard.edu/abs/2021A&A...648A..20T}{\aap, 648, A20}

\bibitem[{{Trifonov} {et~al.}(2021){Trifonov}, {Caballero}, {Morales},
  {Seifahrt}, {Ribas}, {Reiners}, {Bean}, {Luque}, {Parviainen}, {Pall{\'e}},
  {Stock}, {Zechmeister}, {Amado}, {Anglada-Escud{\'e}}, {Azzaro}, {Barclay},
  {B{\'e}jar}, {Bluhm}, {Casasayas-Barris}, {Cifuentes}, {Collins}, {Collins},
  {Cort{\'e}s-Contreras}, {de Leon}, {Dreizler}, {Dressing}, {Esparza-Borges},
  {Espinoza}, {Fausnaugh}, {Fukui}, {Hatzes}, {Hellier}, {Henning}, {Henze},
  {Herrero}, {Jeffers}, {Jenkins}, {Jensen}, {Kaminski}, {Kasper},
  {Kossakowski}, {K{\"u}rster}, {Lafarga}, {Latham}, {Mann}, {Molaverdikhani},
  {Montes}, {Montet}, {Murgas}, {Narita}, {Oshagh}, {Passegger}, {Pollacco},
  {Quinn}, {Quirrenbach}, {Ricker}, {Rodr{\'\i}guez L{\'o}pez}, {Sanz-Forcada},
  {Schwarz}, {Schweitzer}, {Seager}, {Shporer}, {Stangret}, {St{\"u}rmer},
  {Tan}, {Tenenbaum}, {Twicken}, {Vanderspek}, \& {Winn}}]{Trifonov2021}
{Trifonov}, T., {Caballero}, J.~A., {Morales}, J.~C., {et~al.} 2021,
  \href{https://ui.adsabs.harvard.edu/abs/2021Sci...371.1038T}{Science, 371,
  1038}

\bibitem[{{Trifonov} {et~al.}(2018){Trifonov}, {K{\"u}rster}, {Zechmeister},
  {Tal-Or}, {Caballero}, {Quirrenbach}, {Amado}, {Ribas}, {Reiners}, {Reffert},
  {Dreizler}, {Hatzes}, {Kaminski}, {Launhardt}, {Henning}, {Montes},
  {B{\'e}jar}, {Mundt}, {Pavlov}, {Schmitt}, {Seifert}, {Morales}, {Nowak},
  {Jeffers}, {Rodr{\'\i}guez-L{\'o}pez}, {del Burgo}, {Anglada-Escud{\'e}},
  {L{\'o}pez-Santiago}, {Mathar}, {Ammler-von Eiff}, {Guenther}, {Barrado},
  {Gonz{\'a}lez Hern{\'a}ndez}, {Mancini}, {St{\"u}rmer}, {Abril}, {Aceituno},
  {Alonso-Floriano}, {Antona}, {Anwand-Heerwart}, {Arroyo-Torres}, {Azzaro},
  {Baroch}, {Bauer}, {Becerril}, {Ben{\'\i}tez}, {Berdi{\~n}as}, {Bergond},
  {Bl{\"u}mcke}, {Brinkm{\"o}ller}, {Cano}, {C{\'a}rdenas V{\'a}zquez},
  {Casal}, {Cifuentes}, {Claret}, {Colom{\'e}}, {Cort{\'e}s-Contreras},
  {Czesla}, {D{\'\i}ez-Alonso}, {Feiz}, {Fern{\'a}ndez}, {Ferro},
  {Fuhrmeister}, {Galad{\'\i}-Enr{\'\i}quez}, {Garcia-Piquer}, {Garc{\'\i}a
  Vargas}, {Gesa}, {G{\'o}mez Galera}, {Gonz{\'a}lez-Peinado}, {Gr{\"o}zinger},
  {Grohnert}, {Gu{\`a}rdia}, {Guijarro}, {de Guindos}, {Guti{\'e}rrez-Soto},
  {Hagen}, {Hauschildt}, {Hedrosa}, {Helmling}, {Hermelo}, {Hern{\'a}ndez
  Arab{\'\i}}, {Hern{\'a}ndez Casta{\~n}o}, {Hern{\'a}ndez Hernando},
  {Herrero}, {Huber}, {Huke}, {Johnson}, {de Juan}, {Kim}, {Klein},
  {Kl{\"u}ter}, {Klutsch}, {Lafarga}, {Lamp{\'o}n}, {Lara}, {Laun}, {Lemke},
  {Lenzen}, {L{\'o}pez del Fresno}, {L{\'o}pez-Gonz{\'a}lez},
  {L{\'o}pez-Puertas}, {L{\'o}pez Salas}, {Luque}, {Mag{\'a}n Madinabeitia},
  {Mall}, {Mandel}, {Marfil}, {Mar{\'\i}n Molina}, {Maroto Fern{\'a}ndez},
  {Mart{\'\i}n}, {Mart{\'\i}n-Ruiz}, {Marvin}, {Mirabet}, {Moya},
  {Moreno-Raya}, {Nagel}, {Naranjo}, {Nortmann}, {Ofir}, {Oreiro}, {Pall{\'e}},
  {Panduro}, {Pascual}, {Passegger}, {Pedraz}, {P{\'e}rez-Calpena}, {P{\'e}rez
  Medialdea}, {Perger}, {Perryman}, {Pluto}, {Rabaza}, {Ram{\'o}n}, {Rebolo},
  {Redondo}, {Reinhardt}, {Rhode}, {Rix}, {Rodler}, {Rodr{\'\i}guez},
  {Rodr{\'\i}guez Trinidad}, {Rohloff}, {Rosich}, {Sadegi},
  {S{\'a}nchez-Blanco}, {S{\'a}nchez Carrasco}, {S{\'a}nchez-L{\'o}pez},
  {Sanz-Forcada}, {Sarkis}, {Sarmiento}, {Sch{\"a}fer}, {Schiller},
  {Sch{\"o}fer}, {Schweitzer}, {Solano}, {Stahl}, {Strachan}, {Su{\'a}rez},
  {Tabernero}, {Tala}, {Tulloch}, {Veredas}, {Vico Linares}, {Vilardell},
  {Wagner}, {Winkler}, {Wolthoff}, {Xu}, {Yan}, \& {Zapatero
  Osorio}}]{Trifonov2018A&A...609A.117T}
{Trifonov}, T., {K{\"u}rster}, M., {Zechmeister}, M., {et~al.} 2018,
  \href{https://ui.adsabs.harvard.edu/abs/2018A&A...609A.117T}{\aap, 609, A117}

\bibitem[{{Trifonov} {et~al.}(2020{\natexlab{a}}){Trifonov}, {Lee},
  {K{\"u}rster}, {Henning}, {Grishin}, {Stock}, {Tjoa}, {Caballero}, {Wong},
  {Bauer}, {Quirrenbach}, {Zechmeister}, {Ribas}, {Reffert}, {Reiners},
  {Amado}, {Kossakowski}, {Azzaro}, {B{\'e}jar}, {Cort{\'e}s-Contreras},
  {Dreizler}, {Hatzes}, {Jeffers}, {Kaminski}, {Lafarga}, {Montes}, {Morales},
  {Pavlov}, {Rodr{\'\i}guez-L{\'o}pez}, {Schmitt}, {Solano}, \&
  {Barnes}}]{Trifonov2020}
{Trifonov}, T., {Lee}, M.~H., {K{\"u}rster}, M., {et~al.} 2020{\natexlab{a}},
  \href{https://ui.adsabs.harvard.edu/abs/2020A&A...638A..16T}{\aap, 638, A16}

\bibitem[{{Trifonov} {et~al.}(2020{\natexlab{b}}){Trifonov}, {Tal-Or},
  {Zechmeister}, {Kaminski}, {Zucker}, \&
  {Mazeh}}]{Trifonov2020A&A...636A..74T}
{Trifonov}, T., {Tal-Or}, L., {Zechmeister}, M., {et~al.} 2020{\natexlab{b}},
  \href{https://ui.adsabs.harvard.edu/abs/2020A&A...636A..74T}{\aap, 636, A74}

\bibitem[{{Tuomi} {et~al.}(2018){Tuomi}, {Jones}, {Barnes},
  {Anglada-Escud{\'e}}, {Butler}, {Kiraga}, \& {Vogt}}]{Tuomi2018}
{Tuomi}, M., {Jones}, H. R.~A., {Barnes}, J.~R., {et~al.} 2018,
  \href{https://ui.adsabs.harvard.edu/abs/2018AJ....155..192T}{\aj, 155, 192}

\bibitem[{{Tuomi} {et~al.}(2014){Tuomi}, {Jones}, {Barnes},
  {Anglada-Escud{\'e}}, \& {Jenkins}}]{2014MNRAS.441.1545T}
{Tuomi}, M., {Jones}, H. R.~A., {Barnes}, J.~R., {Anglada-Escud{\'e}}, G., \&
  {Jenkins}, J.~S. 2014,
  \href{https://ui.adsabs.harvard.edu/abs/2014MNRAS.441.1545T}{\mnras, 441,
  1545}

\bibitem[{{Tuomi} {et~al.}(2019){Tuomi}, {Jones}, {Butler}, {Arriagada},
  {Vogt}, {Burt}, {Laughlin}, {Holden}, {Shectman}, {Crane}, {Thompson},
  {Keiser}, {Jenkins}, {Berdi{\~n}as}, {Diaz}, {Kiraga}, \&
  {Barnes}}]{Tuomi2019arXiv190604644T}
{Tuomi}, M., {Jones}, H.~R.~A., {Butler}, R.~P., {et~al.} 2019,
  \href{https://ui.adsabs.harvard.edu/abs/2019arXiv190604644T}{arXiv e-prints,
  arXiv:1906.04644}

\bibitem[{{Turtelboom} {et~al.}(2022){Turtelboom}, {Weiss}, {Dressing},
  {Nowak}, {Pall{\'e}}, {Beard}, {Blunt}, {Brinkman}, {Chontos}, {Claytor},
  {Dai}, {Dalba}, {Giacalone}, {Gonzales}, {Harada}, {Hill}, {Holcomb},
  {Korth}, {Lubin}, {Masseron}, {MacDougall}, {Mayo}, {Mo{\v{c}}nik}, {Akana
  Murphy}, {Polanski}, {Rice}, {Rubenzahl}, {Scarsdale}, {Stassun}, {Tyler},
  {Zandt}, {Crossfield}, {Deeg}, {Fulton}, {Gandolfi}, {Howard}, {Huber},
  {Isaacson}, {Kane}, {Lam}, {Luque}, {Mart{\'\i}n}, {Morello}, {Orell-Miquel},
  {Petigura}, {Robertson}, {Roy}, {Van Eylen}, {Baker}, {Belinski}, {Bieryla},
  {Ciardi}, {Collins}, {Cutting}, {Della-Rose}, {Ellingsen}, {Furlan}, {Gan},
  {Gnilka}, {Guerra}, {Howell}, {Jimenez}, {Latham}, {Larivi{\`e}re}, {Lester},
  {Lillo-Box}, {Luker}, {Mann}, {Plavchan}, {Safonov}, {Skinner}, {Strakhov},
  {Wittrock}, {Caldwell}, {Essack}, {Jenkins}, {Quintana}, {Ricker},
  {Vanderspek}, {Seager}, \& {Winn}}]{Turtelboom2022}
{Turtelboom}, E.~V., {Weiss}, L.~M., {Dressing}, C.~D., {et~al.} 2022,
  \href{https://ui.adsabs.harvard.edu/abs/2022AJ....163..293T}{\aj, 163, 293}

\bibitem[{{Wang} {et~al.}(2022){Wang}, {Rice}, {Wang}, {Pu}, {Stef{\'a}nsson},
  {Mahadevan}, {Radzom}, {Giacalone}, {Wu}, {Esposito}, {Dalba}, {Avsar},
  {Holden}, {Skiff}, {Polakis}, {Voeller}, {Logsdon}, {Klusmeyer}, {Schweiker},
  {Wu}, {Beard}, {Dai}, {Lubin}, {Weiss}, {Bender}, {Blake}, {Dressing},
  {Halverson}, {Hearty}, {Howard}, {Huber}, {Isaacson}, {Jackman}, {Llama},
  {McElwain}, {Rajagopal}, {Roy}, {Robertson}, {Schwab}, {Shkolnik}, {Wright},
  \& {Laughlin}}]{Wang2022}
{Wang}, X.-Y., {Rice}, M., {Wang}, S., {et~al.} 2022,
  \href{https://ui.adsabs.harvard.edu/abs/2022ApJ...926L...8W}{\apjl, 926, L8}

\bibitem[{{Wright} {et~al.}(2016){Wright}, {Wittenmyer}, {Tinney}, {Bentley},
  \& {Zhao}}]{Wright2016ApJ...817L..20W}
{Wright}, D.~J., {Wittenmyer}, R.~A., {Tinney}, C.~G., {Bentley}, J.~S., \&
  {Zhao}, J. 2016,
  \href{https://ui.adsabs.harvard.edu/abs/2016ApJ...817L..20W}{\apjl, 817, L20}

\bibitem[{{Wright} \& {Eastman}(2014)}]{Wright2014PASP..126..838W}
{Wright}, J.~T. \& {Eastman}, J.~D. 2014,
  \href{https://ui.adsabs.harvard.edu/abs/2014PASP..126..838W}{\pasp, 126, 838}

\bibitem[{{Wright} {et~al.}(2004){Wright}, {Marcy}, {Butler}, \&
  {Vogt}}]{Wright2004}
{Wright}, J.~T., {Marcy}, G.~W., {Butler}, R.~P., \& {Vogt}, S.~S. 2004,
  \href{https://ui.adsabs.harvard.edu/abs/2004ApJS..152..261W}{\apjs, 152, 261}

\bibitem[{{Yan} {et~al.}(2019){Yan}, {Casasayas-Barris}, {Molaverdikhani},
  {Alonso-Floriano}, {Reiners}, {Pall{\'e}}, {Henning}, {Molli{\`e}re}, {Chen},
  {Nortmann}, {Snellen}, {Ribas}, {Quirrenbach}, {Caballero}, {Amado},
  {Azzaro}, {Bauer}, {Cort{\'e}s Contreras}, {Czesla}, {Khalafinejad}, {Lara},
  {L{\'o}pez-Puertas}, {Montes}, {Nagel}, {Oshagh}, {S{\'a}nchez-L{\'o}pez},
  {Stangret}, \& {Zechmeister}}]{Yan2019}
{Yan}, F., {Casasayas-Barris}, N., {Molaverdikhani}, K., {et~al.} 2019,
  \href{https://ui.adsabs.harvard.edu/abs/2019A&A...632A..69Y}{\aap, 632, A69}

\bibitem[{{Yan} {et~al.}(2021){Yan}, {Wyttenbach}, {Casasayas-Barris},
  {Reiners}, {Pall{\'e}}, {Henning}, {Molli{\`e}re}, {Czesla}, {Nortmann},
  {Molaverdikhani}, {Chen}, {Snellen}, {Zechmeister}, {Huang}, {Ribas},
  {Quirrenbach}, {Caballero}, {Amado}, {Cont}, {Khalafinejad}, {Khaimova},
  {L{\'o}pez-Puertas}, {Montes}, {Nagel}, {Oshagh}, {Pedraz}, \&
  {Stangret}}]{Yan2021}
{Yan}, F., {Wyttenbach}, A., {Casasayas-Barris}, N., {et~al.} 2021,
  \href{https://ui.adsabs.harvard.edu/abs/2021A&A...645A..22Y}{\aap, 645, A22}

\bibitem[{{Zechmeister} {et~al.}(2014){Zechmeister}, {Anglada-Escud{\'e}}, \&
  {Reiners}}]{Zechmeister2014A&A...561A..59Z}
{Zechmeister}, M., {Anglada-Escud{\'e}}, G., \& {Reiners}, A. 2014,
  \href{https://ui.adsabs.harvard.edu/abs/2014A&A...561A..59Z}{\aap, 561, A59}

\bibitem[{{Zechmeister} {et~al.}(2019){Zechmeister}, {Dreizler}, {Ribas},
  {Reiners}, {Caballero}, {Bauer}, {B{\'e}jar}, {Gonz{\'a}lez-Cuesta},
  {Herrero}, {Lalitha}, {L{\'o}pez-Gonz{\'a}lez}, {Luque}, {Morales},
  {Pall{\'e}}, {Rodr{\'\i}guez}, {Rodr{\'\i}guez L{\'o}pez}, {Tal-Or},
  {Anglada-Escud{\'e}}, {Quirrenbach}, {Amado}, {Abril}, {Aceituno},
  {Aceituno}, {Alonso-Floriano}, {Ammler-von Eiff}, {Antona Jim{\'e}nez},
  {Anwand-Heerwart}, {Arroyo-Torres}, {Azzaro}, {Baroch}, {Barrado},
  {Becerril}, {Ben{\'\i}tez}, {Berdi{\~n}as}, {Bergond}, {Bluhm},
  {Brinkm{\"o}ller}, {del Burgo}, {Calvo Ortega}, {Cano}, {Cardona
  Guill{\'e}n}, {Carro}, {C{\'a}rdenas V{\'a}zquez}, {Casal},
  {Casasayas-Barris}, {Casanova}, {Chaturvedi}, {Cifuentes}, {Claret},
  {Colom{\'e}}, {Cort{\'e}s-Contreras}, {Czesla}, {D{\'\i}ez-Alonso}, {Dorda},
  {Fern{\'a}ndez}, {Fern{\'a}ndez-Mart{\'\i}n}, {Fuhrmeister}, {Fukui},
  {Galad{\'\i}-Enr{\'\i}quez}, {Gallardo Cava}, {Garcia de la Fuente},
  {Garcia-Piquer}, {Garc{\'\i}a Vargas}, {Gesa}, {G{\'o}ngora Rueda},
  {Gonz{\'a}lez-{\'A}lvarez}, {Gonz{\'a}lez Hern{\'a}ndez},
  {Gonz{\'a}lez-Peinado}, {Gr{\"o}zinger}, {Gu{\`a}rdia}, {Guijarro}, {de
  Guindos}, {Hatzes}, {Hauschildt}, {Hedrosa}, {Helmling}, {Henning},
  {Hermelo}, {Hern{\'a}ndez Arabi}, {Hern{\'a}ndez Casta{\~n}o}, {Hern{\'a}ndez
  Otero}, {Hintz}, {Huke}, {Huber}, {Jeffers}, {Johnson}, {de Juan},
  {Kaminski}, {Kemmer}, {Kim}, {Klahr}, {Klein}, {Kl{\"u}ter}, {Klutsch},
  {Kossakowski}, {K{\"u}rster}, {Labarga}, {Lafarga}, {Llamas}, {Lamp{\'o}n},
  {Lara}, {Launhardt}, {L{\'a}zaro}, {Lodieu}, {L{\'o}pez del Fresno},
  {L{\'o}pez-Puertas}, {L{\'o}pez Salas}, {L{\'o}pez-Santiago}, {Mag{\'a}n
  Madinabeitia}, {Mall}, {Mancini}, {Mandel}, {Marfil}, {Mar{\'\i}n Molina},
  {Maroto Fern{\'a}ndez}, {Mart{\'\i}n}, {Mart{\'\i}n-Fern{\'a}ndez},
  {Mart{\'\i}n-Ruiz}, {Marvin}, {Mirabet}, {Monta{\~n}{\'e}s-Rodr{\'\i}guez},
  {Montes}, {Moreno-Raya}, {Nagel}, {Naranjo}, {Narita}, {Nortmann}, {Nowak},
  {Ofir}, {Oshagh}, {Panduro}, {Parviainen}, {Pascual}, {Passegger}, {Pavlov},
  {Pedraz}, {P{\'e}rez-Calpena}, {P{\'e}rez Medialdea}, {Perger}, {Perryman},
  {Rabaza}, {Ram{\'o}n Ballesta}, {Rebolo}, {Redondo}, {Reffert}, {Reinhardt},
  {Rhode}, {Rix}, {Rodler}, {Rodr{\'\i}guez Trinidad}, {Rosich}, {Sadegi},
  {S{\'a}nchez-Blanco}, {S{\'a}nchez Carrasco}, {S{\'a}nchez-L{\'o}pez},
  {Sanz-Forcada}, {Sarkis}, {Sarmiento}, {Sch{\"a}fer}, {Schmitt},
  {Sch{\"o}fer}, {Schweitzer}, {Seifert}, {Shulyak}, {Solano}, {Sota}, {Stahl},
  {Stock}, {Strachan}, {Stuber}, {St{\"u}rmer}, {Su{\'a}rez}, {Tabernero},
  {Tala Pinto}, {Trifonov}, {Veredas}, {Vico Linares}, {Vilardell}, {Wagner},
  {Wolthoff}, {Xu}, {Yan}, \& {Zapatero Osorio}}]{Zechmeister2019}
{Zechmeister}, M., {Dreizler}, S., {Ribas}, I., {et~al.} 2019,
  \href{https://ui.adsabs.harvard.edu/abs/2019A&A...627A..49Z}{\aap, 627, A49}

\bibitem[{{Zechmeister} \& {K{\"u}rster}(2009)}]{GLS2009}
{Zechmeister}, M. \& {K{\"u}rster}, M. 2009,
  \href{https://ui.adsabs.harvard.edu/abs/2009A&A...496..577Z}{\aap, 496, 577}

\bibitem[{{Zechmeister} {et~al.}(2009){Zechmeister}, {K{\"u}rster}, \&
  {Endl}}]{Zechmeister2009}
{Zechmeister}, M., {K{\"u}rster}, M., \& {Endl}, M. 2009,
  \href{https://ui.adsabs.harvard.edu/abs/2009A&A...505..859Z}{\aap, 505, 859}

\bibitem[{{Zechmeister} {et~al.}(2018){Zechmeister}, {Reiners}, {Amado},
  {Azzaro}, {Bauer}, {B{\'e}jar}, {Caballero}, {Guenther}, {Hagen}, {Jeffers},
  {Kaminski}, {K{\"u}rster}, {Launhardt}, {Montes}, {Morales}, {Quirrenbach},
  {Reffert}, {Ribas}, {Seifert}, {Tal-Or}, \& {Wolthoff}}]{Zechmeister2018}
{Zechmeister}, M., {Reiners}, A., {Amado}, P.~J., {et~al.} 2018,
  \href{http://adsabs.harvard.edu/abs/2018A%26A...609A..12Z}{\aap, 609, A12}

\end{thebibliography}

\begin{appendix}
\section{Additional tables}

\onecolumn

{
\small
\begin{landscape}
\begin{longtable}{@{}llcclcccccl@{}}
    \caption{\label{tab:planets} Exoplanets in the CARMENES survey target sample with publications using
    CARMENES data.}\\
        \hline\hline
        \noalign{\smallskip}
Karmn & Star name & $N_{\rm CAR}$ & $N_{\rm other}$ & $M_{\star}$ (M$_{\odot}$) & Pl. & $K$ (m\,s$^{-1})$ &
$M_{\rm pl} \sin i$ (M$_{\oplus}$) & $P_{\rm pl}$ (d) & Type\tablefootmark{a} & Ref. \\
        \noalign{\smallskip}
\hline
        \noalign{\smallskip}
\endfirsthead
\caption{Continued.}\\
         \hline\hline
        \noalign{\smallskip}
Karmn & Star name & $N_{\rm CAR}$ & $N_{\rm other}$ & $M_{\star}$ (M$_{\odot}$) & Pl. & $K$ (m\,s$^{-1})$ &
$M_{\rm pl} \sin i$ (M$_{\oplus}$) & $P_{\rm pl}$ (d) & Type\tablefootmark{a} & Ref. \\
        \noalign{\smallskip}
       \hline
        \noalign{\smallskip}
       \endhead
       \hline
       \endfoot
       J00067$-$075 & GJ 1002                &  86 &  53 & $0.120$         & b & $1.31\pm0.14$           & $1.08\pm0.13$          & $10.3465\pm0.0025$                & d & SM22 \\
             &                        &     &     &                 & c & $1.30\pm0.14$           & $1.36\pm0.17$          & $21.202\pm0.013$                  & d & SM22 \\
\noalign{\smallskip}
J00183+440   & GJ 15 A                & 208 & 297 & $0.391\pm0.016$ & b & $1.79\pm0.24$           & $3.98^{+0.38}_{-0.29}$ & $11.436\pm0.001$                  & r & Tri22 \\
             &                        &     &     &                 & c & $1.62^{+0.57}_{-0.36}$  & $50.4^{+6.9}_{-6.8}$   & $6694^{+164}_{-285}$              & d & Tri22 \\
\noalign{\smallskip}
J00403+612   & 2MJ00402129+6112490    &  44 &   0 & $0.471\pm0.011$ & b & $5.83^{+0.86}_{-0.91}$  & $7.52^{+1.12}_{-1.18}$ & $2.527113\pm0.000009$             & f & GA22b \\
             &                        &     &     &                 & c & $3.19^{+1.16}_{-1.07}$  & $7.94^{+2.88}_{-2.64}$ & $18.0881\pm0.0002$                & f & GA22b \\
\noalign{\smallskip}
J01026+623   & GJ 49                  &  80 & 158 & $0.515\pm0.019$ & b & $2.52^{+0.31}_{-0.30}$  & $5.63^{+0.67}_{-0.68}$ & $13.8508^{+0.0053}_{-0.0051}$     & d & Per19 \\
\noalign{\smallskip}
J01066+192   & LSPM J0106+1913        &  65 &  32 & $0.344\pm0.005$ & b & $3.40^{+0.25}_{-0.24}$  & $3.21\pm0.24$          & $1.8805136\pm0.0000025$           & f & Cha22 \\  
             &                        &     &     &                 & c & $3.48^{+0.34}_{-0.35}$  & $6.64^{+0.67}_{-0.68}$ & $15.532482\pm0.000034$            & f & Cha22 \\    
\noalign{\smallskip}
J01125$-$169 & YZ Cet                 & 108 & 332 & $0.142\pm0.010$ & b & $1.31^{+0.15}_{-0.14}$  & $0.70^{+0.09}_{-0.08}$ & $2.02087^{+0.00007}_{-0.00009}$   & r & Sto20a \\ 
             &                        &     &     &                 & c & $1.84^{+0.14}_{-0.15}$  & $1.14^{+0.11}_{-0.10}$ & $3.0599\pm0.0001$                 & r & Sto20a \\ 
             &                        &     &     &                 & d & $1.54^{+0.14}_{-0.15}$  & $1.09\pm0.12$          & $4.65626^{+0.00028}_{-0.00029}$   & r & Sto20a \\ 
\noalign{\smallskip}
J02002+130   & TZ Ari                 &  93 &  79 & $0.150\pm0.010$ & b & $18.8^{+1.3}_{-1.2}$    & $67\pm6$               & $771.4^{+1.3}_{-1.2}$             & d & Qui22 \\
\noalign{\smallskip}
J02489$-$145W& PMJ02489$-$1432W       &  33 &   0 & $0.512\pm0.020$ & b & $4.65^{+0.60}_{-0.64}$  & $6.28^{+0.84}_{-0.88}$ & $2.491986\pm0.000003$             & f & Kos21 \\
\noalign{\smallskip}
J02530+168   & Teegarden's Star       & 238 &   0 & $0.089\pm0.009$ & b & $2.02^{+0.19}_{-0.20}$  & $1.05^{+0.13}_{-0.12}$ & $4.9100\pm0.0014$                 & d & Zec19 \\ 
             &                        &     &     &                 & c & $1.61\pm0.19$           & $1.11^{+0.16}_{-0.15}$ & $11.409\pm0.009$                  & d & Zec19 \\ 
\noalign{\smallskip}
J02573+765   & LHS 1478               &  57 &  13 & $0.236\pm0.012$ & b & $3.13^{+0.23}_{-0.25}$  & $2.33\pm0.20$          & $1.949538\pm0.000004$             & f & Sot21 \\
\noalign{\smallskip}
J03133+047   & CD Cet                 & 106 &   0 & $0.161\pm0.010$ & b & $6.51^{+0.22}_{-0.23}$  & $3.95^{+0.42}_{-0.43}$ & $2.29070\pm0.00012$               & d & Bau20 \\ 
\noalign{\smallskip}
J04167$-$120 & LP714-47               &  34 &  54 & $0.59\pm0.02$   & b & $17.6\pm0.8$            & $30.8\pm1.5$           & $4.052037\pm0.000004$             & f & Dre20 \\
\noalign{\smallskip}
J04343+430   & 2MJ04342248+4302148    &  55 &   0 & $0.495\pm0.019$ & b & $4.41\pm0.73$           & $3.76\pm0.63$          & $0.669140^{+0.000002}_{-0.000002}$& f & Blu21 \\
             &                        &     &     &                 & c & $4.53^{+1.01}_{-1.02}$  & $9.2\pm2.1$            & $9.025^{+0.104}_{-0.119}$         & f & Blu21 \\
\noalign{\smallskip}
J04429+189   & GJ 176                 &  23 & 181 & $0.504\pm0.013$ & b & $4.49^{+1.00}_{-0.23}$  & $9.1^{+1.5}_{-0.7}$    & $8.776^{+0.001}_{-0.002}$         & r & Tri18 \\ 
\noalign{\smallskip}
J06371+175   & HD 260655              &  88 &  92 & $0.439\pm0.011$ & b & $1.69\pm0.27$           & $2.14\pm0.34$          & $2.76953\pm0.00003$               & f & Luq22 \\ 
             &                        &     &     &                 & c & $1.92\pm0.30$           & $3.09\pm0.48$          & $5.70588\pm0.00007$               & f & Luq22 \\
\noalign{\smallskip}
J06548+332   & GJ 251                 & 280 &  75 & $0.360\pm0.015$ & b & $2.24\pm0.18$           & $4.27^{+0.37}_{-0.36}$ & $14.2383\pm0.0018$                & d & Sto20b,TW \\ 
\noalign{\smallskip}
J08023+033   & GJ 3473                &  64 &  88 & $0.360\pm0.016$ & b & $2.21\pm0.35$           & $1.86\pm0.30$          & $1.198003\pm0.000002$             & f & Kem20 \\
             &                        &     &     &                 & c & $3.75^{+0.45}_{-0.42}$  & $7.41^{+0.91}_{-0.86}$ & $15.509\pm0.033$                  & f & Kem20 \\
\noalign{\smallskip}
J08413+594   & GJ 3512                & 182 &   0 & $0.123\pm0.009$ & b & $71.23^{+0.32}_{-0.33}$ & $146\pm7$              & $203.13\pm0.05$                   & d & Mor19,TW \\ 
             &                        &     &     &                 & c & $27.9\pm0.4$            & $143\pm7$              & $2350^{+100}_{-80}$               & d & Mor19,TW \\ 
\noalign{\smallskip}
J09144+526   & GJ 338 B               & 159 &  30 & $0.64\pm0.07$   & b & $3.07\pm0.37$           & $10.27^{+1.47}_{-1.38}$& $24.45\pm0.02$                    & d & GA20 \\
\noalign{\smallskip}
J09360$-$216 & GJ 357                 &  10 & 128 & $0.342\pm0.011$ & b & $1.52\pm0.25$           & $1.84\pm0.31$          & $3.93072^{+0.00008}_{-0.00006}$   & f & Luq19 \\ 
             &                        &     &     &                 & c & $2.13\pm0.28$           & $3.40\pm0.46$          & $9.1247^{+0.0011}_{-0.0010}$      & d & Luq19 \\ 
             &                        &     &     &                 & d & $2.09^{+0.34}_{-0.35}$  & $6.1\pm1.0$            & $55.661\pm0.055$                  & d & Luq19 \\ 
\noalign{\smallskip}
J10088+692   & TYC4384-1735-1         &  40 &  21 & $0.630\pm0.024$ & b & $3.40^{+0.35}_{-0.34}$  & $5.90^{+0.62}_{-0.61}$ & $3.444717^{+0.000040}_{-0.000042}$& f & Blu20 \\ 
\noalign{\smallskip}
J10185$-$117 & LTT 3780               &  52 &   4 & $0.379\pm0.016$ & b & $3.13^{+0.31}_{-0.3}$   & $2.34^{+0.24}_{-0.23}$ & $0.768377\pm0.000001$             & f & Now20 \\
             &                        &     &     &                 & c & $3.36^{+0.32}_{-0.31}$  & $6.29^{+0.63}_{-0.61}$ & $12.25213^{+0.00007}_{-0.00006}$  & f & Now20 \\
\noalign{\smallskip}
J10289+008   & GJ 393                 &  84 & 250 & $0.426\pm0.017$ & b & $1.01\pm0.14$           & $1.71\pm0.24$          & $7.02679^{+0.00082}_{-0.00085}$   & d & Ama21 \\
\noalign{\smallskip}
J11033+359   & Lalande 21185          & 321 & 416 & $0.390\pm0.011$ & b & $1.39\pm0.14$           & $2.69\pm0.25$          & $12.946\pm0.005$                  & r & Sto20b \\ 
\noalign{\smallskip}
J11302+076   & K2-18                  &  58 &   0 & $0.359\pm0.047$ & b & $3.38^{+0.76}_{-0.75}$  & $8.49^{+1.97}_{-2.08}$ & $32.9396\pm0.00010$               & f & Sar18 \\ 
\noalign{\smallskip}
J11417+427   & GJ 1148                &  76 & 125 & $0.354\pm0.015$ & b & $38.54^{+0.43}_{-0.37}$ & $96.6^{+1.0}_{-1.3}$   & $41.380^{+0.001}_{-0.002}$        & r & Tri20 \\
             &                        &     &     &                 & c & $12.26^{+0.59}_{-0.56}$ & $72.1^{+3.2}_{-7.0}$   & $532.6^{+1.0}_{-1.1}$             & d & Tri20 \\
\noalign{\smallskip}
J11421+267   & GJ 436                 & 113 & 525 & $0.436\pm0.012$ & b & $17.38\pm0.17$          & $21.36^{+0.20}_{-0.21}$& $2.644^{+0.001}_{-0.001}$         & r & Tri18 \\ 
\noalign{\smallskip}
J11509+483   & GJ 1151                &  97 &  71 & $0.164\pm0.009$ & b & $3.10^{+0.38}_{-0.43}$  & $10.6^{+1.3}_{-1.5}$   & $389.7^{+5.4}_{-6.5}$             & d & Bla22 \\
\noalign{\smallskip}
J12123+544S  & GJ 458 A               & 108 &   0 & $0.578\pm0.021$ & b & $2.85^{+0.38}_{-0.39}$  & $6.89^{+0.92}_{-0.95}$ & $13.671^{+0.011}_{-0.010}$        & d & Sto20b \\
\noalign{\smallskip}
J12479+097   & GJ 486                 &  76 &  65 & $0.323\pm0.015$ & b &$3.370^{+0.078}_{-0.080}$& $2.82^{+0.11}_{-0.12}$ & $1.467119^{+0.000031}_{-0.000030}$& d & Tri21 \\ 
\noalign{\smallskip}
J13229+244   & GJ 3779                & 104 &   0 & $0.27\pm0.02$   & b & $8.62\pm0.39$           & $8.0\pm0.5$            & $3.0232\pm0.0004$                 & d & Luq18 \\ 
\noalign{\smallskip}
J13299+102   & GJ 514                 & 274 & 266 & $0.510\pm0.051$ & b & $1.15^{+0.21}_{-0.19}$  & $5.2\pm0.9$            & $140.43\pm0.41$                   & d & Dam22 \\
\noalign{\smallskip}
J13255+688   & 2MJ13253177+6850106    &  55 &   0 & $0.59\pm0.02$   & b & $3.74^{+1.03}_{-0.99}$  & $3.75^{+1.14}_{-1.06}$ & $0.764597^{+0.000013}_{-0.000011}$& f & GA22a \\ 
             &                        &     &     &                 & c & $5.10^{+1.02}_{-1.06}$  & $8.32^{+1.90}_{-1.88}$ & $3.294736^{+0.000034}_{-0.000036}$& f & GA22a \\ 
\noalign{\smallskip}
J14010$-$026 & GJ 536                 &  28 & 277 & $0.530\pm0.011$ & b & $3.12^{+0.36}_{-0.19}$  & $6.52^{+0.69}_{-0.40}$ & $8.708^{+0.002}_{-0.001}$         & r & Tri18 \\ 
\noalign{\smallskip}
J14342$-$125 & GJ 555                 &  94 &  48 & $0.291\pm0.014$ & b & $5.46\pm0.75$           & $2.59\pm0.34$          & $36.116^{+0.027}_{-0.029}$        & d & GA22c \\
\noalign{\smallskip}
J15194$-$077 & GJ 581                 &  20 & 664 & $0.323\pm0.013$ & b & $12.35^{+0.18}_{-0.20}$ & $15.20^{+0.22}_{-0.27}$& $5.368^{+0.001}_{-0.001}$         & r & Tri18 \\ 
             &                        &     &     &                 & c & $3.28^{+0.22}_{-0.12}$  & $5.65^{+0.39}_{-0.24}$ & $12.919^{+0.003}_{-0.002}$        & r & Tri18 \\ 
             &                        &     &     &                 & e & $1.55^{+0.22}_{-0.13}$  & $1.66^{+0.24}_{-0.16}$ & $3.153^{+0.001}_{-0.006}$         & r & Tri18 \\ 
\noalign{\smallskip}
J15583+354   & GJ 3929                &  73 &   0 & $0.309\pm0.014$ & b & $1.29^{+0.47}_{-0.46}$  & $1.27\pm0.46$          & $2.616267^{+0.000005}_{-0.000005}$& f & Kem22 \\ 
\noalign{\smallskip}
J16167+672S  & HD 147379              & 186 &  30 & $0.58\pm0.08$   & b & $4.87^{+0.41}_{-0.37}$  & $23.0^{+2.8}_{-2.9}$   & $86.57\pm0.06$                    & d & Rei18a,TW \\
\noalign{\smallskip}
J17378+185   & GJ 686                 & 100 & 198 & $0.426\pm0.033$ & b & $3.02^{+0.18}_{-0.20}$  & $6.64^{+0.53}_{-0.54}$ & $15.5314^{+0.0015}_{-0.0014}$     & r & Lal19 \\ 
\noalign{\smallskip}
J17578+046   & Barnard's Star         & 196 & 575 & $0.163\pm0.022$ & b & $1.20\pm0.12$           & $3.23\pm0.44$          & $232.80^{+0.38}_{-0.41}$          & d & Rib18 \\ 
\noalign{\smallskip}
J18580+059   & GJ 740                 &  32 & 129 & $0.58\pm0.06$   & b & $2.13^{+0.34}_{-0.32}$  & $2.96^{+0.50}_{-0.48}$ & $2.37756^{+0.00013}_{-0.00011}$   & d & TP21 \\
\noalign{\smallskip}
J19169+051N  & HD 180617              & 123 & 296 & $0.45\pm0.04$   & b & $2.85^{+0.16}_{-0.25}$  & $12.2^{+1.0}_{-1.4}$   & $105.90^{+0.09}_{-0.10}$          & d & Kam18 \\
\noalign{\smallskip}
J20260+585   & Wolf 1069              & 268 &   0 & $0.167\pm0.011$ & b & $1.07\pm0.17$           & $1.26\pm0.21$          & $15.564\pm0.015$                  & d & Kos23 \\
\noalign{\smallskip}
J20450+444   & GJ 806                 &  67 & 160 & $0.413\pm0.011$ & b & $2.25\pm0.20$           & $1.90\pm0.17$          & $0.9263237\pm0.0000009$           & f & Pal22 \\
             &                        &     &     &                 & c & $3.55\pm0.17$           & $5.80\pm0.30$          & $6.64064\pm0.00025$               & f & Pal22 \\
\noalign{\smallskip}
J20451$-$313 & AU Mic                 & 100 & 310 & $0.50\pm0.03$   & b & $10.23^{+0.88}_{-0.91}$ & $20.12^{+1.72}_{-1.57}$& $8.463000\pm0.000002$             & f & Cal21 \\
             &                        &     &     &                 & c & $<7.7$                  & $<20.1$                & $18.859019\pm0.000016$            & f & Cal21 \\
\noalign{\smallskip}
J21164+025   & LSPM J2116+0234        &  72 &   0 & $0.430\pm0.031$ & b & $6.31^{+0.44}_{-0.43}$  & $13.4\pm1.1$           & $14.4433^{+0.0079}_{-0.0086}$     & d & Lal19 \\ 
\noalign{\smallskip}
J21221+229   & TYC 2187-512-1         &  94 &   0 & $0.498\pm0.019$ & b & $12.02^{+0.47}_{-0.46}$ & $105\pm6$              & $691.9^{+8.8}_{-8.6}$             & d & Qui22 \\
\noalign{\smallskip}
J21466+668   & G264-012               & 159 &   0 & $0.297\pm0.024$ & b & $2.72^{+0.28}_{-0.29}$  & $2.50^{+0.29}_{-0.30}$ & $2.30538\pm0.00031$               & d & Ama21 \\ 
             &                        &     &     &                 & c & $2.69^{+0.31}_{-0.30}$  & $3.75^{+0.48}_{-0.47}$ & $8.0518\pm0.0034$                 & d & Ama21 \\ 
\noalign{\smallskip}
J21474+627   & TYC 4266-00736-1       &  57 &   0 & $0.606\pm0.020$ & b & $3.64^{+0.50}_{-0.51}$  & $10.8\pm1.5$           & $18.85019\pm0.00013$              & f & Esp22 \\ 
\noalign{\smallskip}
J22137$-$176 & GJ 1265                &  87 &  11 & $0.178\pm0.018$ & b & $9.82^{+0.51}_{-0.52}$  & $7.4\pm0.5$            & $3.6511\pm0.0001$                 & d & Luq18 \\
\noalign{\smallskip}
J22252+594   & GJ 4276                & 100 &   0 & $0.406\pm0.030$ & b & $8.79\pm0.27$           & $16.57^{+0.94}_{-0.95}$& $13.352\pm0.003$                  & d & Nag19 \\
\noalign{\smallskip}
J22532$-$142 & GJ 876                 &  28 & 594 & $0.350\pm0.013$ & b & $212.07^{+0.27}_{-0.26}$& $760.9^{+1.0}_{-1.0}$  & $61.082^{+0.006}_{-0.010}$        & r & Tri18 \\ 
             &                        &     &     &                 & c & $88.34^{+0.23}_{-0.25}$ & $241.5^{+0.7}_{-0.6}$  & $30.126^{+0.011}_{-0.003}$        & r & Tri18 \\ 
             &                        &     &     &                 & d & $6.14^{+0.23}_{-0.22}$  & $6.91^{+0.22}_{-0.27}$ & $1.938^{+0.001}_{-0.001}$         & r & Tri18 \\ 
             &                        &     &     &                 & e & $3.39^{+0.29}_{-0.28}$  & $15.4\pm1.3$           & $124.4^{+0.3}_{-0.7}$             & r & Tri18 \\ 

\end{longtable}
    \tablefoot{$N_{\rm CAR}$ is the number of CARMENES measurements used in the analysis to determine the
    listed parameters (i.e. it may not coincide with the number of measurements released as part of DR1).
    $N_{\rm other}$ is the number of measurements from other spectrographs used in the analysis. The stellar
    mass column, $M_{\star}$, lists the value used by the quoted publication and is consistent with the
    planet's minimum mass, $M_{\rm pl} \sin i$. Small differences may exist with the values tabulated in
    Table~\ref{tab:props}.\\
    \tablefoottext{a}{d: Planet discovered by CARMENES; f: Transiting planet confirmed with CARMENES follow-up
    observations; r: Re-analysis with CARMENES data.}
    \tablebib{
    Ama21: \citet{Amado2021};
    Bau20: \citet{Bauer2020};
    Bla22: \citet{BlancoPozo2022};
    Blu20: \citet{Bluhm2020};
    Blu21: \citet{Bluhm2021};
    Cal21: \citet{Cale2021};
    Cha22: \citet{Chaturvedi2022};
    Dam22: \citet{Damasso2022};
    Dre20: \citet{Dreizler2020};
    Esp22: \citet{Espinoza22};
    GA20: \citet{Gonzalez2020};
    GA22a: \citet{Gonzalez2022a};
    GA22b: Gonz\'alez-\'Alvarez et al. (in prep.);
    GA22c: Gonz\'alez-\'Alvarez et al. (in prep.);
    Kam18: \citet{Kaminski2018};
    Kem20: \citet{Kemmer2020};
    Kem22: \citet{Kemmer2022};
    Kos21: \citet{Kossakowski2021};
    Kos23: \citet{Kossakowski2022b};
    Lal19: \citet{Lalitha2019};
    Luq18: \citet{Luque2018};
    Luq19: \citet{Luque2019};
    Luq22: \citet{Luque2022};
    Mor19: \citet{Morales2019};
    Nag19: \citet{Nagel2019};
    Now20: \citet{Nowak2020};
    Pal22: \citet{Palle2022};
    Per19: \citet{Perger2019};
    Qui22: \citet{Quirrenbach2022};
    Rei18a: \citet{Reiners2018a};
    Rib18: \citet{Ribas2018};
    Sar18: \citet{Sarkis2018AJ....155..257S};
    Sot21: \citet{Soto2021};
    Sto20a: \citet{Stock2020a};
    Sto20b: \citet{Stock2020b};
    SM22: \citet{Suarez2022};
    TP21: \citet{Toledo2021};
    Tri18: \citet{Trifonov2018A&A...609A.117T};
    Tri20: \citet{Trifonov2020};
    Tri21: \citet{Trifonov2021};
    Tri22: Trifonov et al. (in prep.);
    TW: This work;
    Zec19: \citet{Zechmeister2019}.
    }
    }
\end{landscape}
}

\twocolumn

\begin{table}[ht]
\centering
\caption{Periodic signals with FAP $<1$\,\% in the 238-target sample for occurrence rate analysis.}
\label{tab:signals}
\begin{tabular}{@{}lccl@{}}
\hline\hline
\noalign{\smallskip}
Karmn & $P$ (d) & FAP & Remark \\
\hline
\noalign{{\smallskip}}
J00051+457 & ~ & ~ & no signal \\
J00067$-$075 & ~ & ~ & no signal \\
J00162+198E & ~ & ~ & no signal \\
J00183+440$^a$ & 40.68 & 0.0912\% &  rotation\\
J00184+440 &10136.69 & $ < 10^{-6} $ &  $P$ > $\frac{2}{3}$ time baseline \\
J00286$-$066 & ~ & ~ & no signal \\
J00389+306 &20.17 & 0.0203\% &  unsolved \\
J00570+450 & ~ & ~ & no signal \\
J01013+613 & ~ & ~ & no signal \\
J01025+716 &43.53 & $ < 10^{-6} $ &  dLW \\
J01026+623$^a$ & 18.9 & 0.0359\% &   H$\alpha$ \\
J01026+623$^a$ & 9.35 & 0.0036\% &  rotation \\
J01048$-$181 & ~ & ~ & no signal \\
J01125$-$169$^a$ &3.06 & 0.0108\%  &  planet \\
J01125$-$169$^a$ &4.7 & 0.0346\%  &  planet \\
J01125$-$169$^a$ &80.77 & 0.0004\% &   dLW \\
J01339$-$176 & ~ & ~ & no signal \\
J01433+043 & ~ & ~ & no signal \\
J01518+644 & ~ & ~ & no signal \\
J02002+130 &1.95 & $ < 10^{-6} $ &   H$\alpha$, CRX, dLW \\
J02002+130 &782.52 & $ < 10^{-6} $ &  planet \\
J02015+637 & ~ & ~ & no signal \\
J02070+496 & ~ & ~ & no signal \\
J02123+035 & ~ & ~ & no signal \\
J02222+478$^d$ &28.29 & 0.0048\% &   dLW \\
J02336+249 & ~ & ~ & no signal \\
J02358+202 & ~ & ~ & no signal \\
J02362+068 & ~ & ~ & no signal \\
J02442+255 & ~ & ~ & no signal \\
J02530+168 &11.41 & $ < 10^{-6} $  &  planet \\
J02530+168 &174.09 & $ < 10^{-6} $ &   CRX \\
J02530+168 &4.91 & $ < 10^{-6} $  &  planet \\
J02565+554W & ~ & ~ & no signal \\
J03133+047 &2.29 & $ < 10^{-6} $  &  planet \\
J03133+047 &67.52 & 0.2525\%  &  rotation \\
J03181+382 & ~ & ~ & no signal \\
J03213+799 & ~ & ~ & no signal \\
J03217$-$066 & ~ & ~ & no signal \\
J03463+262 & ~ & ~ & no signal \\
J03531+625 & ~ & ~ & no signal \\
J04153$-$076 &1.8 & $ < 10^{-6} $ &   CRX \\
J04225+105 & ~ & ~ & no signal \\
J04290+219 &12.54 & 0.0036\% &  rotation \\
J04290+219 &170.18 & 0.0489\% &  CRX \\
J04290+219 &24.99 & 0.0326\% &   H$\alpha$, dLW \\
J04376$-$110 & ~ & ~ & no signal \\
J04376+528 &16.32 & 0.3865\% &   H$\alpha$, dLW \\
J04376+528 &419.62 & 0.7739\% &   CaIRT \\
J04376+528 &7.9 & 0.3618\% &  unsolved \\
J04429+189$^c$ & ~ & ~ & no signal \\
J04429+214 & ~ & ~ & no signal \\
J04520+064$^b$ &10582.5 & 0.1972\% &  $P$ > $\frac{2}{3}$ time baseline \\
J04538$-$177$^a$ $^c$ & ~ & ~ & no signal \\
J04588+498 &8.89 & 0.0140\% &  unsolved \\
J05033$-$173 &1.92 & 0.0015\% &  candidate \\
J05033$-$173 &73.78 & 0.4055\% &  unsolved \\
J05127+196 & ~ & ~ & no signal \\
J05280+096 & ~ & ~ & no signal \\
J05314$-$036 &1362.47 & $ < 10^{-6} $ &  $P$ > $\frac{2}{3}$ time baseline \\
J05314$-$036 &34.09 & 0.0022\% &   H$\alpha$ \\
J05348+138 & ~ & ~ & no signal \\
J05360$-$076 & ~ & ~ & no signal \\
J05365+113 &11.76 & $ < 10^{-6} $ &   H$\alpha$, dLW \\
J05365+113 &12.45 & 0.0043\% &   H$\alpha$, dLW \\
\noalign{{\smallskip}}
\hline
\end{tabular}
\end{table}

\addtocounter{table}{-1}
\begin{table}[ht]
\centering
\caption{Continued.}
\begin{tabular}{@{}lccl@{}}
\hline\hline
\noalign{\smallskip}
Karmn & $P$ (d) & FAP & Remark \\
\hline
\noalign{{\smallskip}}
J05365+113 &6.31 & 0.0836\% &  activity \\
J05366+112 & ~ & ~ & no signal \\
J05415+534 &9.03 & 0.0070\% &   CaIRT \\
J05421+124 & ~ & ~ & no signal \\
J06011+595 &44.0 & 0.0268\% &  dLW \\
J06011+595 &82.97 & 0.0821\% &   dLW \\
J06024+498 & ~ & ~ & no signal \\
J06103+821 &409.8 & 0.0081\% &   CaIRT \\
J06105$-$218 &2621.33 & 0.0001\% &  $P$ > $\frac{2}{3}$ time baseline \\
J06371+175$^a$ & ~ & ~ & no signal \\
J06421+035 & ~ & ~ & no signal \\
J06548+332 &120.37 & $ < 10^{-6} $  &  rotation \\
J06548+332 &14.24 & $ < 10^{-6} $  &  planet \\
J06548+332 &53.65 & 0.0024\%  &  rotation \\
J06594+193 & ~ & ~ & no signal \\
J07033+346 & ~ & ~ & no signal \\
J07044+682 & ~ & ~ & no signal \\
J07274+052 &19.62 & 0.0813\% &  planet \\
J07274+052 &461.91 & $ < 10^{-6} $ &  unsolved \\
J07274+052 &95.0 & $ < 10^{-6} $ &  rotation \\
J07287$-$032 &616.27 & 0.5847\% &  unsolved \\
J07319+362N &4.78 & 0.6855\% &  unsolved \\
J07393+021 &29.78 & 0.0121\% &  activity \\
J07403$-$174 &6.61 & 0.1475\% &  unsolved \\
J07582+413 & ~ & ~ & no signal \\
J08119+087 & ~ & ~ & no signal \\
J08126$-$215 & ~ & ~ & no signal \\
J08161+013 &22.45 & 0.1773\% &  rotation \\
J08293+039 & ~ & ~ & no signal \\
J08315+730 & ~ & ~ & no signal \\
J08358+680 &1.73 & 0.1084\% &  unsolved \\
J08409$-$234 &710.89 & $ < 10^{-6} $ &  planet \\
J08413+594 &10118.37 & $ < 10^{-6} $ &  $P$ > $\frac{2}{3}$ time baseline \\
J08413+594 &203.24 & $ < 10^{-6} $  &  planet \\
J08413+594 &2204.2 & $ < 10^{-6} $ &  $P$ > $\frac{2}{3}$ time baseline \\
J08526+283 & ~ & ~ & no signal \\
J09028+680 & ~ & ~ & no signal \\
J09143+526 &1210.67 & 0.1987\% &  $P$ > $\frac{2}{3}$ time baseline \\
J09143+526 &16.28 & $ < 10^{-6} $ &   H$\alpha$, dLW \\
J09144+526 &16.66 & $ < 10^{-6} $ &   H$\alpha$, dLW \\
J09144+526 &24.43 & 0.0003\%  &  planet \\
J09144+526 &3971.21 & $ < 10^{-6} $ &  $P$ > $\frac{2}{3}$ time baseline \\
J09307+003 &294.04 & 0.5392\% &  unsolved \\
J09360$-$216$^c$ & ~ & ~ & no signal \\
J09411+132 & ~ & ~ & no signal \\
J09423+559 & ~ & ~ & no signal \\
J09425+700 &677.39 & 0.7290\% &  unsolved \\
J09428+700 &2.48 & 0.0303\% &   H$\alpha$ \\
J09447$-$182 & ~ & ~ & no signal \\
J09468+760 & ~ & ~ & no signal \\
J09511$-$123 & ~ & ~ & no signal \\
J09561+627$^d$ &18.69 & $ < 10^{-6} $ &   H$\alpha$, dLW \\
J10023+480 &3.82 & 0.7102\% &  planet \\
J10122$-$037 &10.66 & 0.0008\% &  rotation \\
J10122$-$037 &21.43 & 0.0048\% &   H$\alpha$ \\
J10167$-$119 & ~ & ~ & no signal \\
J10196+198 &2.24 & $ < 10^{-6} $ &   CRX \\
J10251$-$102 & ~ & ~ & no signal \\
J10289+008$^a$ &317.21 & 0.0051\% &  unsolved \\
J10350$-$094 & ~ & ~ & no signal \\
J10482$-$113 &1.52 & 0.0415\% &   dLW \\
J10482$-$113 &2.93 & 0.1639\% &  rotation \\
J10508+068 & ~ & ~ & no signal \\
J10564+070$^d$ &2.7 & $ < 10^{-6} $ &   CRX, dLW \\
J10584$-$107 &1.27 & $ < 10^{-6} $ &   CRX \\
\noalign{{\smallskip}}
\hline
\end{tabular}
\end{table}

\addtocounter{table}{-1}
\begin{table}[ht]
\centering
\caption{Continued.}
\begin{tabular}{@{}lccl@{}}
\hline\hline
\noalign{\smallskip}
Karmn & $P$ (d) & FAP & Remark \\
\hline
\noalign{{\smallskip}}
J11000+228 & ~ & ~ & no signal \\
J11026+219 &13.76 & 0.0030\% &   CRX \\
J11026+219 &13.93 & 0.2527\% &   CRX \\
J11033+359 &12.94 & $ < 10^{-6} $  &  planet \\
J11033+359 &2017.54 & $ < 10^{-6} $ &  $P$ > $\frac{2}{3}$ time baseline \\
J11054+435 &1026.37 & 0.0001\% &  $P$ > $\frac{2}{3}$ time baseline \\
J11055+435 & ~ & ~ & no signal \\
J11110+304W & ~ & ~ & no signal \\
J11126+189 & ~ & ~ & no signal \\
J11306$-$080 & ~ & ~ & no signal \\
J11417+427 &45.9 & $ < 10^{-6} $  &  planet \\
J11417+427 &639.31 & $ < 10^{-6} $ &   H$\alpha$ \\
J11417+427 &835.19 & 0.0736\% &  $P$ > $\frac{2}{3}$ time baseline \\
J11421+267 &2.64 & $ < 10^{-6} $  &  planet \\
J11467$-$140 & ~ & ~ & no signal \\
J11476+786 & ~ & ~ & no signal \\
J11477+008$^a$ & ~ & ~ & no signal \\
J11509+483 &316.73 & $ < 10^{-6} $ &  rotation \\
J11511+352 &11.12 & 0.0014\% &  rotation \\
J12100$-$150 & ~ & ~ & no signal \\
J12111$-$199 & ~ & ~ & no signal \\
J12123+544S &13.67 & $ < 10^{-6} $  &  planet \\
J12230+640 &15.13 & 0.0005\% &   CRX \\
J12230+640 &73.06 & 0.0586\% &   CRX \\
J12230+640 &8254.83 & $ < 10^{-6} $ &  $P$ > $\frac{2}{3}$ time baseline \\
J12248$-$182 & ~ & ~ & no signal \\
J12312+086 & ~ & ~ & no signal \\
J12373$-$208 & ~ & ~ & no signal \\
J12479+097 &1.42 & $ < 10^{-6} $  &  planet \\
J12479+097 &650.94 & 0.3179\%  &  unsolved \\
J13102+477 & ~ & ~ & no signal \\
J13209+342 &5962.88 & $ < 10^{-6} $ &  $P$ > $\frac{2}{3}$ time baseline \\
J13229+244 &3.02 & $ < 10^{-6} $  &  planet \\
J13229+244 &87.79 & 0.0015\% &   H$\alpha$, CRX, dLW \\
J13299+102$^a$ &15.42 & 0.3861\% &  rotation \\
J13299+102$^a$ &15.83 & 0.0023\% &  rotation \\
J13299+102$^a$ &512.48 & 0.0589\% &  unsolved \\
J13427+332 & ~ & ~ & no signal \\
J13450+176 & ~ & ~ & no signal \\
J13457+148 &105.98 & 0.0074\% &  rotation \\
J13457+148 &306.51 & 0.0020\% &  unsolved \\
J13457+148 &33.89 & 0.0007\% &  unsolved \\
J13458$-$179 & ~ & ~ & no signal \\
J13582+125 & ~ & ~ & no signal \\
J14010$-$026$^c$ & ~ & ~ & no signal \\
J14082+805 & ~ & ~ & no signal \\
J14251+518 & ~ & ~ & no signal \\
J14257+236E & ~ & ~ & no signal \\
J14257+236W &10002.44 & 0.2462\% &  $P$ > $\frac{2}{3}$ time baseline \\
J14307$-$086 &252.02 & 0.2434\% &  unsolved \\
J14342$-$125 &113.37 & $ < 10^{-6} $ &   CaIRT \\
J14342$-$125 &36.11 & 0.0015\% &  planet \\
J14524+123 &26.71 & 0.0010\% &  rotation? \\
J14544+355 & ~ & ~ & no signal \\
J15013+055 & ~ & ~ & no signal \\
J15095+031 & ~ & ~ & no signal \\
J15194$-$077 &12.92 & 0.0148\%  &  planet \\
J15194$-$077 &5.37 & $ < 10^{-6} $  &  planet \\
J15598$-$082 & ~ & ~ & no signal \\
J16028+205 & ~ & ~ & no signal \\
J16167+672N &42.22 & 0.0208\% &  activity \\
J16167+672S &21.99 & 0.0641\% &   rotation \\
J16167+672S &365.13 & $ < 10^{-6} $ &   CaIRT, CRX \\
J16167+672S &86.43 & $ < 10^{-6} $ &   planet \\
J16254+543$^a$ & ~ & ~ & no signal \\
\noalign{{\smallskip}}
\hline
\end{tabular}
\end{table}

\addtocounter{table}{-1}
\begin{table}[ht]
\centering
\caption{Continued.}
\begin{tabular}{@{}lccl@{}}
\hline\hline
\noalign{\smallskip}
Karmn & $P$ (d) & FAP & Remark \\
\hline
\noalign{{\smallskip}}
J16303$-$126 &1.26 & 0.3514\% &   planet \\
J16303$-$126 &17.88 & $ < 10^{-6} $  &  planet \\
J16303$-$126 &1.84 & 0.3134\%  &  unsolved \\
J16327+126 & ~ & ~ & no signal \\
J16462+164 & ~ & ~ & no signal \\
J16554$-$083N & ~ & ~ & no signal \\
J16581+257 &12.42 & 0.0500\% &  rotation \\
J16581+257$^c$ &661.35 & 0.0033\% &  $P$ > $\frac{2}{3}$ time baseline \\
J17033+514 &6.94 & 0.0752\% &  candidate \\
J17052$-$050 & ~ & ~ & no signal \\
J17071+215 & ~ & ~ & no signal \\
J17115+384 &5.58 & 0.8902\% &  unsolved \\
J17166+080 & ~ & ~ & no signal \\
J17198+417 &23.13 & 0.3605\% &  unsolved \\
J17303+055 & ~ & ~ & no signal \\
J17355+616$^c$ & ~ & ~ & no signal \\
J17364+683 &38.56 & $ < 10^{-6} $ &   planet \\
J17378+185 &15.52 & 0.0001\%  &  planet \\
J17378+185 &40.28 & 0.4268\% &   H$\alpha$, dLW \\
J17378+185 &499.08 & 0.0479\%  &  planet \\
J17542+073 &1.55 & 0.2591\% &  activity \\
J17578+046$^a$ &287.23 & 0.0056\% &  rotation \\
J17578+046$^a$ &387.44 & 0.2243\% &   CaIRT, H$\alpha$, dLW \\
J17578+046$^a$ &652.12 & 0.0225\% &  unsolved \\
J17578+465 & ~ & ~ & no signal \\
J18027+375 & ~ & ~ & no signal \\
J18051$-$030 & ~ & ~ & no signal \\
J18075$-$159 & ~ & ~ & no signal \\
J18165+048 &23239.74 & 0.0095\% &  $P$ > $\frac{2}{3}$ time baseline \\
J18174+483 &16.04 & 0.0332\% &  activity \\
J18174+483 &7.96 & 0.4972\% &   CRX \\
J18180+387E & ~ & ~ & no signal \\
J18221+063 & ~ & ~ & no signal \\
J18224+620 &2.06 & 0.7429\% &   CaIRT \\
J18319+406 &2.93 & 0.2034\% &  unsolved \\
J18346+401 &5786.99 & $ < 10^{-6} $ &  $P$ > $\frac{2}{3}$ time baseline \\
J18353+457$^c$ &2.62 & 0.0842\% &   H$\alpha$ \\
J18363+136 &8189.81 & 0.9454\% &  $P$ > $\frac{2}{3}$ time baseline \\
J18409$-$133 &5.1 & 0.2652\% &  candidate \\
J18419+318 & ~ & ~ & no signal \\
J18427+596N &11.2 & 0.0018\% &  unsolved \\
J18427+596N &27647.04 & $ < 10^{-6} $ &  $P$ > $\frac{2}{3}$ time baseline \\
J18427+596N &64.8 & 0.0871\% &  unsolved \\
J18427+596S &10000.0 & $ < 10^{-6} $ &  unsolved \\
J18427+596S &117.9 & 0.0053\% &  unsolved \\
J18427+596S &186.6 & 0.0209\% &  unsolved \\
J18480$-$145 & ~ & ~ & no signal \\
J18482+076 &1.4 & 0.8063\% &  rotation \\
J18498$-$238 &1.43 & 0.0294\% &   CRX, dLW \\
J18498$-$238 &2.89 & $ < 10^{-6} $ &   dLW \\
J18580+059$^c$ & ~ & ~ & no signal \\
J19070+208 & ~ & ~ & no signal \\
J19072+208 & ~ & ~ & no signal \\
J19084+322 & ~ & ~ & no signal \\
J19098+176 & ~ & ~ & no signal \\
J19169+051N$^e$ &132.39 & $ < 10^{-6} $ &   CRX \\
J19216+208 & ~ & ~ & no signal \\
J19251+283 & ~ & ~ & no signal \\
J19346+045 &2.52 & 0.6353\% &  unsolved \\
J20260+585 &16.21 & 0.0080\% &  planet \\
J20260+585 &14.95 & 0.5846\% &  unsolved \\
J20260+585 &403.71 & $ < 10^{-6} $ &  tellurics \\
J20305+654 & ~ & ~ & no signal \\
J20336+617 &175.74 & 0.6727\% &   CRX \\
J20405+154 &153.66 & 0.7230\% &  unsolved \\
\noalign{{\smallskip}}
\hline
\end{tabular}
\end{table}

\addtocounter{table}{-1}
\begin{table}[ht]
\centering
\caption{Continued.}
\begin{tabular}{@{}lccl@{}}
\hline\hline
\noalign{\smallskip}
Karmn & $P$ (d) & FAP & Remark \\
\hline
\noalign{{\smallskip}}
J20450+444 & ~ & ~ & no signal \\
J20525$-$169 & ~ & ~ & no signal \\
J20533+621 &118.6 & 0.7191\% &   CaIRT, CRX \\
J20556$-$140S & ~ & ~ & no signal \\
J20567$-$104 &1.25 & 0.1960\% &  unsolved \\
J21019$-$063 & ~ & ~ & no signal \\
J21152+257 & ~ & ~ & no signal \\
J21164+025 &14.46 & $ < 10^{-6} $  &  planet \\
J21164+025 &43.71 & 0.0006\% &   H$\alpha$ \\
J21221+229 &39.91 & 0.0001\% &   H$\alpha$, dLW \\
J21221+229 &686.81 & $ < 10^{-6} $ &  planet \\
J21348+515 &26.33 & 0.7167\% &  rotation \\
J21463+382 & ~ & ~ & no signal \\
J21466$-$001 & ~ & ~ & no signal \\
J21466+668 &2.31 & $ < 10^{-6} $  &  planet \\
J21466+668 &8.05 & 0.0001\%  &  planet \\
J21466+668 &92.49 & $ < 10^{-6} $  &  rotation \\
J22021+014 &10.95 & 0.0373\% &   H$\alpha$ \\
J22057+656 &122.52 & $ < 10^{-6} $ &   CRX \\
J22096$-$046$^b$ &3998.46 & $ < 10^{-6} $ &  $P$ > $\frac{2}{3}$ time baseline \\
J22114+409 &15.22 & 0.5957\% &  rotation \\
J22115+184 &374.96 & $ < 10^{-6} $ &   CRX \\
J22115+184 &39.02 & 0.0015\% &   dLW \\
J22125+085 &2911.44 & $ < 10^{-6} $ &  $P$ > $\frac{2}{3}$ time baseline \\
J22137$-$176 &3.65 & $ < 10^{-6} $  &  planet \\
J22137$-$176 &588.11 & $ < 10^{-6} $ &  $P$ > $\frac{2}{3}$ time baseline \\
J22231$-$176 & ~ & ~ & no signal \\
J22252+594 &13.35 & $ < 10^{-6} $  &  planet \\
J22298+414 &5.78 & 0.0280\% &   dLW \\
J22330+093 &34.06 & 0.6282\% &  unsolved \\
J22503$-$070 & ~ & ~ & no signal \\
J22532$-$142 &30.07 & $ < 10^{-6} $  &  planet \\
J22532$-$142 &60.85 & $ < 10^{-6} $  &  planet \\
J22559+178 & ~ & ~ & no signal \\
J22565+165 &18.61 & $ < 10^{-6} $ &  rotation \\
J22565+165 &1845.72 & $ < 10^{-6} $ &  $P$ > $\frac{2}{3}$ time baseline \\
J22565+165 &39.28 & $ < 10^{-6} $ &   H$\alpha$, dLW \\
J23216+172 & ~ & ~ & no signal \\
J23245+578 & ~ & ~ & no signal \\
J23340+001 & ~ & ~ & no signal \\
J23351$-$023 & ~ & ~ & no signal \\
J23381$-$162 & ~ & ~ & no signal \\
J23419+441 &175.09 & 0.0001\% &  unsolved \\
J23431+365 & ~ & ~ & no signal \\
J23492+024 &372.77 & 0.0001\% &   CaIRT \\
J23492+024 &53.83 & 0.4878\% &   H$\alpha$ \\
J23505$-$095 & ~ & ~ & no signal \\
\noalign{{\smallskip}}
\hline
\end{tabular}
\tablefoot{
\tablefoottext{a}{Targets with known planets below detection limit.}
\tablefoottext{b}{Targets with long period planets.}
\tablefoottext{c}{Targets with known planets that fall below the CARMENES detection limit but that are included in the analysis anyway (semi-amplitude $K > 2$\,m\,s$^{-1}$ and $N_\text{AVC}<50$).}
\tablefoottext{d}{Targets with claimed planets that are not supported by CARMENES data.}
\tablefoottext{e}{Known planet host but at a different period.}
}
\end{table}

\end{appendix}

\end{document}